\documentclass[fleqn,usenatbib]{mnras}

\usepackage{newtxtext,newtxmath}
\usepackage[para,online,flushleft]{threeparttable}
\usepackage{caption}
\usepackage{subcaption}
\usepackage{adjustbox}
\usepackage{rotating}

\def\gs{\mathrel{\raise0.35ex\hbox{$\scriptstyle >$}\kern-0.6em \lower0.40ex\hbox{{$\scriptstyle \sim$}}}}
\def\ls{\mathrel{\raise0.35ex\hbox{$\scriptstyle <$}\kern-0.6em \lower0.40ex\hbox{{$\scriptstyle \sim$}}}}
\newcommand{\orcid}[1]{\textsuperscript{\,\,\href{https://orcid.org/#1}{\includegraphics[scale=0.06]{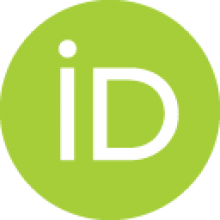}}}}

\usepackage[T1]{fontenc}

\DeclareRobustCommand{\VAN}[3]{#2}
\let\VANthebibliography\thebibliography
\def\thebibliography{\DeclareRobustCommand{\VAN}[3]{##3}\VANthebibliography}

\usepackage{graphicx}	
\usepackage{amsmath}	

\title[{\it DEIMOS} spectroscopy of protocluster candidate]{{\it DEIMOS} spectroscopy of $z=6$ protocluster candidate in COSMOS\\ -- A massive protocluster embedded in a large scale structure?}

\author[M. Brinch et al.]{
Malte Brinch,\orcid{0000-0002-0245-6365}$^{1,2}$\thanks{E-mail: malbr@space.dtu.dk}
Thomas R. Greve,\orcid{0000-0002-2554-1837}$^{1,2,3}$
David B. Sanders,\orcid{0000-0002-1233-9998}$^{4}$
Conor J. R. McPartland,\orcid{0000-0003-0639-025X}$^{1,4,5,6}$
\and
Nima Chartab,\orcid{0000-0003-3691-937X}$^{7}$
Steven Gillman,\orcid{0000-0001-9885-4589}$^{1,2}$
Aswin P. Vijayan,\orcid{0000-0002-1905-4194}$^{1,2}$
Minju M. Lee,\orcid{0000-0002-2419-3068}$^{1,2}$
Gabriel Brammer,\orcid{0000-0003-2680-005X}$^{1,5}$
\and
Caitlin M. Casey,\orcid{0000-0002-0930-6466}$^{8}$
Olivier Ilbert,\orcid{0000-0002-7303-4397}$^{9}$
Shuowen Jin,\orcid{0000-0002-8412-7951}$^{1,2}$
Georgios Magdis,\orcid{0000-0002-4872-2294}$^{1,2,5,10}$
H. J. McCracken,\orcid{0000-0002-9489-7765}$^{11}$
\and
Nikolaj B. Sillassen,\orcid{0000-0002-4517-3998}$^{1,2}$
Sune Toft,\orcid{0000-0003-3631-7176}$^{1,5}$
Jorge A. Zavala,\orcid{0000-0002-7051-1100}$^{12}$
\\\\
$^{1}$Cosmic Dawn Center (DAWN)\\
$^{2}$DTU-Space, National Space Institute, Technical University of Denmark, Elektrovej 327, 2800 Kgs.~Lyngby, Denmark\\
$^{3}$ Department of Physics and Astronomy, University College London, Gower Street, London WC1E 6BT, United Kingdom\\
$^{4}$Institute for Astronomy, University of Hawaii, 2680 Woodlawn Drive, Honolulu, HI 96822, USA\\
$^{5}$Niels Bohr Institute, University of Copenhagen, Jagtvej 128, 2200 Copenhagen, Denmark\\
$^{6}$Physics and Astronomy Department, University of California, 900 University Ave., Riverside, CA 92521, USA\\
$^{7}$The Observatories of the Carnegie Institution for Science, 813 Santa Barbara St., Pasadena, CA 91101, USA\\
$^{8}$Department of Astronomy, The University of Texas at Austin, Austin, TX, USA\\
$^{9}$Aix Marseille Univ, CNRS, CNES, LAM, Marseille, France\\
$^{10}$Institute for Astronomy, Astrophysics, Space Applications and Remote Sensing, National Observatory of Athens, 15236 Athens, Greece\\
$^{11}$Institut d’Astrophysique de Paris, UMR 7095, CNRS, and Sorbonne Universit´e, 98 bis boulevard Arago, 75014 Paris, France\\
$^{12}$National Astronomical Observatory of Japan, 2-21-1 Osawa, Mitaka, Tokyo 181-8588, Japan\\
}

\date{Accepted XXX. Received YYY; in original form ZZZ}

\pubyear{2023}

\begin{document}
\label{firstpage}
\pagerange{\pageref{firstpage}--\pageref{lastpage}}

\maketitle

\begin{abstract}
We present the results of our {\it Keck}/{\it DEIMOS} spectroscopic follow-up of candidate galaxies of {\sc i}-band-dropout protocluster candidate galaxies at $z\sim6$ in the COSMOS field. We securely detect Lyman-$\alpha$ emission lines in 14 of the 30 objects targeted, 10 of them being at $z=6$ with a signal-to-noise ratio of $5-20$, the remaining galaxies are either non-detections or interlopers with redshift too different from $z=6$ to be part of the protocluster. The 10 galaxies at $z\approx6$ make the protocluster one of the riches at $z>5$. The emission lines exhibit asymmetric profiles with high skewness values ranging from 2.87 to 31.75, with a median of 7.37. This asymmetry is consistent with them being Ly$\alpha$, resulting in a redshift range of $z=5.85-6.08$. Using the spectroscopic redshifts, we re-calculate the overdensity map for the COSMOS field and find the galaxies to be in a significant overdensity at the $4\sigma$ level, with a peak overdensity of $\delta=11.8$ (compared to the previous value of $\delta=9.2$). The protocluster galaxies have stellar masses derived from {\sc Bagpipes} SED fits of $10^{8.29}-10^{10.28} \rm \,M_{\rm \odot}$ and star formation rates of $2-39\,\rm M_{\rm \odot}\rm\,yr^{-1}$, placing them on the main sequence at this epoch. Using a stellar-to-halo-mass relationship, we estimate the dark matter halo mass of the most massive halo in the protocluster to be $\sim 10^{12}\rm M_{\rm \odot}$. By comparison with halo mass evolution tracks from simulations, the protocluster is expected to evolve into a Virgo- or Coma-like cluster in the present day.
\end{abstract}




\section{Introduction} \label{sec:intro}
Galaxy clusters are the most massive gravitationally-bound virialised objects in the Universe. While star formation activity and stellar mass build-up in clusters must have peaked at earlier times ($z \gs 2$) \citep[following the results of][]{Poggianti1999,Elbaz2007,Cooper2008,Scoville2013}, we do not know how and when such protoclusters appeared, nor how they affected the formation and evolution of galaxies within them \citep[e.g.,][]{Overzier2016}. Studies of present-day galaxy clusters have had some success in addressing these questions, but are limited by the fact that key signatures of a cluster’s formation history are erased in its final stage of assembly as it undergoes dynamical relaxation and virialization \citep{Zabludoff1996,Kodama2001}.

To better understand the emergence of clusters, we need to study their progenitors (i.e., protoclusters) in the distant Universe, where we can directly observe them during their formative stages. Observations must be pushed back to the Cosmic Dawn era ($z \gs 6$), when the Universe was less than a billion years old, since according to state-of-the-art simulations of cosmic structure formation \citep[e.g.,][]{Chiang2013,Chiang2017,Lovell2021}, protoclusters were assembled from the peaks of dark matter density distribution during the Cosmic Dawn. Simulations suggest that protoclusters form in an ”inside-out” manner, where the central galaxies sitting in more massive dark matter halos form first, and then the outer parts form later. Consequently, we expect the central galaxies to be more massive and to have an older stellar population and higher gas-phase metallicity than the outer galaxies. 

Overdensities of galaxies at $z \sim 6$ and beyond have been observed in increasingly large numbers in the last few years \citep[e.g.,][]{Harikane2019,Brinch2023}. While such galaxy overdensities are possible protoclusters that will eventually evolve into today's massive galaxy clusters, unambiguous spectroscopic confirmation is essential to establishing these overdensities as genuine protoclusters. It has been demonstrated that a couple of nights (8-16 hours) of deep spectroscopy on $8-10\,{\rm m}$-class telescopes can confirm the spectroscopic redshift of galaxies out to $z\sim7$ \citep{Stanway2004,Stark2011}.
To date, however, little more than a handful of spectroscopically confirmed protoclusters at $z \gs 6$ have been found 
\citep[e.g.,][]{Toshikawa2012, Toshikawa2014, Chanchaiworawit2019, Harikane2019, Hu2021,Endsley2021,Laporte2022,Helton2023,Scholtz2023,Tacchella2023}, and very few of these have the deep multi-band optical/near-IR (OIR) data required to accurately characterize their galaxy populations (e.g., stellar masses, star-formation rates, and ages).
Properly constraining the stellar mass of the protocluster galaxies is essential, as it directly ties into the estimation of the dark matter halo mass of the protocluster, which is one of the key pieces of evidence pointing towards these structures evolving into present-day clusters \citep[e.g.,][]{Long2020,Shuntov2022,Brinch2023}. With the launch of the James Webb Space Telescope (JWST) and large-scale surveys like COSMOS-WEB \citep{Casey2022}, it is possible to search for overdensities at even higher redshifts and with the possibility of spectroscopic follow-up with instruments like NIRSpec \citep[see][]{Laporte2022,Helton2023}.

In this paper, we present spectroscopic follow-up observations with {\it Keck} of a protocluster candidate at $z=6$, first discovered by \cite{Brinch2023}. 
We present spectroscopic confirmation of 10 protocluster galaxies, which allows us to unambiguously confirm that the structure is one of the most massive and galaxy-rich protoclusters known in the Cosmic Dawn epoch. Moreover, it has a well-characterised galaxy population due to unprecedented OIR multi-wavelength data.

The paper is organised as follows. Section \ref{section:data} discusses how the protocluster candidate galaxies were selected and followed up with the DEep Imaging Multi-Object Spectrograph ({\it DEIMOS}) instrument on the {\it Keck} telescope. We further describe the data reduction of the 1-D and 2-D spectra in section \ref{section:data}. Section \ref{section:results} presents the emission line fitting routine and characterisation of the galaxy spectra, how the quality of each spectrum was graded and the distribution of parameters from the line fit. Section \ref{section:discussion} discusses the implications of the protocluster in the wider context of protocluster and galaxy evolution, showing an updated overdensity map updated with new weights given the spectroscopic redshifts and updated estimates for parameters such as stellar mass and star formation rate. Section \ref{section:conclusions} presents our conclusions.
\\\\
Throughout the paper, we have adopted a standard $\Lambda$CDM cosmology
with $H_{\rm 0}=70\,{\rm km\,s^{-1}\,Mpc^{-1}}$, $\Omega_{\rm m} = 0.3$, 
and $\Omega_{\rm \Lambda} = 0.70$. All magnitudes are expressed in the
AB system \citep{Oke1974}. A \citet{Chabrier2003} stellar Initial Mass
Function (IMF) is used to present our results. The results are reported with 68\% confidence interval uncertainties.
\section{Target selection, Observations and Data Reduction}\label{section:data}
\subsection{Target selection}
The photometric selection of candidate sources was made based on the results of the overdensity analysis done in the COSMOS field \citep{Brinch2023} using the COSMOS2020 catalogue \citep{Weaver2021}, with {\sc Lephare} \citep{Ilbert2006} photometric redshifts. We identified a significantly overdense ($\delta_{\rm peak} = 9.2$, where $\delta=\frac{\Sigma-\langle \Sigma\rangle}{\langle \Sigma \rangle}$ and $\Sigma$ is the galaxy surface density) and galaxy-rich (19 galaxies) protocluster candidate within a (photometric) redshift bin of $z=6.05\pm0.1$. The 19 galaxies that make up the overdensity, hereafter referred to as PC$z$6.05-1, were the main targets of our spectroscopic follow-up. {\it DEIMOS}, being a multiobject spectrometer, allows us to have up to 130 slitlets. To maximize the utility of {\it DEIMOS}, we included a number of extra sources in our sample.
\\\\
Our target selection is divided into three separate categories under the following criteria:\\\\
Priority 1: The 19 protocluster candidate galaxies found in \cite{Brinch2023}.
\\

\noindent Priority 2: Galaxies that are also in the $z=6.05\pm0.1$ redshift bin and within the {\it DEIMOS} footprint but outside the $4\sigma$ overdensity contour. These targets constitute five sources in total.
\\

\noindent Priority 3: Galaxies obtained by increasing the size of our redshift bin size to $\Delta z=\pm 0.5$ and by relaxing the sample selection criteria we had in \cite{Brinch2023}, letting the galaxies inside the bin have
their maximum p($z$) value deviate more than 5\% of their median value. These targets constitute six sources in total. 
\\\\
On top of this, some slits have multiple sources on them, we report these extra serendipitous sources in appendix \ref{sec:other objects}. There were 8 extra sources found on our slits.
In total, we have 30 sources across our three priorities.
For the priority 1 sources, 14 of them are classified as i-dropouts according to the criteria in \cite{Ono2018}, with Ks = $25.08-28.08$ mag. The remaining slitlets were used to target other lower redshift galaxies.
The many filters that have been used to observe the COSMOS field allow for well-characterised spectral energy distribution (SEDs) with well-constrained photometric redshifts, resulting in a high probability that the protocluster candidate galaxies are close to their median photometric redshift. 
\subsection{Observations}
The spectroscopic follow-up was performed using the DEep Imaging Multi-Object Spectrograph \citep[{\it DEIMOS};][]{Faber2003} at the Nasmyth focus of the $10\rm\, m$ {\it Keck} II telescope. Observations took place on 2022 February $6-7$ UT, with all our Priority $1-3$ targets observed.   
The seeing was in the range 0\arcsec.75 – 1\arcsec.00.
Two {\it DEIMOS} masks were used to observe the targets, one for each night, as seen in Table \ref{tab:SpecObs}. The first mask H250-1 used the OG550 filter and the 900 lines mm$^{-1}$
grating (900ZD), with a blaze wavelength of $5500\,$Å, tilted to place a central wavelength of $8000\,$Å on the detectors. This configuration provided a spectral coverage between $6193.8\,$Å and $9799.6\,$Å. The second mask H250-2 used the OG550 filter and the 600 lines mm$^{-1}$
grating (600ZD), with a blaze wavelength of $7500\,$Å, tilted to place a central wavelength of $7500\,$Å on the detectors. This configuration provided a spectral coverage between $5500.0\,$Å and $10118.9\,$Å. 
\\
The spatial pixel scale was 0\arcsec.1185 pixel$^{-1}$ (0.68 kpc pixel$^{-1}$ at z=6), with spectral dispersion of
$0.44\,$Å$\,$pixel$^{-1}$ ($16\,km\,s^{-1}$$\,$pixel$^{-1}$) for the 900 lines $\rm mm^{-1}$ and $0.65\,$Å$\,$pixel$^{-1}$ ($16\,km\,s^{-1}$$\,$pixel$^{-1}$) for the 600 lines $\rm mm^{-1}$ grating. Slit widths were 1\arcsec. The FWHM resolution for the 900ZD was $\simeq\,2.1\,$Å and $\simeq\,3.1\,$Å for the 600ZD grating ($74\,\rm km\,\rm s^{-1}$ and $109\,\rm km\,\rm s^{-1}$ respectively for a Lyman-$\alpha$ line at $z=6$). The specifics of the {\it DEIMOS} observations are detailed in Table \ref{tab:SpecObs}\footnote{More info about {\it DEIMOS} can also be found at \url{https://www2.keck.hawaii.edu/inst/DEIMOS/specs.html}.}.

\begin{table}
\caption{Summary of observation with {\it Keck}/{\it DEIMOS} in the COSMOS field. 
}
\label{tab:SpecObs}
\centering
\begin{threeparttable}
\resizebox{\columnwidth}{!}{%
\begin{tabular}{l l l l l l l l}
\hline
\hline
Mask ID & Date (UT) & Exp time & N\tnote{a} & Grating & $\lambda_{\rm central}$ & Filter & Seeing\tnote{b}\\ 
 &  & $[{\rm s}]$ &   & $[{\rm mm^{-1}}]$ & [Å]  &  & $[{\rm arcsec}]$\\ \hline
H250-1 & 2022 Feb 7 & 24000 & 11 & 900 & 8000 & OG550 & 1\\
H250-2 & 2022 Feb 8 & 25200 & 19 & 600 & 7500 & OG550 & 0.75\\
\hline
\end{tabular}
}
   \begin{tablenotes}
      \small
      \item[a] Number of observed targets.\\
      \item[b] Average seeing over the night.
    \end{tablenotes}
\end{threeparttable}
\end{table}

\subsection{Data Reduction}\label{subsection:data-reduction}
The {\it DEIMOS} slits are constructed as a blue-red pair between two detectors, with four pairs in total.
For the reduction of the {\it DEIMOS} spectra, {\ttfamily PypeIt}\footnote{\url{https://pypeit.readthedocs.io/en/latest/}} \citep{pypeit:zenodo,pypeit:joss_pub}, a \textsc{python} package for semi-automated reduction of astronomical slit-based spectroscopy was used. The individual observations of each 2-D slit were reduced by {\ttfamily PypeIt} (including coadding) and then fluxed calibrated to convert the units from counts to $10^{-17}\,{\rm erg/s/cm^{2}}$. Then each slit was visually inspected for possible sources. To flag sky regions, which can affect the detection of sources on the slit, we locate noise spikes in the 1-D spectrum that are equal to or larger than the FWHM resolution of {\it DEIMOS}, which are indicative of sky regions. Data in sky regions are masked when fitting potential emission lines, except in cases where there is a strong detection within a sky region that extends beyond it so that there is no ambiguity as to whether or not the detection is due to a sky region. The flagged sky regions were also checked against the skylines from \cite{Rousselot2000}, with vacuum line wavelengths computed from laboratory data and a spectra range $0.997$-$2.25\,\rm\mu$m with a resolution of $R\sim8000$.
Slits can have multiple detections that are not part of our original sample. To identify our main targets, as well as all other sources in the slit, we construct true-colour RGB images of the Y, z, i and K, H, J bands with the {\it DEIMOS} slit overplotted. We also compare this to galaxies in the COSMOS2020 catalogue to determine the counterpart of each source detected in the slits. The true color images are found next to each spectrum for comparison (see for example figure \ref{fig:A protocluster members}). 
The 1-D spectra were manually extracted by collapsing the spectra around the 2-D line observation. This was done to account for the multiple objects in the slits. The noise is calculated as the RMS of all the pixels in each column that are not part of the pixels used to create the signal for the 1-D spectra (i.e. away from the sky and emission lines). If there are multiple sources on the slit, they are masked when calculating the noise. 

\section{Results \& Analysis}\label{section:results}
With the spectroscopic data reduced, we now turn to fitting the Lyman-$\alpha$ lines to determine their characteristics and the galaxies' redshifts.
The Lyman-$\alpha$ emission line is associated with the hydrogen atom and is emitted when an electron transitions from an excited state ($n=2$) to the ground state ($n=1$), with a rest frame wavelength of $1216\,$Å. The emission line is used to observe distant galaxies due to its relatively high line strength, which comes from Star-forming galaxies that produce huge amounts of Lyman-$\alpha$ photons by recombination in H{\sc ii} regions that are ionised by young stars \citep{Tapken2007}. 
\subsection{Emission line fitting}
The Lyman-$\alpha$ line profile at $z\sim 6$ is expected to be observed as one of two profiles \citep{Childs2018,Calvi2019,Mason2020}. 
The shape of the Lyman-$\alpha$ line profile arises due to the path Lyman-$\alpha$ photons take to escape their local environment. As the photons travel they experience strong
resonant scattering from neutral hydrogen in the interstellar medium (ISM) and circumgalactic medium (CGM). The scattering through an optically thick cloud of HI gas means only the blue- and red-shifted emission is observed leading to a double-peaked Gaussian line profile, with the central wavelength being strongly absorbed and re-scattered.  It is expected that the blueward peak of the profile is lower than the redward one, due to the intervening Inter Galatic Medium (IGM) between us and the galaxy. It is possible for the blue part of the line to be preferentially absorbed, or for the red part to be boosted by
back-scattering from an expanding medium. The resulting line profile in that case is an asymmetric P-Cygni or more generally skewed Gaussian line profile.   
The skewed Gaussian is parameterised using the following formula,
\begin{equation}
    f(x;A,\mu,\sigma,\gamma)=\frac{A}{\sigma\sqrt{2\pi}}e^{[-(x-\mu)^2/\sigma^2]}\{1+\rm erf[\frac{\gamma(x-\mu)}{\sigma\sqrt{2}}]\},
\end{equation}
with erf() being the error function and the four parameters being the Amplitude (A), centre ($\mu$), sigma ($\sigma$) and skew ($\gamma$). We expect the skew from the fit to be positive due to the attenuation in the blue part of the line \citep{Calvi2019}. Note that in the case of zero skew, the skewed Gaussian becomes a standard Gaussian.  
We use the \texttt{LMfit} package \citep{Newville2014} to fit the emission lines, which uses a form of least-squares fitting.
To test whether there is significant continuum detection in conjunction with line detection, a linear baseline is fitted together with the emission line profile(s). For all the detections except two, the line fits have no continuum detection above the noise. The spectra were also visually inspected to check for any continuum detection between the Lyman $\alpha$ line and the supposed Lyman break given the redshift of each object.
\\\\
To accurately quantify the uncertainty on the emission-line fitting, we apply Monte Carlo (MC) sampling \citep{Kroese2011}, with N=1000 iterations\footnote{Higher iteration counts were tested and gave virtually identical results.}. The noise at each wavelength is used to sample Gaussian noise which is then added to each individual flux value before the lines are fit. This is done N times to obtain a distribution of line widths and fluxes. We report the median, upper and lower $1\sigma$ ($68\%$) confidence interval for the MC sampling results. 
\\\\
It is important to note for the skewed Lyman-$\alpha$ line profile that the wavelength obtained when fitting, and consequently the redshift, will be at the position of the lower wavelength edge (as marked by the blue lines in Figure \ref{fig:A protocluster members}) and not at the centre of the measured line flux. Since the line is asymmetrically attenuated, it is not possible to know how much of the line is attenuated and hence where the true wavelength centre of the un-attenuated line would be. This means that there is a possibility that the redshift of the galaxy is lower. The error reported in Table \ref{tab:specz} is from the line fitting and does not include inherent negative uncertainty due to the nature of the skewed Lyman-$\alpha$ line profile.
\subsection{Final Spectra}
To determine the quality of the final spectra as seen in Figures \ref{fig:A protocluster members}-\ref{fig:C protocluster members}, we adopt the – excellent (A), acceptable (B), marginal (C) and non-detection (D) categories, adapted from \cite{Calvi2019}.
The grading criteria were based on 
\begin{itemize}
\item[\textbf{(1)}] the shape of their spectra resembling either a P-Cygni or double-peaked Gaussian profile in the 1D spectra; 
\item[\textbf{(2)}] the sizes of Ly$\alpha$ emission features are about 1 arcsec ($\sim$ seeing of a spectroscopic night), but neither too extended nor too compact in 2D (to discern the real detection from the sky and cosmic ray residuals) 
\item[\textbf{(3)}] The locations of emission lines are fairly far away from strong sky emission lines or their wings. 
\end{itemize}
If a detection satisfies all of the three criteria, it is given the letter grade
‘A’. If a detection fails any one criterion, it is given grade
‘B’. If a detection fails two criteria it is given the grade ‘C’. If a detection fails all three criteria, or no detection is found on the slit, the source is labeled as a non-detection and is given grade ‘D’. For a line to be considered a detection, its peak signal-to-noise has to be above $3\sigma$. Every line is also checked to see whether it is larger than the FWHM resolution of the instrument and treated as a non-detection if this is not the case. In total we found 3 grade A (Figure \ref{fig:A protocluster members}), 7 grade B (Figure \ref{fig:B protocluster members}) and 2 grade C (Figure \ref{fig:C protocluster members}) sources at $z\approx6$. Sources that are interlopers are shown in appendix \ref{sec:interlopers}. 
We chose not to include the redshifts from grade C detections in our further analysis due to their uncertainty.
\subsection{Spectroscopic Protocluster membership}
\begin{figure}
    \centering
    \includegraphics[width=0.5\textwidth]{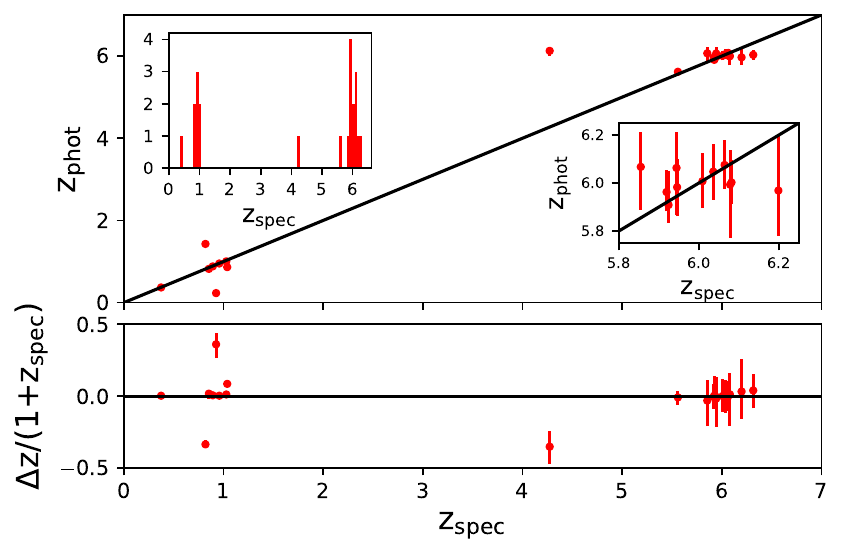}
    \caption{Comparison between the photometric redshift from {\sc LePhare} and the spectroscopic redshift from the {\it DEIMOS} campaign. The inset in the upper left panel shows the histogram of spectroscopic redshifts for all detections, while the lower right panel shows a zoom-in at $z\approx6$. The lower panel shows the residuals $\Delta z/(1+z_{\rm spec})$, where $\Delta z=z_{\rm spec}-z_{\rm phot}$.}
    \label{fig:zcompare}
\end{figure}

The 10 galaxies have a median redshift of $z=5.98$ and a $\Delta\,z=0.23$.
it should be noted that 2/10 galaxies  
are situated outside the main overdensity peak (see Figure \ref{fig:smoothing_kernel}). 
We address the size of $\Delta\,z$ in \S\ref{sec:PCsize}.
The remaining sources are either grade C detections or interlopers where the redshift deviates too much from $z=6$ 
to be part of the protocluster or non-detections.\\
Overall 14/30 (47\%) of the galaxies had a grade A/B detection.
Figure \ref{fig:zcompare} shows a comparison with the photometric redshift from {\sc LePhare}. We have chosen not to show the grade C detections in this plot, due to their uncertainty.
\subsubsection{Low-z objects}
The [O\textsc{iii}] lines at $4959\,$Å, $5007\,$Å, as well as the [O\textsc{ii}] doublet at $3927\,$Å, $3929\,$Å and H${\upbeta}$ 4861 Å line for galaxies at $z\sim1$ fall within the wavelength range probed by {\it DEIMOS}. The [OIII]5007Å line is expected to be the brightest, while the [O\textsc{iii}]4959Å line is expected to be $\sim 3\times$ fainter \citep{Storey2000}. Meanwhile the [O\textsc{ii}] doublet around $3928\,$Å can look similar to a double peaked Lyman-$\alpha$ profile. We would like to highlight a potential candidate for such a low-$z$ interloper in galaxy 70505 and explain why we believe it to be a genuine high-z detection. Firstly, if the two peaks that appear to be present were due to [O\textsc{ii}], we should expect their peaks to be separated by $\approx 3\,$Å rest frame, and the galaxies redshift would be $z=1.26$ meaning the peaks should be separated by $6.3\,$Å observed frame. Given that the separation is $5.7\,$Å, it leads us to conclude that is not the [O\textsc{ii}] doublet. Secondly, since one part of the line is over a skyline, there is a part of the 2D spectrum which is over-subtracted. This leads to a dip in the 1D spectrum when collapsed, which can make it appear as if the line is two lines. For these reasons, we have chosen to set the line as a Lyman-$\alpha$ detection and to extrapolate over the missing part of the line. It is possible that we are missing flux from the fit, though this should not affect the determination of the spectroscopic redshift.
Generally, it is important to keep these lines in mind when investigating other sources found within the slit. The spectroscopic redshifts of these low redshift galaxies are shown in appendix \ref{sec:other objects} and Figure \ref{fig:zcompare}.
\begin{figure*}
    \centering
\includegraphics[width=\textwidth, trim={0.1cm 0.1cm 0.1cm 0.1cm},clip]{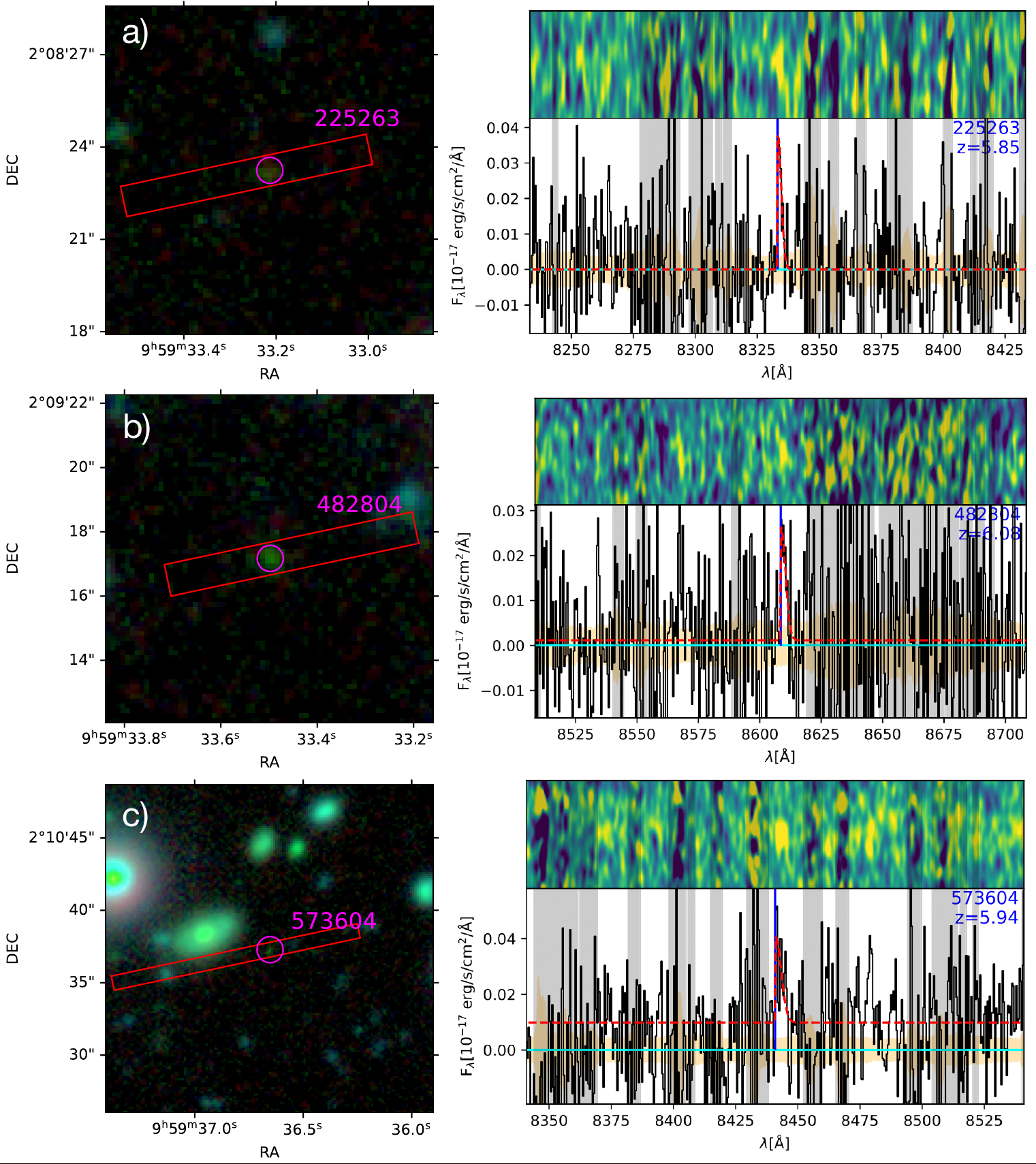}
    \caption{2D, 1D and Yzi RGB true-colour images of the protocluster members with grade A as seen in Table \ref{tab:specz}. The left side shows VISTA Yzi RBG true colour images using the COSMOS2020 catalogue photometry. The slit used to observe each object is plotted on top of the image and a magenta circle with an accompanying object ID is shown to highlight the position of the object in the image. The right side shows a two-panel figure of the spectrum in 2D (top), and 1D (bottom). A Gaussian filter has been applied to the 2D spectrum to visually highlight the detection. The black solid line is the extracted 1D spectrum. The vertical blue line indicated the spectroscopic redshift of the object. The shaded yellow area is the noise. The grey-shaded regions in all three panels highlight the sky regions, where skylines are prevalent.}
    \label{fig:A protocluster members}
\end{figure*}
\begin{figure*}
    \centering
    \includegraphics[width=0.9\textwidth, trim={0.1cm 0.1cm 0.1cm 0.1cm},clip]{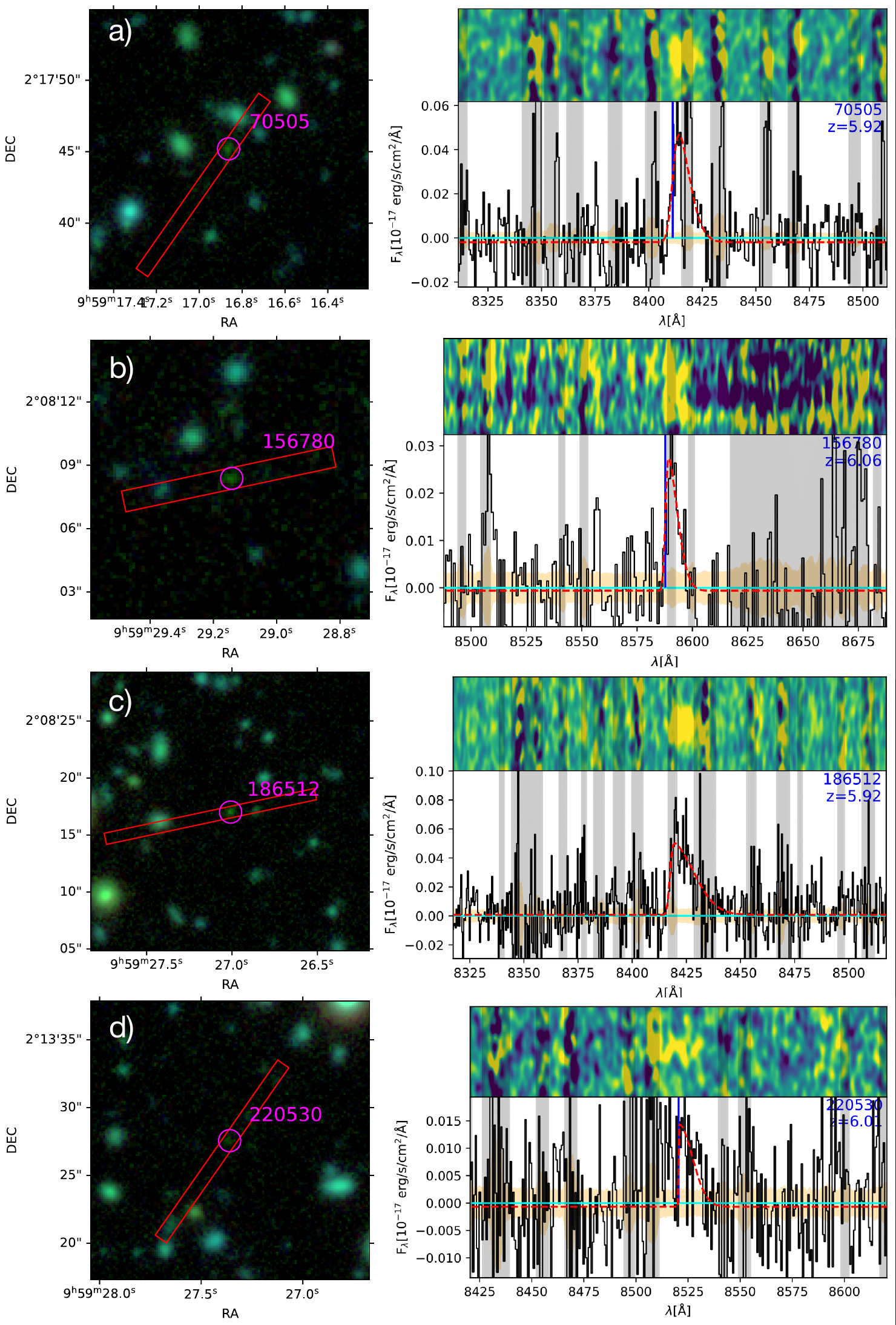}
    \caption{Continued on the next page.}
\end{figure*}
\begin{figure*}\ContinuedFloat
    \centering
    \includegraphics[width=\textwidth, trim={0.1cm 0.1cm 0.1cm 0.1cm},clip]{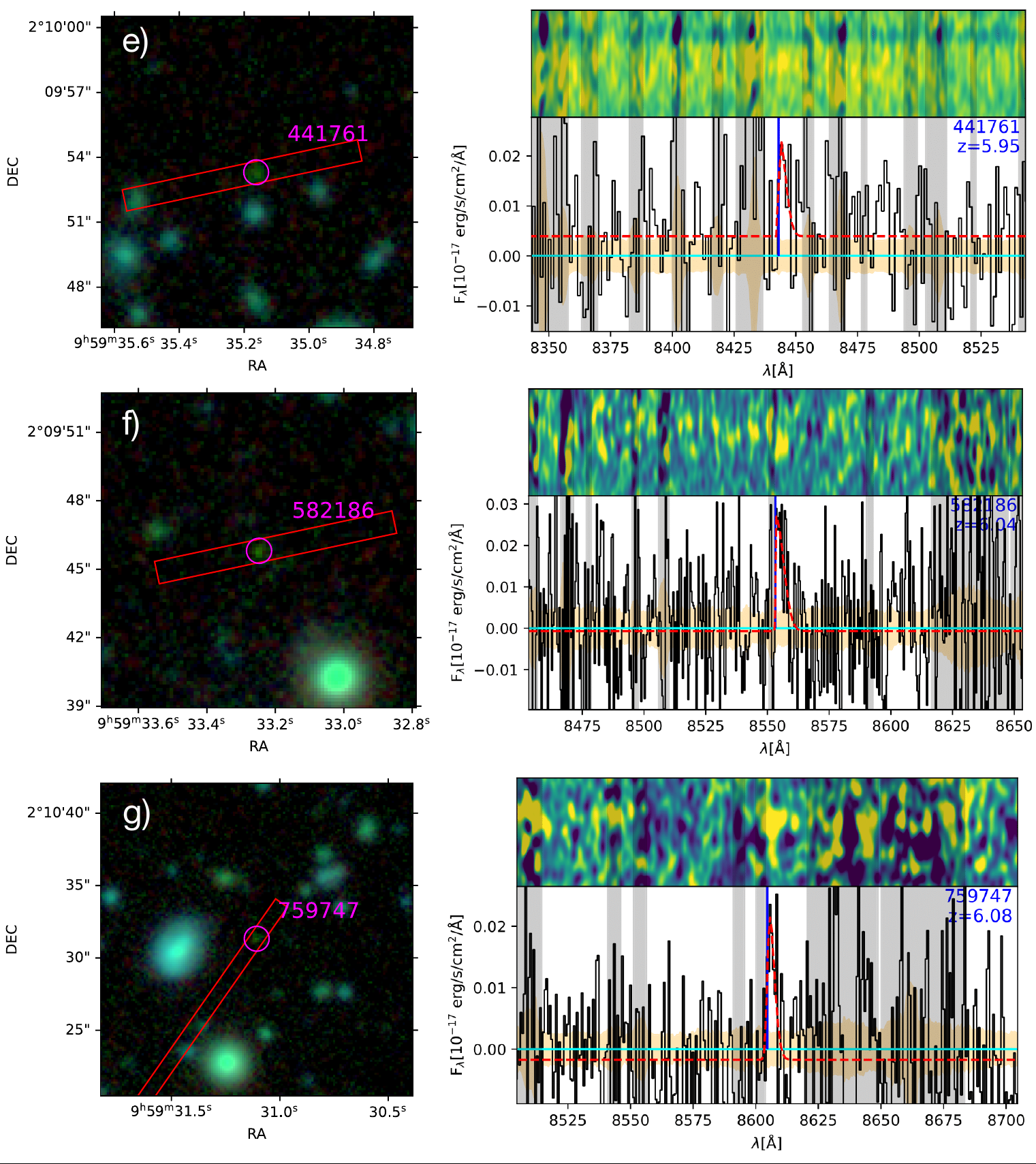}
    \caption{Same as Figure \ref{fig:A protocluster members} but for those protocluster members ranked B.}
    \label{fig:B protocluster members}
\end{figure*}
\begin{figure*}
    \centering
    \includegraphics[width=\textwidth, trim={0.1cm 0.1cm 0.1cm 0.1cm},clip]{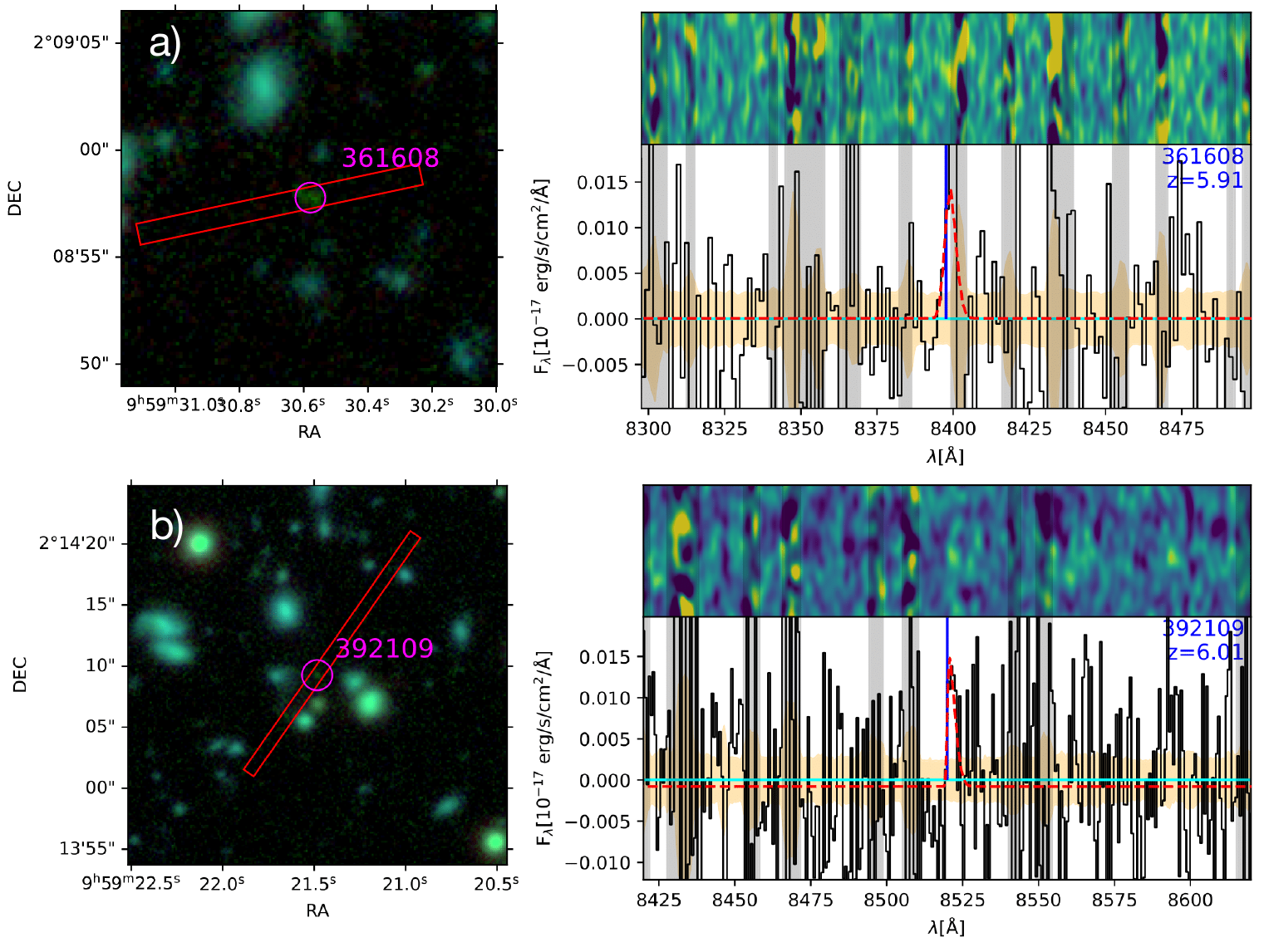}
    \caption{Same as Figure \ref{fig:A protocluster members} but for those protocluster members ranked C.}
    \label{fig:C protocluster members}
\end{figure*}

\subsection{Ly-$\alpha$ Line Characteristics}
\begin{figure}
    \centering
    \includegraphics[width=0.50\textwidth]{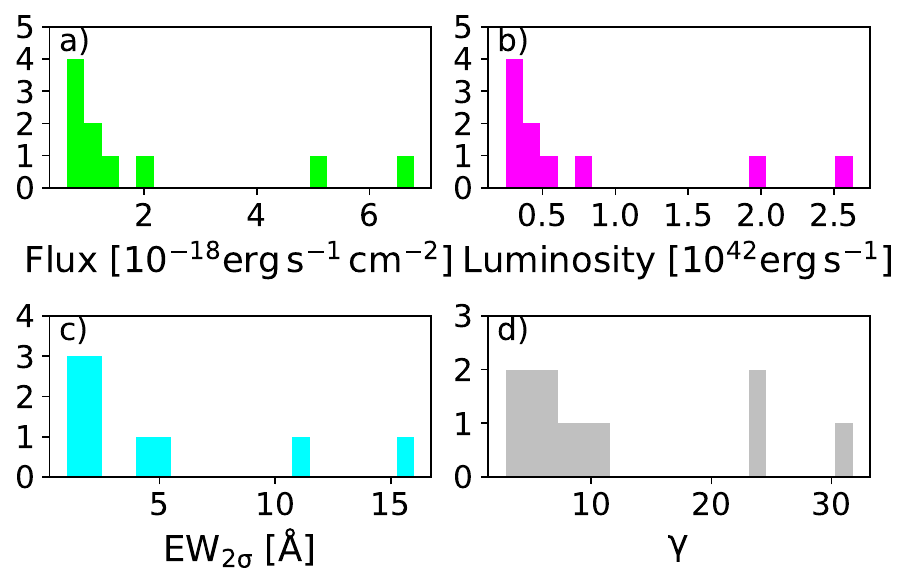}
    \caption{Histograms showing the distribution of {\bf a)} Ly$\alpha$ fluxes, {\bf b)} Ly$\alpha$ luminosities, {\bf c)} $2\sigma$ lower limits on the equivalent widths, and {\bf d)} line skewness parameters.}
    \label{fig:PC histograms}
\end{figure}
Figure \ref{fig:PC histograms} shows the distribution of the Lyman-$\alpha$ emission line properties for the emission lines we have detected. The flux range is $F=0.63-6.79\times10^{-18}\,{\rm erg\,s^{-1}\,cm^{-2}}$ with a median of $F_{\rm med}=1.09\times10^{-18}\,{\rm erg\,s^{-1}\,cm^{-2}}$ and the corresponding luminosity range is $L=0.246-2.628\times10^{42}\,{\rm erg\,s^{-1}}$ with a median of $L_{\rm med}=0.431\times10^{42}\,{\rm erg\,s^{-1}}$. We find that the peak signal-to-noise of the line detection for these 10 sources is between $5-20$.
\\\\
Many of the sources we observe show no clear continuum detection above the noise. In those cases we calculate the $2\sigma$ lower limit for the equivalent width as follows, 
\begin{equation}
    {\rm EW}^{\rm lower}_{\rm 0, Ly\alpha}=\frac{1}{1+z}\times\frac{F_{\rm Ly\alpha}}{f^{2\,\sigma}_{\rm cont}},
\end{equation}
where ${\rm EW}^{\rm lower}_{\rm 0, Ly\alpha}$, Ly$\alpha$, $z$, $F_{\rm Ly\alpha}$, and $f^{2\sigma}_{\rm cont}$ are the lower limit of Ly-$\alpha$ restframe equivalent width, redshift of a galaxy, total Ly-$\alpha$ emission
flux, and the $2\sigma$ non-detected flux density level, respectively. We
measured the value of $f^{2\sigma}_{\rm cont}$
cont by averaging $2\sigma$ RMS noise levels $\sim 50-100\,$Å redward of the Ly-$\alpha$ emission lines. 
In the case where there is a sufficiently high continuum above the noise, we calculate the equivalent width using the average continuum flux, $f^{\rm avg}_{\rm cont}\sim 50-100\,$Å redward of the line instead. Two of the $z=6$ protocluster galaxies have a continuum above the noise, 441761 and 573604 (see table \ref{tab:EW}). Their equivalents widths are $3\,$Å and $1\,$Å, respectively. We have chosen to compare their $2\sigma$ lower limit, to consistently compare with the other sources.
The distribution of the equivalent width lower limits 
are shown in Figure \ref{fig:PC histograms}c. We find an EW range of $1-16\,$Å with a median of $2$Å. None of the detections would definitively qualify the objects as being Lyman alpha emitters (LAEs), since that would require ${\rm EW}>20\,$Å \citep{Guaita2015}. This does not mean that the galaxies are not LAEs, but rather that our observations are not deep enough to draw this conclusion. The ${\rm EW}>20\,$Å cutoff between Lyman break galaxies (LBGs) and LAEs is also debated \citep{Kerutt2022}.
\\\\
Finally, the gamma (skew) distribution for our detection has a range of $2.87-31.75$ with a median of $7.37$. Since we expect the skewed Ly$\alpha$ line to have a positive skew, it leads further credence to the validity of the detection that all of them are significantly positively skewed. It should also be noted that while the uncertainty on the determination of the skew tends towards being large (see Table \ref{tab:specz}), the uncertainty tends towards higher values, not lower or negative values. An important point to keep in mind when fitting a skewed Gaussian is that the skew value is sensitive to the binning and it is possible to get both positive and negative values for the skew depending on the amount of binning done. Therefore, we have chosen not to bin the data for any of the grade A and B detection, to not bias our skew values.
\\\\
For comparison, \cite{Calvi2019} did multi-object spectroscopy using {\it OSIRIS} at the
{\it Gran Telescopio Canarias} of 16 LAE candidates discovered in the {\it Subaru}/{\it XMM Newton Deep} Survey as part of a $z\sim6.5$ protocluster candidate. Compared to the galaxies studied in \cite{Calvi2019}, their sources are generally brighter, more luminous and broader (higher EW) than our galaxies, with the skew being comparable. It should be noted that their OSIRIS observation has $\sim3\times$ the amount of observation time as our objects. Given that none of their and few of our objects have continuum detections it makes sense that their observations are generally brighter and broader since they have had more time to uncover more of the line flux.

\section{Discussion}\label{section:discussion}
\subsection{A revised overdensity estimate of PC$z$6.05-1}
Given our new spectroscopic redshifts, we can calculate an improved estimate of how overdense of an environment PC$z$6.05-1 resides in. We do this by re-creating an overdensity map with updated weights, following the Weighted Adaptive Kernel method laid out in \cite{Brinch2023}. 
All galaxies with grade A or B detections will have their redshift updated and their weight set to one. Grade C or D detections will not have their redshift updated and their photometric redshift probability distribution will still be used. To analyse the distribution of overdensities in the COSMOS field we make a redshift bin which includes galaxies whose spectroscopic redshift falls within the bin, as well as galaxies with photometric redshifts that fall within the bin \citep[see][for a list of selection criteria used]{Brinch2023}. 
Since the redshift distribution of the spectroscopically confirmed galaxies is skewed slightly lower than the original photometric distribution, we chose a slightly lower centre of $z=5.97$ for our bin, with a $\Delta z=0.30$ to incorporate all the galaxies with spectroscopic redshifts around $z=6$. 
Figure \ref{fig:smoothing_kernel} shows the resulting overdensity map using the spectroscopic redshift, including those from the literature \citep{Mallery2012,Hasinger2018,Casey2019,Jin2019,Williams2019,Zavala2021,Priv_Comm_Olivier}. 
We find that the overdensity associated with PC$z$6.05-1 has significantly increased, from the previous peak overdensity value of $\rm \delta_{\rm OD}=9.2\,(8.0\sigma)$ reported in \citet{Brinch2023} to a revised value presented here of $\rm \delta_{\rm OD}=11.6\,(10.9\sigma)$. This unequivocally puts PC$z$6.05-1 on par with the most overdense protoclusters identified at these redshifts \citep{Harikane2019}, although, we caution that overdensity calculations vary across different studies \citep[and references therein]{Harikane2019}, making direct comparisons between protoclusters challenging.
Richness, that is the number of galaxies in an overdensity, is a simpler and more easily comparable measure between different protocluster studies. There are also indications that richness is among the better tracers of which overdensities will end up as massive ($10^{15}\,\rm M_{\rm \odot}$) a galaxy cluster at $z=0$ \citep{Remus2022}.  
\subsection{Discovery of a $z\simeq 6$ large-scale structure in COSMOS}
\begin{figure*}
    \centering
    \includegraphics[width=\textwidth]{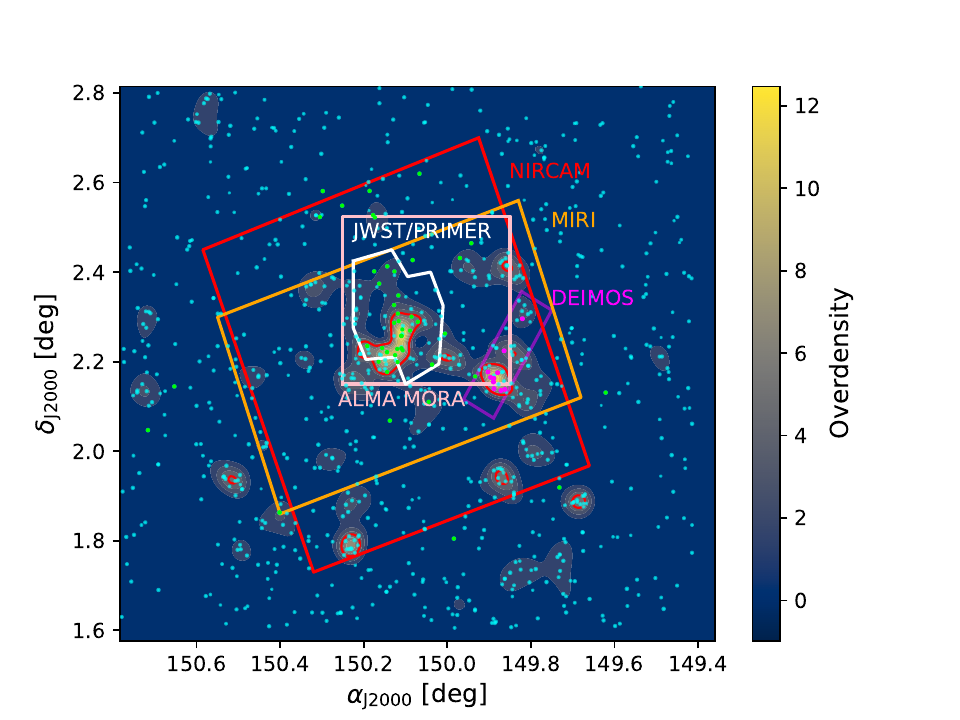}
    \caption[\protect]{The galaxy overdensity field at $z=5.97\pm0.15$ in COSMOS, derived from the Weighted Adaptive Kernel method described in \cite{Brinch2023}. Contours are in steps of $1\sigma$, where $\sigma$ is calculated as the standard deviation of the overdensity values in the field. The $4\sigma$ contours have been highlighted in red. Individual galaxies 
    are indicated as cyan dots, except for the magenta dots which indicate the galaxies with secure (grade A and B) Ly$\alpha$ detections presented in this paper, and the green dots which show other spectroscopically confirmed galaxies within the redshift bin \citep{Mallery2012,Hasinger2018,Casey2019,Jin2019,Williams2019,Zavala2021,Priv_Comm_Olivier}. The {\it DEIMOS} field of view is shown as the magenta rectangle. The JWST/PRIMER (white), ALMA MORA (pink), COSMOS-WEB NIRCAM (red) and COSMOS-WEB MIRI (orange) footprints are also shown.}
    \label{fig:smoothing_kernel}
\end{figure*}

From Figure \ref{fig:smoothing_kernel} it is seen that PC$z$6.05-1 appears to be embedded in a large-scale structure that spans across the central parts of the COSMOS field. PC$z$6.05-1 is connected via a filament to an even stronger and more extended overdensity located within the JWST/PRIMER field (ID: 1837; PI: J. Dunlop). 
This second $z\simeq 6$ protocluster is hitherto unknown, and we report its discovery here. Its centroid location is at (RA, DEC)~$=(150.113^{\circ}, 2.255^{\circ})$ and it contains 22 galaxies with spectroscopic redshifts within the $4\sigma$ overdensity contour. The median spectroscopic redshift of these 22 galaxies is $\overline{z}=5.89$ with a $z$-range of $\Delta z=0.27$. The peak overdensity is $\delta=12.5$ ($11.8\sigma$), i.e., larger albeit comparable to that of PC$z$6.05-1. The overdensity seems to be made up of at least two more compact 'cores', with one containing 13 spectroscopically confirmed galaxies (${\overline z}=5.86\pm 0.07$) and the other with 9 spectroscopically confirmed galaxies ($\overline{z}=6.01\pm 0.13$).   
In addition to these highly significant overdensities and their connecting filaments in the central regions of the COSMOS field, four smaller, yet significant ($\geq 4\sigma$), overdensity peaks are seen in the lower half of the overdensity map in Figure \ref{fig:smoothing_kernel}. It is possible that they trace out a large structure that connects to PC$z$6.05-1 and the second central overdensity in the map. 
Taking the $1\sigma$ contour as a crude measure of the size, the overall structure has an of $84\, \rm cMpc$ across, with 4 peaks in the lower parts of the field indicating the structure could have an even larger extent.
\subsection{Properties of the galaxies in PC$z$6.05-1}
\subsubsection{SED fitting}\label{subsubsection:SED-fitting}
To analyse the physical properties of the protocluster galaxies in PC$z$6.05-1, we employ the
Bayesian Analysis of Galaxies for Physical Inference and Parameter
EStimation ({\sc bagpipes}), which is a Bayesian SED fitting code \citep{Carnall2018}. {\sc bagpipes} allows for the fitting of both photometry and spectra to obtain estimates of physical properties such as stellar mass, star formation rate (SFR), Dust obscuration (A$_{\rm V}$) and the mass-weighted age of the galaxy. {\sc bagpipes} also includes the treatment of nebular emission lines from {\sc cloudy} modelling \citep{Ferland2013,Ferland2017}, which has been shown to be increasingly relevant for high-$z$ galaxies, especially with the advent of JWST \citep{Endsley2023}. 
Although {\sc bagpipes} allows for the spectrum to be included as part of the fitting process, it encounters challenges in accurately fitting the Ly$\alpha$ line due to its variable dampening and as a result, we do not include the spectra in the fitting process. The redshifts of the PCz6.05-1 galaxies were fixed to their spectroscopic redshift values.
We use the following parameter ranges when fitting: formation mass $\rm log(M_{\rm \star})\in[6.0:12.0]\,log(M_{\rm \odot})$, metallicity $\rm Z\in[0.01:2.00]\,log(Z_{\rm \odot})$, reddening $\rm A_{\rm v}\in[0.0:4.0]$ (utilizing a \cite{Calzetti2000} dust law) and the $U$ parameter, which is the strength of the nebular radiation field to $\log(U) \in[-4.0:-1.0]$.
To fit the galaxies we use all available {\sc FARMER} photometry from the COSMOS2020 catalogue, which includes observations from the ultraviolet to the far infrared (see \cite{Weaver2021} for details).
\\
{\sc bagpipes} allows for different star formation histories (SFH) when fitting.  We tested the following SFHs: constant, double power law, delayed, lognormal, exponential, constant+exponential model and a non-parametric SFH \citep{Leja2019}. 
We found that parametric models resulted in an artificially tight relationship for the galaxy main sequence and yielded unrealistic ages, all of which were $< 10\,{\rm Myr}$.
We have, therefore, chosen to use the non-parametric SFH when fitting our galaxies. 
It is worth noting that \citet{Endsley2023} argue that galaxies, especially those with very young ages ($\lesssim10\,\rm Myr$), tend to produce stellar masses $\sim0.5-1\,\rm dex$ higher when utilizing a non-parametric SFH with a continuity prior, as opposed to employing a constant SFH. A non-parametric SFH that disfavors extremely rapid changes in SFR, like the continuity prior used in {\sc Bagpipes}, can increase the masses up to $\approx2$dex higher than using a constant SFH with a very young SED age.
We argue that the use of a non-parametric SFH is appropriate for our galaxies for several reasons. Firstly, when inspecting the SFH for the constant and non-parametric models for the galaxies with very young ages, all of the star formation for the constant model happens in a single short burst, as opposed to the non-parametric model where the shape of the SFH tends to be lognormal for these galaxies (see Figure appendix \ref{SFH comparison}). Secondly, the stellar mass provided by the non-parametric model is at most $0.5\,\rm dex$ greater, and in most cases $0.2\,\rm dex$ greater. This difference in stellar mass does not change any conclusion in our further analysis. Thirdly, the shape of the non-parametric SFH is akin to a (truncated) log-normal distribution, which is in line with what is found from simulations of high-$z$ galaxies within our mass range \citep{Wilkins2023}. For these reasons, we have opted to stick with the non-parametric SFH. {\sc Bagpipes} fits to the galaxies in PC$z$6.05-1 are shown in Appendix \ref{Bagpipes fits}, and
the resulting physical properties from their best-fit SED models are listed in Table \ref{tab:Bagpipes}.  Overall, there is good agreement with the physical properties presented in \citet{Brinch2023}, which were based on SED fits with {\sc Lephare} and without spectroscopic redshifts. The main difference is with the stellar masses where {\sc bagpipes} generally yields somewhat lower masses.
\begin{table}
\caption[\protect]{{\sc bagpipes} results for PC$z$6.05-1 protocluster galaxies with spectroscopic redshift. Columns: (1) COSMOS2020 catalogue ID; (2) The mass-weighted age; (3) the stellar mass; (4) the star formation rate; (5) reddening using a \cite{Calzetti2000} dust law.}
\label{tab:Bagpipes}
\centering
\begin{tabular}{l l l l l l}
\hline
\hline
ID & age & $\log {\rm M_{\rm \star}/M_{\rm \odot}}$ & SFR & $\rm A_{\rm V}$ & Z\\ 
 & [Myr] & [dex] & [M$_{\rm \odot}$/yr$^{-1}$] & & [Z$_{\rm \odot}$] \\
(1) & (2) & (3) & (4) & (5) & (6)\\  
\hline
70505 & 100$^{+142}_{-72}$ & 8.64$^{+0.23}_{-0.2}$ & 3$^{+1}_{-1}$ & 0.29$^{+0.05}_{-0.05}$ & 0.24$^{+0.06}_{-0.03}$\\
156780 & 133$^{+120}_{-91}$ & 8.60$^{+0.23}_{-0.2}$ & 3$^{+1}_{-1}$ & 0.36$^{+0.06}_{-0.06}$ & 0.24$^{+0.06}_{-0.04}$\\
186512 & 96$^{+85}_{-60}$ & 8.56$^{+0.18}_{-0.15}$ & 3$^{+1}_{-1}$ & 0.04$^{+0.04}_{-0.03}$ & 0.03$^{+1.35}_{-0.01}$\\
220530 & 77$^{+74}_{-47}$ & 8.29$^{+0.12}_{-0.16}$ & 2$^{+1}_{-1}$ & 0.04$^{+0.04}_{-0.03}$ & 0.02$^{+0.04}_{-0.01}$\\
225263 & 101$^{+119}_{-67}$ & 9.01$^{+0.24}_{-0.22}$ & 8$^{+4}_{-2}$ & 0.75$^{+0.05}_{-0.05}$ & 0.04$^{+0.02}_{-0.01}$\\
441761 & 134$^{+114}_{-73}$ & 9.65$^{+0.19}_{-0.18}$ & 35$^{+13}_{-9}$ & 1.11$^{+0.04}_{-0.05}$ & 0.04$^{+0.02}_{-0.01}$\\
482804 & 213$^{+109}_{-110}$ & 9.18$^{+0.12}_{-0.16}$ & 8$^{+2}_{-2}$ & 0.28$^{+0.12}_{-0.13}$ & 0.53$^{+0.19}_{-0.18}$\\
573604 & 31$^{+67}_{-24}$ & 9.09$^{+0.12}_{-0.07}$ & 12$^{+2}_{-1}$ & 1.08$^{+0.04}_{-0.04}$ & 0.22$^{+0.02}_{-0.01}$\\
582186 & 386$^{+66}_{-81}$ & 10.28$^{+0.06}_{-0.07}$ & 39$^{+18}_{-13}$ & 0.73$^{+0.15}_{-0.18}$ & 1.37$^{+0.4}_{-0.48}$\\
759747 & 235$^{+112}_{-116}$ & 8.91$^{+0.18}_{-0.23}$ & 4$^{+1}_{-1}$ & 0.24$^{+0.11}_{-0.12}$ & 0.3$^{+0.18}_{-0.11}$\\
\hline
\end{tabular}
\end{table} 
\begin{figure*}
    \centering
    \includegraphics[width=\textwidth]{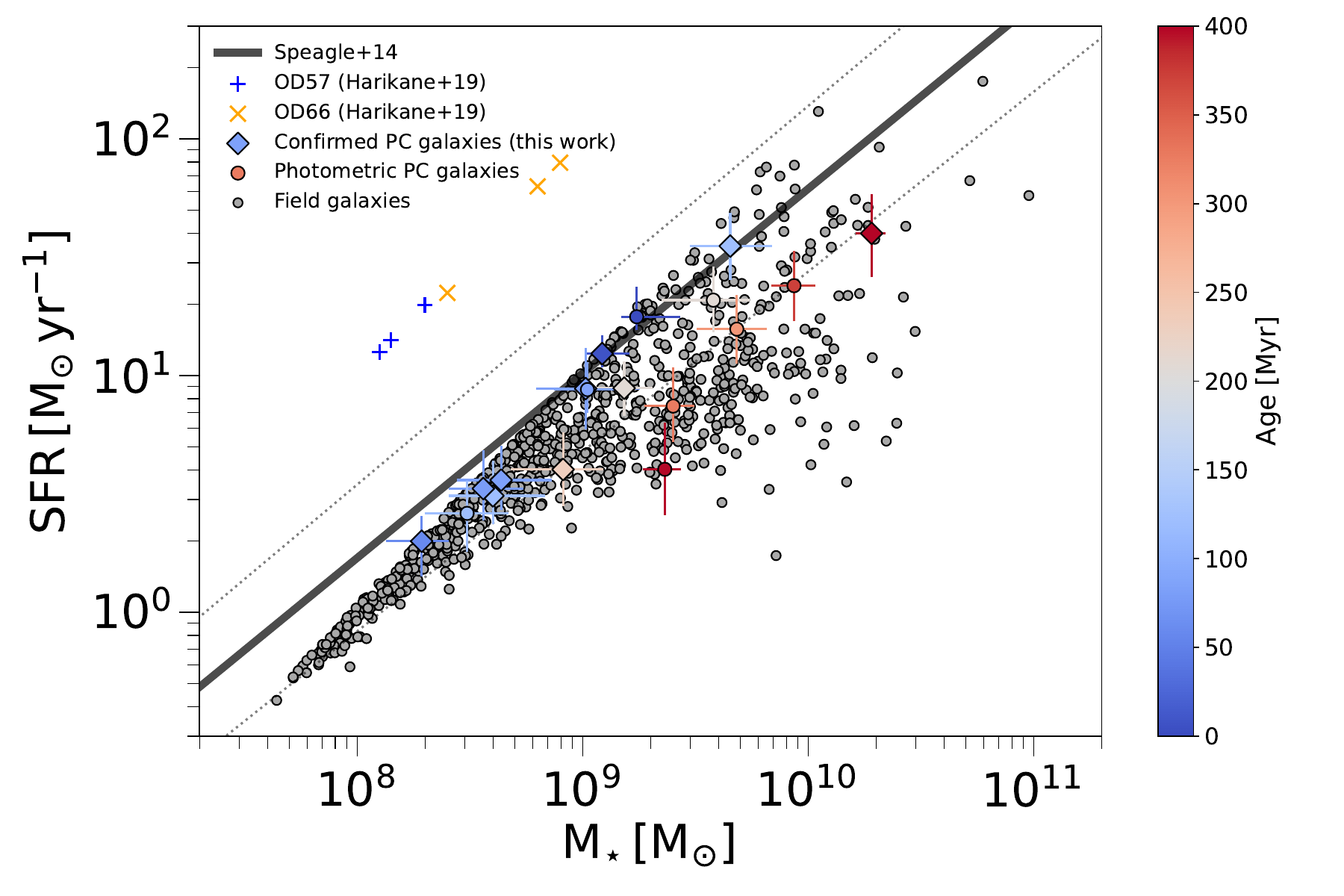}
    \caption{The ${\rm SFR - M_{\star}}$ plane as derived from {\sc bagpipes} (\S\ref{subsubsection:SED-fitting}) for all $z_{\rm phot} = 5.97\pm 0.15$ galaxies in COSMOS2020 (grey circles) and for the protocluster galaxies with squares for galaxies with spec-$z$ (grade A \& B) and circles for galaxies with photo-z (grade C \& D) that were part of the original sample of 19 galaxies from \citet{Brinch2023}, both coloured according to their age from {\sc bagpipes}. The galaxy main sequence of \citet{Speagle2014} at $z\simeq 6$ is plotted as a grey line, with the dotted lines on both sides being the $1\sigma$ uncertainty. For comparison, the $z=5.7$ and $z=6.6$ spectroscopically confirmed LAE-overdensities reported by \citep{Harikane2019} are also shown (shown as blue $+$ and yellow $\times$ symbols, respectively).}
    \label{fig:BagpipesMS}
\end{figure*}
\begin{figure}
    \centering
    \includegraphics[width=0.5\textwidth]{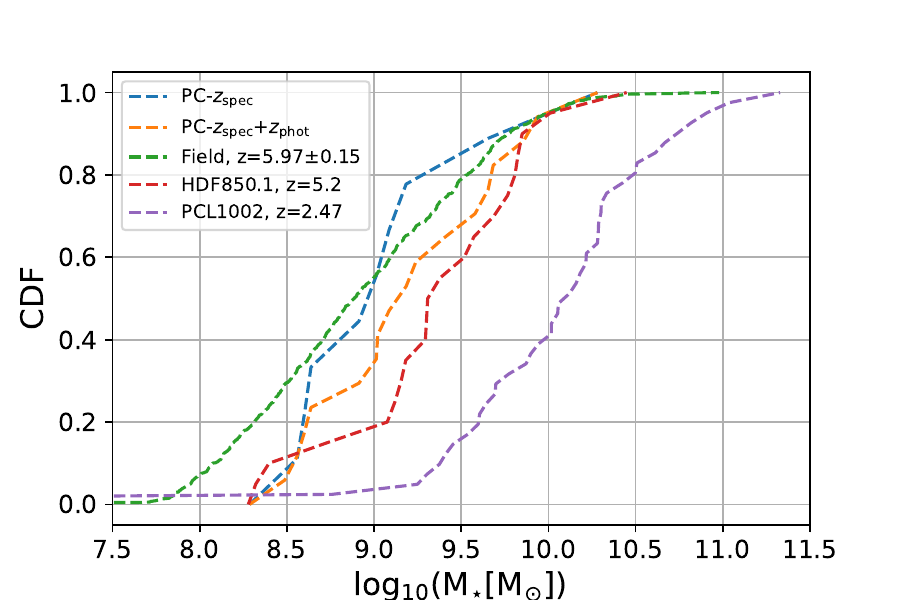}
    \caption{The stellar mass cumulative distribution function (CDF) of the spec-$z$ confirmed galaxies (blue line), the spec-$z$+photo-$z$ includes 8 galaxies with photo-$z$ from our original 19 galaxy sample (orange line) and of the field (green line) consisting of 876 galaxies at $z_{\rm phot} = 5.97\pm 0.15$. For comparison, we also show the stellar mass CDF from the spectroscopically verified 
    protoclusters HDF850.1 at $z=5.2$ \citep{Calvi2021} (red line) and PCL1002 at $z=2.47$ \citep{Casey2015} (purple line).}
    \label{fig:CDF}
\end{figure}

To compare with other galaxies in the field at a similar redshift, we also fit all the galaxies in the $z=5.97\pm 0.15$ redshift bin (shown as cyan dots in Figure \ref{fig:smoothing_kernel}). Because all of our galaxies with spectroscopic redshifts were best fit with a non-parametric SFH, we also adopted this when fitting the galaxies with only photometric redshifts. 
Assuming that all these galaxies are within the redshift bin, we have therefore opted to fix it to the photometric redshifts from {\sc Lephare}. 

\subsubsection{The $SFR - M_{\star}$ plane}
Figure \ref{fig:BagpipesMS} shows the location of all the galaxies in the $z_{\rm phot} = 5.97\pm 0.15$ bin based on the results from the {\sc bagpipes} fitting in the previous section. We confirm the result found in \citet{Brinch2023} that the bulk of the galaxies at $z\sim6$ appear to be main sequence galaxies.
We observe that the $SFR - M_{\star}$ appear to be separated into two populations. A very young, star-forming population and an older, less star-forming population. The very young ages of some galaxies can be explained by the presence of a recent star formation burst. We also observe a tight relation for the main sequence for these young galaxies especially around a stellar mass of $10^{8}\,\rm M_{\rm \odot}$, which we believe is due to limitations of the templates used by the {\sc bagpipes} modeling\footnote{Note that this is not due to the choice of star formation history, as other models were tried and gave an even tighter relationship at low stellar masses.}. We find that the galaxies in PC$z$6.05-1 (diamonds and circles) fall on this upper part of the main sequence, with some notable exceptions. 
The galaxy with the highest mass (ID 582186)
has a stellar mass of $10^{10.3}\,\rm M_{\odot}$ and an age of $386\,{\rm Myr}$, making it the oldest galaxy in our spec-$z$ sample. It is possible that this galaxy is the progenitor of the brightest central cluster galaxies (BCG) that we find at the centres of galaxy clusters at lower redshifts, given its evolved nature and substantial mass at $z=6$. This is suggestive of downsizing being at play, where, on average, the central BCGs of the most massive (proto)clusters assemble at earlier epochs than those in less massive counterparts \citep{Rennehan2020}. BCGs are formed in high
galaxy density protocluster cores, which fits with 582186 being the most massive galaxy in our sample with a central position in the protocluster. One difference between this proto-BCG and the lower-$z$ cluster BCGs is its star formation. 582186 does not appear to be quiescent or off the main sequence, though this is to be expected, as the galaxy is still in its star-forming phase and the difference between it (and protocluster galaxies in general) and the field galaxies is less severe at high-$z$.
Another galaxy not highlighted due to being a C-grade detection in Figure \ref{fig:BagpipesMS} is 392109, with a stellar mass of $10^{9.9}\,\rm M_{\rm \odot}$ and an SFR of $1.7\,\rm M_{\rm \odot}/yr$ ($\log {\rm sSFR} = -9.6\,{\rm dex}$), the galaxy appears to be below the main sequence and could be classified as quiescent. Further study would allow us to investigate the causes behind the quenching. Is the galaxy being quenched due to its environment already at $z=6$? or is it internal processes in the galaxy that suppress its star formation?

The \citet{Harikane2019} galaxies selected as narrowband LAEs appear to be $\rm\sim1$ dex above our main sequence, but we appear to have more massive galaxies in our structure, with about half of the protocluster galaxies being more massive than any of the ones found in OD57 or OD66. The difference is less severe compared to our findings in \citet{Brinch2023} since our protocluster galaxies have generally gotten lower masses when fitting with {\sc bagpipes} than with {\sc Lephare}.\\
{\sc bagpipes} also allows for the dust obscuration to be estimated in the form of the $A_{\rm V}$ value. The 10 protocluster galaxies have $A_{\rm V}$ values in the range $0.04-1.11$ with a median of $0.32$. None of the galaxies appear to be heavily dust-obscured, which fits with our expectations from \citep{Brinch2023} where we saw no strong detection in the available mid- to far-IR data such as MIPS at $24\,{\rm \mu m}$ and SCUBA-2 at $850\,{\rm \mu m}$ and ALMA \citep{Ferrara2022}. Since we expect some galaxies to have recently had a starburst due to their young ages from {\sc bagpipes} and their SFH as seen in Appendix \ref{SFH comparison}, we would also expect great amounts of dust to be produced. It is clear that further study is necessary to understand the dust production in these high-$z$ galaxies.
\subsection{Stellar mass function and dark matter halo mass}
\begin{figure*}
      \includegraphics[width=0.57\textwidth]{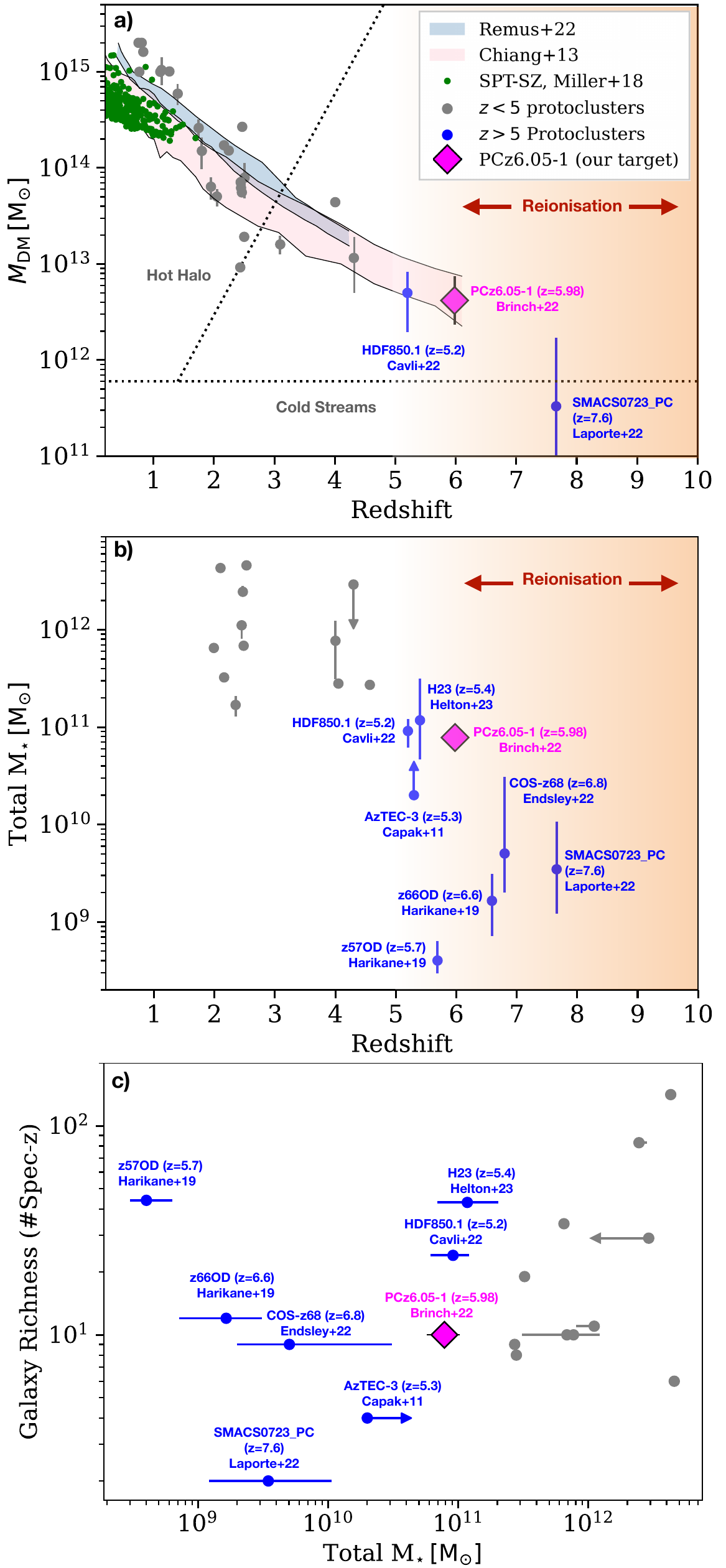}
  \caption{PC$z$6.05-1 (magenta diamond), is the
          most massive, and one of the richest, spectroscopically
          confirmed protoclusters at $z\gs 6$, and has far better OIR
          multi-wavelength data than other protoclusters at this epoch. 
          {\bf a)} Dark matter halo mass estimates for spectroscopically
          confirmed protoclusters  \citep[compiled from][]{Miller2018,Brinch2023} and simulations \citep{Chiang2013,Remus2022} 
          across cosmic time. 
          The dashed lines mark different regions of gas inflow and cooling
          mechanisms on massive halos \citep{Dekel2013}. 
          In the ‘cold streams in hot media’ regime, where PC$z$6.05-1 is located, halo-penetrating cold gas flows can help sustain galaxy growth.
          {\bf b)} Same as a) but for the total stellar mass of the protoclusters. 
          {\bf c)} Galaxy richness, defined by the number of spectroscopically confirmed cluster-member galaxies, vs the total stellar mass of protocluster.
      } \label{fig:PCcomparisons}
\end{figure*}

To investigate the distribution of stellar masses in PC$z$6.05-1, we show the stellar mass cumulative distribution function (CDF) in Figure \ref{fig:CDF}. For comparison, we also show the CDFs for the protoclusters HDF850.1 \citep{Calvi2021} and PCL1002 \citep{Casey2015}. We see that the spec-$z$ sample follows the field distribution, being skewed slightly towards higher masses below $9.0\,{\rm dex}$ and slightly towards lower masses above $9.0\,{\rm dex}$, compared to the field. Adding the remaining galaxies with non-detections in our 19 galaxy sample that is within the redshift bin (8 galaxies) the CDF is skewed towards higher masses up until $9.8\,{\rm dex}$, where it aligns with the field. This is similar to what we found for this protocluster in \cite{Brinch2023}. 
\\\\
With the spectroscopic redshifts and the updated stellar masses, we are able to update our estimate of the dark matter halo mass. For simplicity, we choose the \citep{Behroozi2013} abundance matching method to estimate the halo mass. For a discussion of other possible ways to estimate the halo mass, we refer the reader to \S4.3 in \citep{Brinch2023}. We take the most massive galaxy in our sample, 582186 with a stellar mass of $10^{10.27}\,{\rm M_{\odot}}$, and estimate the dark matter halo mass using a stellar mass to halo mass relationship. We estimate a halo mass of $10^{12.10^{+0.20}_{-0.16}}\,{\rm M_{\odot}}$. Following the dark matter halo mass evolution curves of \citep{Chiang2013}, this would suggest that the protocluster would evolve into a Virgo- ($\sim10^{14} \rm M_{\rm \odot}$) or Coma-like ($\sim10^{15} \rm M_{\rm \odot}$) cluster in the present day. With the spectroscopic redshift confirmation of multiple galaxies in close proximity and in conjunction with an overall overdense structure that the protocluster lies in as seen in  Figure \ref{fig:smoothing_kernel}, we argue this is likely to be the case. 
The protocluster position in a large-scale structure allows the protocluster to be fed galaxies away from the central core over time, which will allow it to grow in mass.
\\\\
To compare with other (proto-)clusters from the literature \citep{Miller2018,Brinch2023}, we have shown their dark matter halo mass, total stellar mass and richness (i.e. the number of spectroscopically confirmed galaxies in the protocluster) in Figure \ref{fig:PCcomparisons}. We see from Figure \ref{fig:PCcomparisons}a that our protocluster occupies the ‘cold streams in hot media’ regime where halo-penetrating cold gas flows can help sustain galaxy growth. We also see that our dark matter halo mass estimate is similar to what is found in HDF850.1 \citep{Calvi2021} and both are in line with the dark matter halo mass evolution track from \citep{Chiang2013}. Figure \ref{fig:PCcomparisons}b shows the total stellar mass with redshift, and we see that our protocluster falls within the expected scatter of our protocluster sample and is again close to HDF850.1 at $\sim10^{11}\,\rm M_{\odot}$. Figure \ref{fig:PCcomparisons}c shows the richness of protoclusters with the total stellar mass, and we see that our protocluster falls in the middle at 10 galaxies for protocluster at $z>5$.
\subsection{Protocluster size and evolution scenarios}\label{sec:PCsize}
\begin{figure*}
\centering
\includegraphics[width=\textwidth]{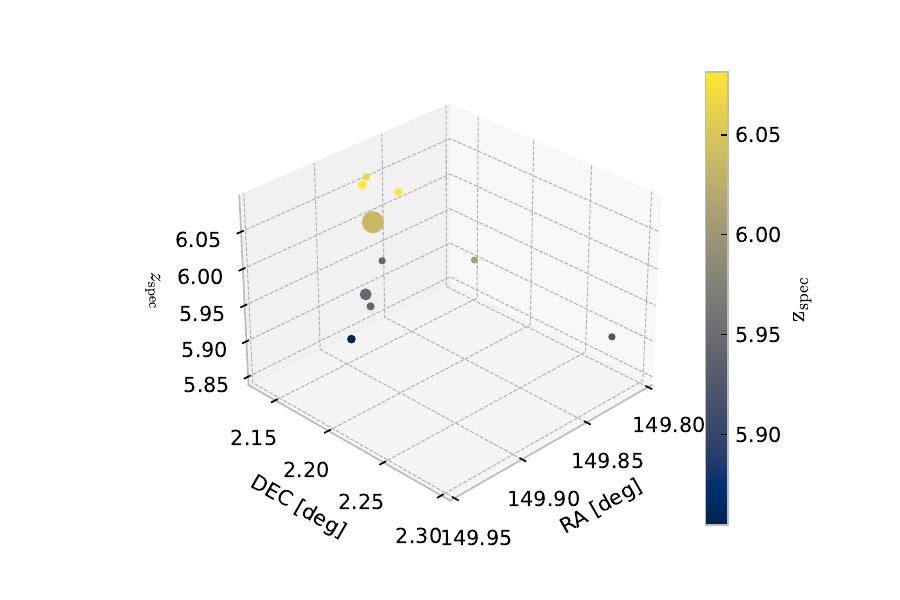}
\caption{3D distribution of galaxies with spectroscopic redshift in RA, DEC, redshift. points with larger sizes have larger stellar masses and are coloured according to their spectroscopic redshift.} 
      \label{fig:3Dplot}
\end{figure*}

If we consider the 8 galaxies within the $4\sigma$ contour to be the centre of the protocluster, they cover a $\sim5.8\times5.9\,\rm cMpc$ area, in line with what is expected from a protocluster at $z=6$ \citep{Chiang2013,Chiang2017,Lovell2018}. 
The structure is extended more in the $z$ direction, with a distance of $96\,\rm cMpc$ between the lowest and highest redshift galaxy. The reason for this length can be explained by multiple different factors. Firstly is the negative uncertainty due to the nature of the skewed Lyman-$\alpha$ line means that the galaxies could be closer together. Without knowing the Lyman-$\alpha$ escape fraction and being able to model the line before it was attenuated, it is difficult to access where the true line centre is. Secondly is the possibility of the presence of substructures in the protocluster, one above $z=6$ and one below as seen in figure \ref{fig:3Dplot}. Considering the eight $4\sigma$ galaxies, the distance between the 4 galaxies above $z=6$ is $18.9\,\rm cMpc$, while those below $z=6$ have a distance of $39.0\,\rm cMpc$. Thirdly, and perhaps most interestingly, is the scenario that this could be two separate protoclusters, which could eventually merge over time. The most massive galaxy below $z=6$ would have a halo mass of $10^{11.40^{+0.07}_{-0.08}}$. 
If we consider the two groups of galaxies as separate we can calculate their overdensity by making smaller redshift bins. For the galaxies at $z<6$ we use a bin of $z=5.90\pm0.05$ and find that 4 of the galaxies with spec-$z$ occupy a $4\sigma$ overdensity with a peak of $\delta=8.2$ ($5.7\sigma$). If the procedure is repeated with the $z>6$ galaxies using a $z=6.06\pm0.05$ bin, we find that 6 of the galaxies with spec-$z$ in the bin occupy a $4\sigma$ overdensity with a peak of $\delta=16.7$ ($12.8\sigma$). the overdensity maps for both bins are shown in figure \ref{fig:JWSTsmoothing_kernels}. we also note the presence of the second protocluster hitherto reported on is present in both bins. We know for any protocluster dark matter halo to grow to become a cluster in the present day, it must merge with other halos. This is how dark matter halo evolution is tracked in simulation, though a cluster merger tree \citep[see][]{Chiang2013}. If this third scenario is true, we are witnessing a snapshot of the growth of protoclusters into becoming the massive galaxy-rich clusters of today.\\
\begin{figure*}
    \centering
    \includegraphics[width=\textwidth, trim={0.3cm 9.3cm 0.3cm 0.2cm},clip]{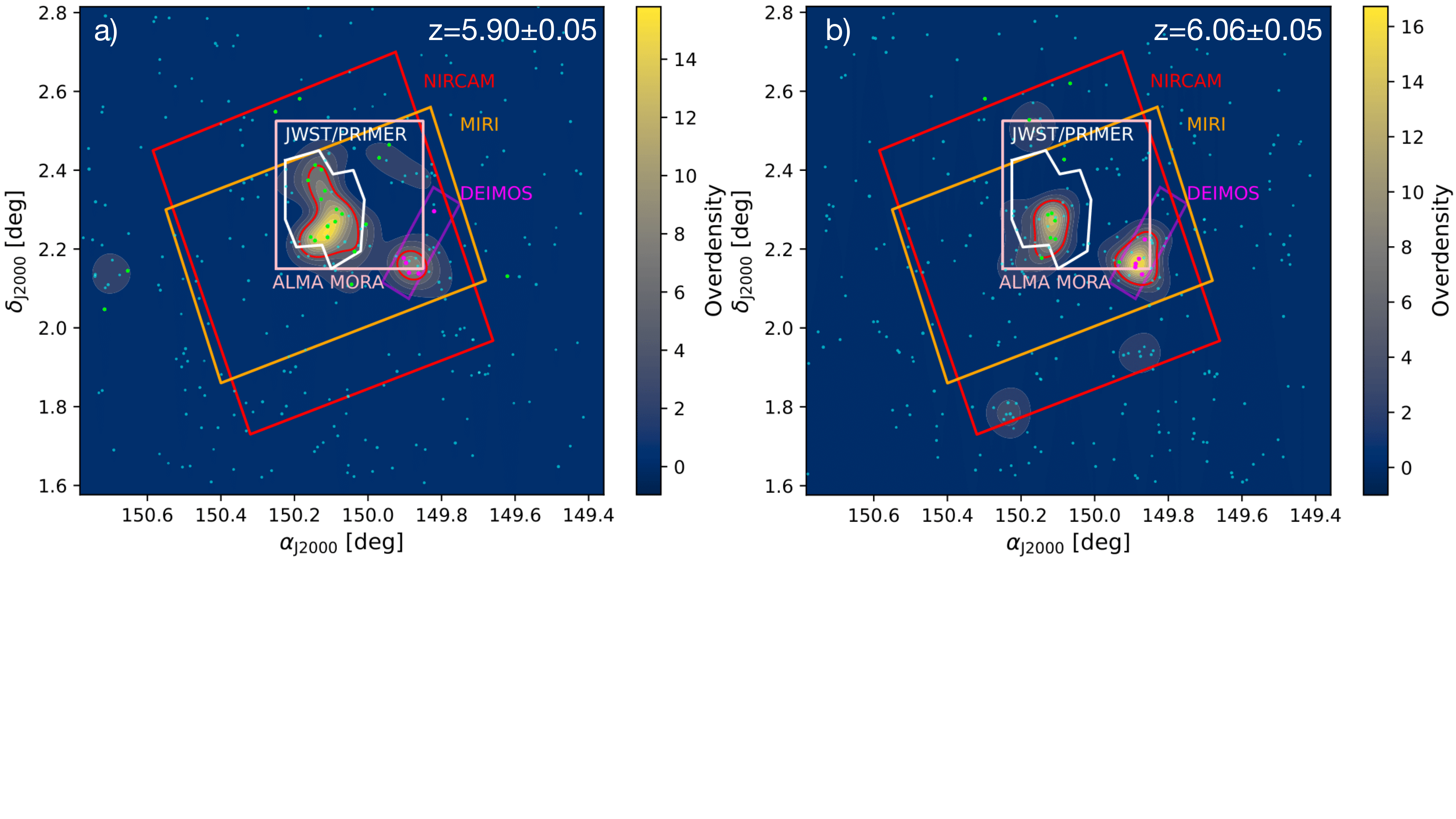}
    \caption[\protect]{Galaxy overdensity maps using a \textbf{a)} $z=5.90\pm0.05$ bin and \textbf{b)} $z=6.06\pm0.05$ bin. The setup is the same as figure \ref{fig:smoothing_kernel}.}
    \label{fig:JWSTsmoothing_kernels}
\end{figure*}
\begin{table*}
\caption{List of all confirmed Ly$\alpha$ detections with their parameters; Columns: (1) COSMOS2020 catalogue ID.; (2) \& (3) R.A. and Dec.; (4) spectroscopic redshift; (5) spectroscopically measured Ly$\alpha$ flux; (6) spectroscopically measured Ly$\alpha$ luminosity; (7) Skewness (gamma parameter), galaxies without have a double-peaked Ly$\alpha$ profile; (8) detection grade: A = secured detection, B = sound detection, C = marginal detection and D=non-detection. The numbers in the parenthesis show the criteria fulfilled.}
\label{tab:specz}
\centering
\begin{tabular}{l l l l l l l l l}
\hline
\hline
ID & RA & Dec  & $z_{\rm spec}$ & F$_{\rm Ly\alpha, spec}$ & L$_{\rm Ly\alpha, spec}$ & Skewness & Grade\\ 
& {\it hh:mm:ss.ss} & {\it dd:mm:ss.ss} & & ${\rm [10^{-18}erg\,s^{-1}\,cm^{-2}]}$ & ${\rm [10^{42}erg\,s^{-1}]}$ & ${\rm [\gamma]}$ & \\ 
(1) & (2) & (3) & (4) & (5) & (6) & (7) & (8)\\ 
\hline
Protocluster members\\
\hline
70505 & 09:59:16.86 & 02:17:45.20 & 5.9191$^{+0.0003}_{-0.0002}$ & 5.21$^{+0.19}_{-0.17}$ & 2.014$^{+0.974}_{-0.067}$ & 2.87$^{+0.43}_{-0.53}$ & B(1,2)\\
156780 & 09:59:29.14 & 02:08:08.38 & 6.0642$^{+0.0003}_{-0.0003}$ & 1.92$^{+0.22}_{-0.24}$ & 0.785$^{+0.088}_{-0.097}$ & 6.10$^{+4.19}_{-2.62}$ & B(1,2)\\
186512 & 09:59:27.01 & 02:08:17.01 & 5.9240$^{+0.0001}_{-0.0001}$ & 6.79$^{+0.25}_{-0.25}$ & 2.628$^{+0.098}_{-0.098}$ & 10.77$^{+4.00}_{-1.54}$ & B(1,2) \\
220530 & 09:59:27.36 & 02:13:27.56 & 6.0088$^{+0.0001}_{-0.0002}$ & 1.28$^{+0.09}_{-0.10}$ & 0.512$^{+0.038}_{-0.039}$ & 8.52$^{+53.26}_{-2.94}$ & B(1,3)\\
225263 & 09:59:33.21 & 02:08:23.24 & 5.8547$^{+0.0001}_{-0.0001}$ & 0.79$^{+0.07}_{-0.06}$ & 0.296$^{+0.025}_{-0.022}$ & 23.19$^{+1675.38}_{-14.73}$ & A(1,2,3)\\
361608 & 09:59:30.58 & 02:08:57.78 & 5.9070$^{+0.0022}_{-0.0010}$ & 0.73$^{+0.50}_{-0.39}$ & 0.282$^{+0.192}_{-0.150}$ & 4.87$^{+29.17}_{-4.74}$ & C(1)\\
392109 & 09:59:21.48 & 02:14:09.26 & 6.0090$^{+0.0004}_{-0.0003}$ & 0.47$^{+0.05}_{-0.05}$ & 0.189$^{+0.021}_{-0.020}$ & 6.22$^{+35.17}_{-4.47}$ & C(3)\\
441761 & 09:59:35.16 & 02:09:53.31 & 5.9455$^{+0.0004}_{-0.0003}$ & 0.63$^{+0.10}_{-0.08}$ & 0.246$^{+0.038}_{-0.032}$ & 5.26$^{+313.37}_{-2.74}$ & B(1,3)\\
482804 & 09:59:33.50 & 02:09:17.17 & 6.0814$^{+0.0017}_{-0.0002}$ & 0.64$^{+0.08}_{-0.07}$ & 0.265$^{+0.0.033}_{-0.030}$ & 9.84$^{+30.96}_{-11.89}$ & A(1,2,3)\\
573604 & 09:59:36.65 & 02:10:37.28 & 5.9440$^{+0.0002}_{-0.0003}$ & 1.05$^{+0.08}_{-0.07}$ & 0.409$^{+0.030}_{-0.029}$ & 23.48$^{+1209.28}_{-7.83}$ & A(1,2,3)\\
582186 & 09:59:33.25 & 02:09:45.81 & 6.0359$^{+0.0002}_{-0.0001}$ & 1.12$^{+0.12}_{-0.10}$ & 0.453$^{+0.048}_{-0.038}$ & 31.75$^{+1052.30}_{-27.03}$ & B(1,2)\\
759747 & 09:59:31.11 & 02:10:31.32 & 6.0781$^{+0.0004}_{-0.0001}$ & 0.83$^{+0.07}_{-0.11}$ & 0.341$^{+0.030}_{-0.046}$ & 3.06$^{+952.19}_{-3.02}$ & B(1,2)\\
\hline
Interlopers\\
\hline
169797 & 09:59:16.21 & 02:18:17.85 & 6.3156$^{+0.0001}_{-0.0001}$ & 6.62$^{+0.16}_{-0.15}$ &2.974$^{+0.072}_{-0.065}$ & 7.97$^{+16.99}_{-1.98}$ & B(1,2)\\
345111 & 09:59:35.75 & 02:09:18.69 & 4.7571$^{+0.0015}_{-0.0006}$ & 0.73    $^{+0.08}_{-0.08}$ & 0.169$^{+0.019}_{-0.020}$ & 1.61$^{+1.21}_{-1.57}$ & C(1)\\
350720 & 09:59:34.48 & 02:13:54.03 & 5.5594$^{+0.0014}_{-0.0002}$ & 0.62$^{+1.41}_{-0.11}$ & 0.207$^{+0.470}_{-0.038}$ & 3.46$^{+28.84}_{-6.60}$ & A(1,2,3)\\
369661 & 09:59:23.33 & 02:14:04.42 & 6.1322$^{+0.0037}_{-0.0034}$ & 1.96$^{+1.96}_{-0.25}$ & 0.825$^{+0.817}_{-0.106}$ & 5.51$^{+102.56}_{-7.80}$ & C(2)\\
444487 & 09:59:30.56 & 02:09:10.48 & 4.2729$^{+0.0004}_{-0.0004}$ & 5.57$^{+0.09}_{-0.08}$ & 1.001$^{+0.017}_{-0.015}$ & -0.31$^{+0.31}_{-0.33}$ & A(1,2,3) \\
654354 & 09:59:16.82 & 02:20:15.24 & 6.3200$^{+0.0002}_{-0.0003}$ & 0.57$^{+0.14}_{-0.13}$ & 0.258$^{+0.062}_{-0.059}$ & 17.58$^{+666.68}_{-16.11}$ & C(1)\\ 
958367 & 09:59:24.54 & 02:12:27.15 & 6.1987$^{+0.0013}_{-0.0005}$ & 0.80$^{+0.07}_{-0.07}$ & 0.342$^{+0.029}_{-0.028}$ & 1.45$^{+1.96}_{-1.36}$ & B(1,3)\\
\hline
Non-detections\\
\hline
127337 & 09:59:22.49 & 02:12:53.54 & & & & & D\\
142959 & 09:59:20.56 & 02:13:10.62 & & & & & D\\
187778 & 09:59:23.08 & 02:13:20.10 & & & & & D\\
324132 & 09:59:31.67 & 02:13:50.86 & & & & & D\\
339126 & 09:59:06.74 & 02:18:56.87 & & & & & D\\
413243 & 09:59:23.09 & 02:09:04.73 & & & & & D\\
694706 & 09:59:38.29 & 02:11:12.90 & & & & & D\\
742465 & 09:59:35.88 & 02:11:29.06 & & & & & D\\
783817 & 09:59:27.96 & 02:10:39.35 & & & & & D\\
858263 & 09:59:26.26 & 02:16:28.20 & & & & & D\\
901963 & 09:59:44.44 & 02:07:10.92 & & & & & D\\
\hline
\end{tabular}
\end{table*} 

\begin{table*}
\caption{List of fit parameters for Ly$\alpha$ detections; Columns: (1) name; (2) Baseline from fit (3) lower limit of the Ly$\alpha$ rest-frame equivalent width, followed by the width if continuum is detected, both using values noise/flux $100\,$Å on the red end of the line.}
\label{tab:EW}
\centering
\begin{tabular}{r r l}
\hline
\hline
ID & Baseline & EW$_{\rm 0,Ly\alpha}$\\ 
& ${\rm [10^{-18}erg\,s^{-1}\,cm^{-2}]}$ & [Å]\\ 
(1) & (2) & (3)\\  
\hline
70505 & -0.029$^{+0.002}_{-0.002}$ & $>16$\\
156780 & -0.006$^{+0.004}_{-0.004}$ & $>5$\\
169797 & 0.030$^{+0.003}_{-0.003}$ & $>12/18$\\
186512 & 0.007$^{+0.003}_{-0.003}$ & $>11$\\
220530 & -0.007$^{+0.003}_{-0.002}$ & $>4$\\
225263 & 0.000$^{+0.003}_{-0.004}$ & $>1$\\
345111 & -0.015$^{+0.002}_{-0.002}$ & $>2$\\ 
350720 & 0.022$^{+0.005}_{-0.006}$ & $>2$\\ 
361608 & 0.000$^{+0.005}_{-0.004}$ & $>2$\\
369661 & -0.033$^{+0.004}_{-0.012}$ & $>4$\\
392109 & -0.007$^{+0.003}_{-0.002}$ & $>1$\\
441761 & 0.040$^{+0.004}_{-0.004}$ & $>1/3$\\
444487 & -0.015$^{+0.004}_{-0.004}$ & $>15$\\
482804 & 0.010$^{+0.006}_{-0.006}$ & $>1$\\
573604 & 0.081$^{+0.004}_{-0.004}$ & $>2/1$\\
582186 & -0.007$^{+0.004}_{-0.004}$ & $>2$\\
654354 & -0.001$^{+0.003}_{-0.002}$ & $>1$\\ 
759747 & -0.017$^{+0.005}_{-0.003}$ & $>2$\\
958367 & -0.019$^{+0.003}_{-0.003}$ & $>3$\\
\hline
\end{tabular}
\end{table*} 
\section{Conclusions}\label{section:conclusions}
Using the {\it DEIMOS} spectrograph on the {\it Keck} telescope, we have detected the redshifted Lyman-$\alpha$ line toward 10 protocluster galaxies at $z\simeq 6$ identified in the COSMOS field \citep{Brinch2023}.
This unequivocally confirms the discovery of a massive and rich protocluster at this early epoch. 
We obtained an overall spectroscopic success rate of 
47\% when considering the full sample.
\\
We applied skewed Gaussian fits to the spectral lines and determined their fluxes, equivalent widths and skewness values. Notably, the galaxies exhibited relatively faint line emission, characterized by narrow to intermediate equivalent widths and positive skewness. Additionally, we observed possible continuum detections in two galaxies within our sample, suggesting we have not reached the necessary depth to fully uncover the complete line profiles in these galaxies.
\\
Incorporating the new spectroscopic redshifts into our overdensity analysis, we found a notable increase in the peak overdensity for the protocluster, now measuring $\rm \delta=11.6\,(10.9\sigma)$.
\\
We find that the protocluster is part of a large-scale structure in the centre of the COSMOS field, with multiple significant overdensity peaks connected by filaments. We also report the discovery of a hitherto unknown $z\simeq 6$ protocluster that is part of the same large-scale structure as PC$z$6.05-1.  
\\
We fitted the SEDs of the galaxies using {\sc bagpipes} and found them to be dust-poor main sequence galaxies, in some cases with young ages ($\rm \lesssim100$ Myr), indicative of a recent starburst. The most massive galaxy (ID 582186) also has the oldest age of $386\,{\rm Myr}$ and is possibly the progenitor of what will become the brightest cluster galaxy (BCG) at later times.
\\
We estimated the dark matter halo mass and found it to be $\sim10^{12}\rm M_{\rm \odot}$. This suggests an evolution into a cluster resembling the Virgo or Coma clusters in the present-day Universe.
\\
We discuss the redshift extent of the galaxies' possible evolution scenarios, ranging from possible substructure in the protocluster to the possibility that the galaxies inhabit two separate protocluster that could merge over time.
\section{Acknowledgements}
The Cosmic Dawn Center (DAWN) is funded by the Danish National Research Foundation under grant No. 140. T.R.G. and M.B. are grateful for support from the Carlsberg Foundation via grant No.~CF20-0534. The data presented herein were obtained at the W. M. Keck Observatory, which is operated as a scientific partnership among the California Institute of Technology, the University of California and the National Aeronautics and Space Administration. The Observatory was made possible by the generous financial support of the W. M. Keck Foundation. The authors wish to recognize and acknowledge the very significant cultural role and reverence that the summit of Maunakea has always had within the indigenous Hawaiian community.  We are most fortunate to have the opportunity to conduct observations from this mountain. This work was partially supported by DeiC National HPC (g.a. DeiC-DTU-L-20210103) and by the “dtu\_00026” project.

\section*{Data Availability}
The original sample of galaxies came from the COSMOS2020 catalogue \citep{Weaver2021} and is available from \url{https://cosmos2020.calet.org/}. The original spectra are available upon reasonable request.



\bibliographystyle{mnras}
\bibliography{mnras_template} 




\appendix
\section{Interlopers galaxies images and spectra}\label{sec:interlopers}
Figure \ref{fig:appendix 1 protocluster members} show the true color images and spectra for the interloper galaxies.
\begin{figure*}
     \centering
     \begin{subfigure}{\textwidth}
     \centering
         \includegraphics[width=0.37\textwidth]{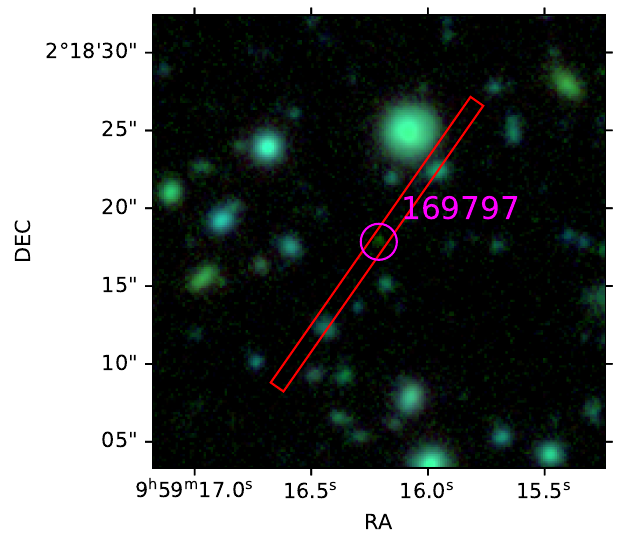}
         \centering
         \includegraphics[width=0.49\textwidth]{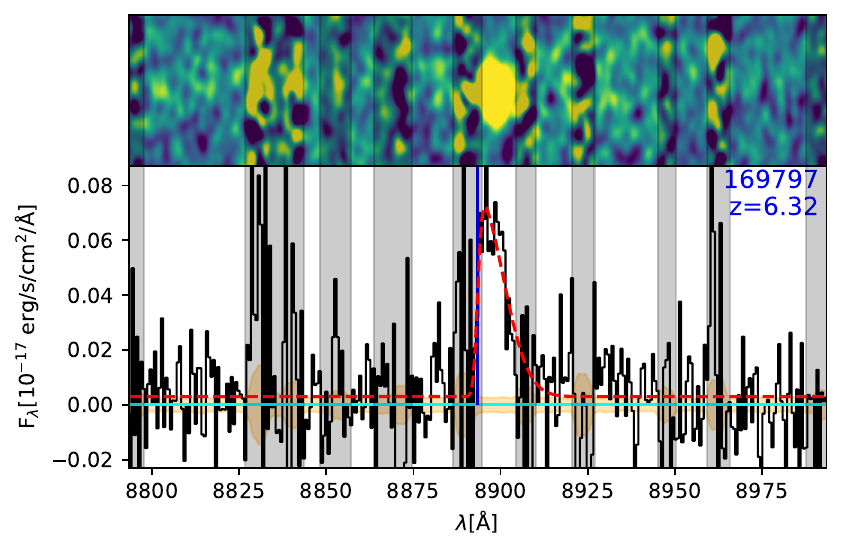}
         \caption{169797}
         \label{fig:169797}
     \end{subfigure}
     \hfill
     \begin{subfigure}{\textwidth}
     \centering
         \includegraphics[width=0.37\textwidth]{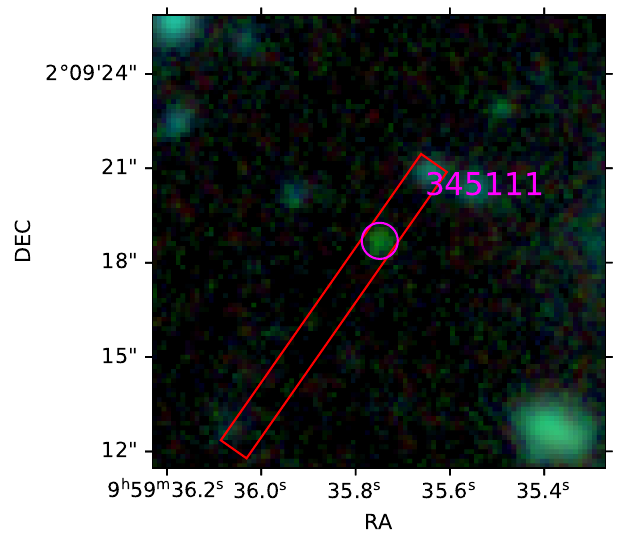}
         \centering
         \includegraphics[width=0.49\textwidth]{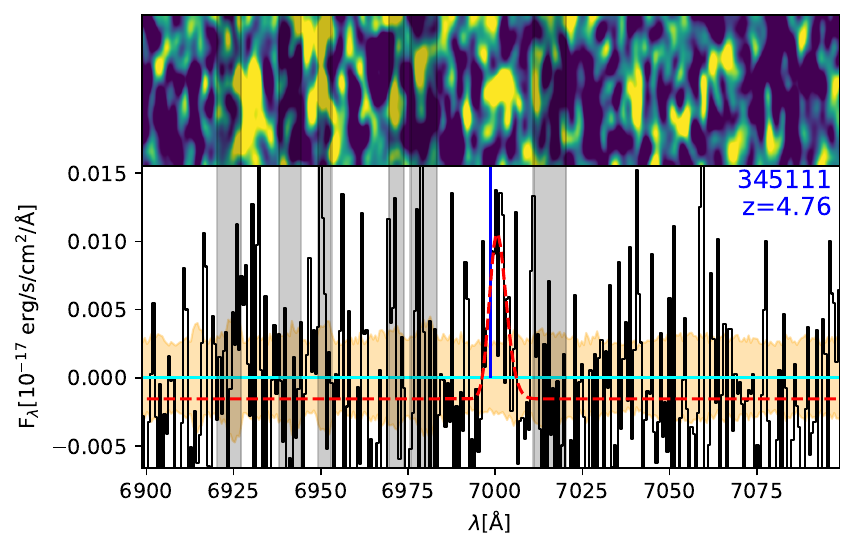}
         \caption{345111}
         \label{fig:345111}
     \end{subfigure}
     \begin{subfigure}{\textwidth}
     \centering
         \includegraphics[width=0.37\textwidth]{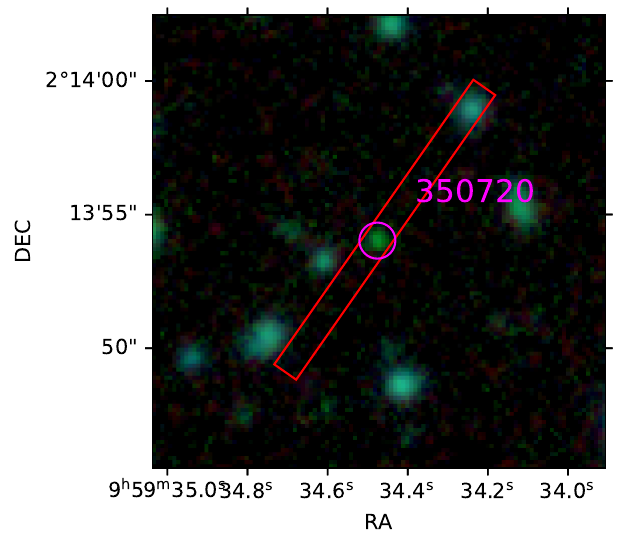}
         \centering
         \includegraphics[width=0.49\textwidth]{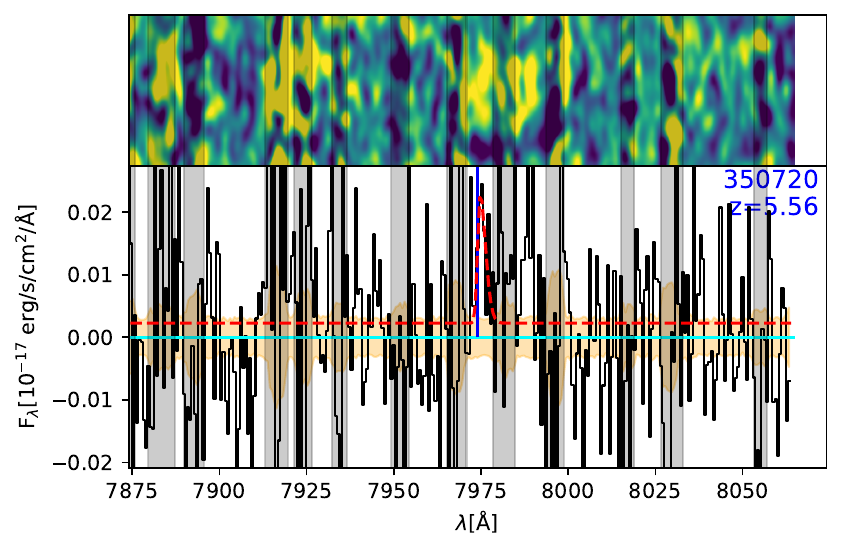}
         \caption{350720}
         \label{fig:350720}
     \end{subfigure}
        \caption{Continued on next page.}
        \label{fig:appendix 1 protocluster members}
\end{figure*}

\begin{figure*}\ContinuedFloat
     \centering
     \begin{subfigure}{\textwidth}
     \centering
         \includegraphics[width=0.37\textwidth]{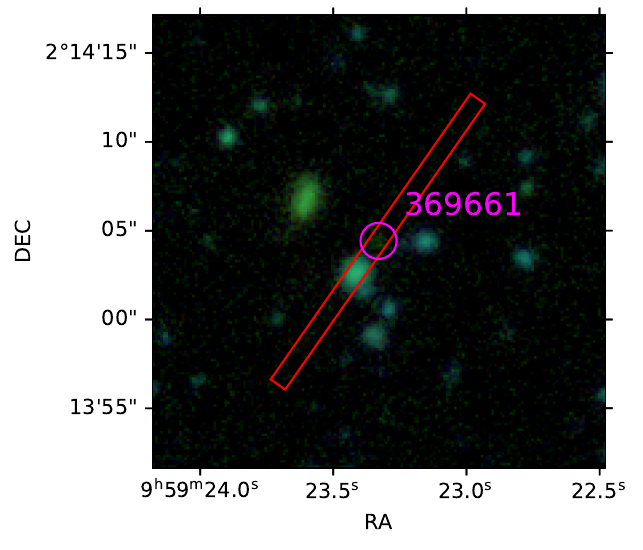}
         \centering
         \includegraphics[width=0.49\textwidth]{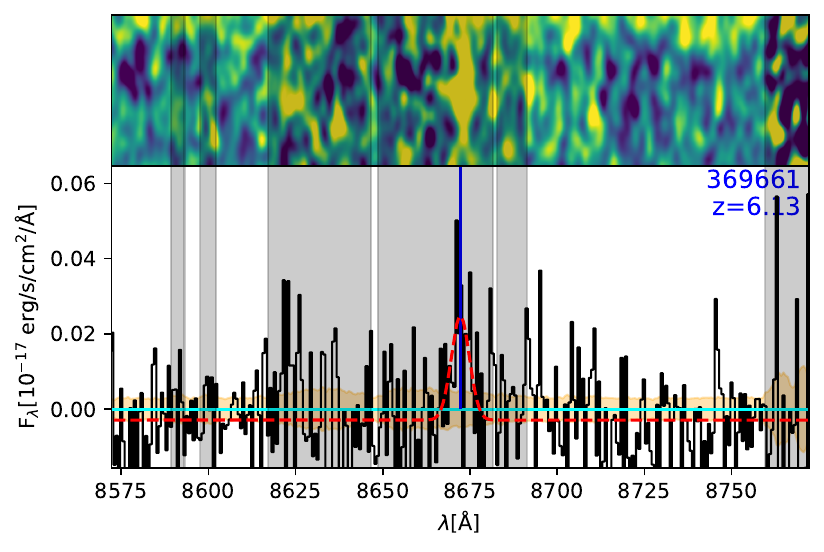}
         \caption{369661}
         \label{fig:369661}
     \end{subfigure}
     \hfill
     \begin{subfigure}{\textwidth}
     \centering
         \includegraphics[width=0.37\textwidth]{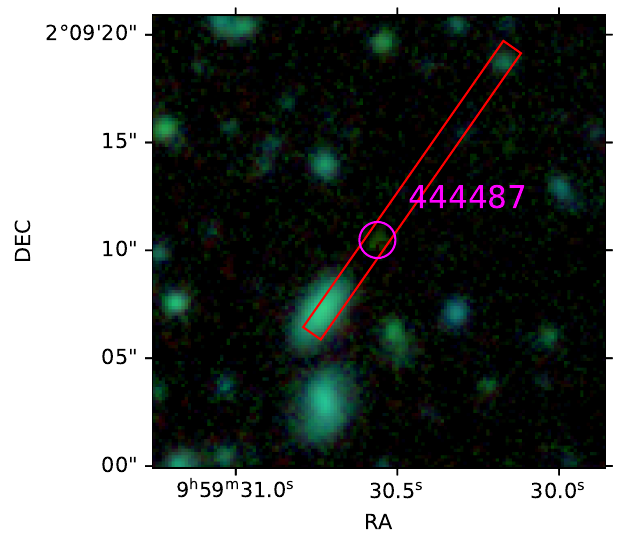}
         \centering
         \includegraphics[width=0.49\textwidth]{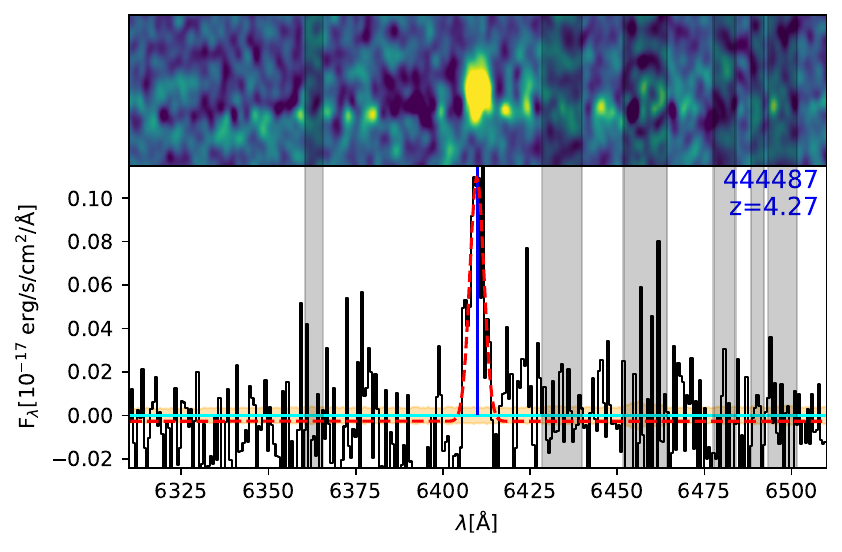}
         \caption{444487}
         \label{fig:444487}
     \end{subfigure}
     \begin{subfigure}{\textwidth}
     \centering
         \includegraphics[width=0.37\textwidth]{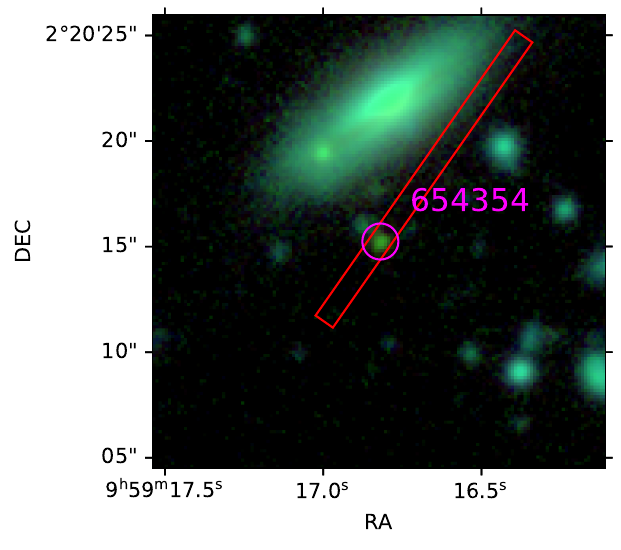}
         \centering
         \includegraphics[width=0.49\textwidth]{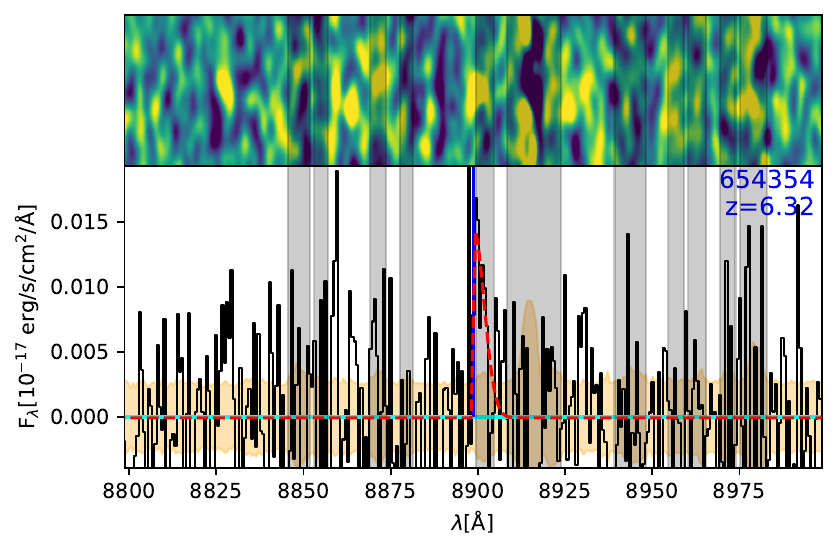}
         \caption{654354}
         \label{fig:654354}
     \end{subfigure}
        \caption{Continued on next page.}
        \label{fig:appendix 2 protocluster members}
\end{figure*}

\begin{figure*}\ContinuedFloat
     \centering
     \begin{subfigure}{\textwidth}
     \centering
         \includegraphics[width=0.37\textwidth]{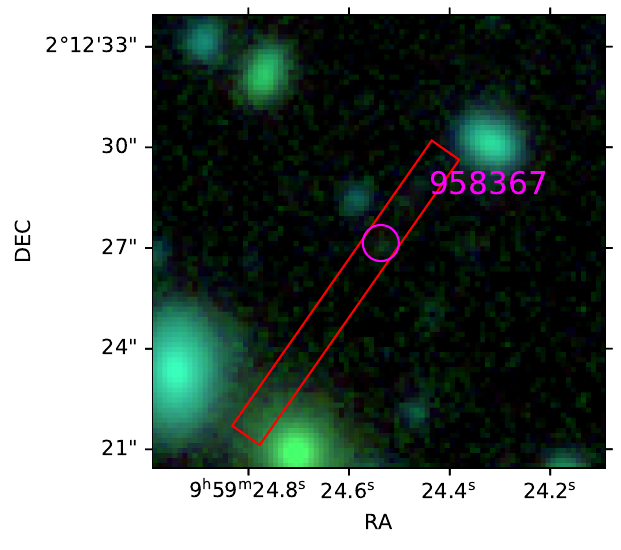}
         \centering
         \includegraphics[width=0.49\textwidth]{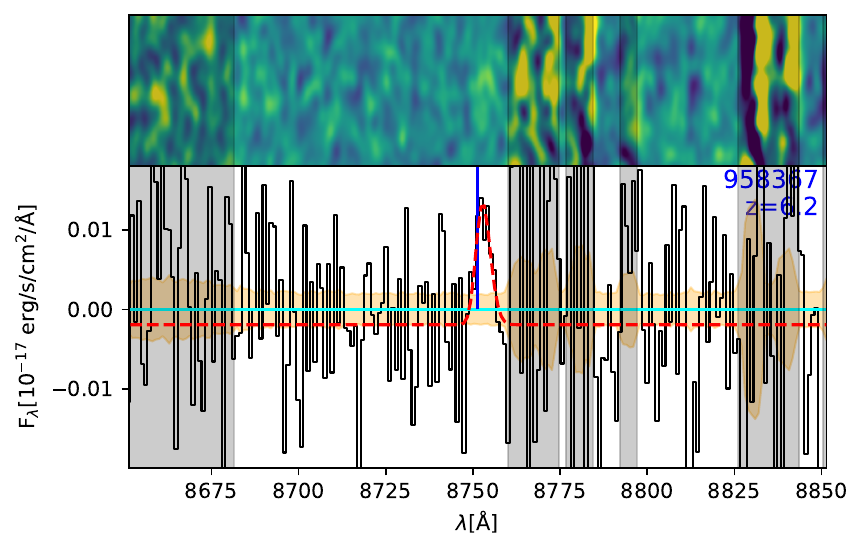}
         \caption{958367}
         \label{fig:958367}
     \end{subfigure}
        \caption{2D, 1D and Yzi RGB true-colour images of the interloper galaxies. The left side shows VISTA Yzi RBG true colour images using the COSMOS2020 catalogue photometry. The slit used to observe each object is plotted on top of the image and a magenta circle with an accompanying object ID is shown to highlight the position of the object in the image. The right side shows three-panel figures of the 2D (top), 1D (middle) and Signal-to-Noise Ratio (SNR) (bottom) for the line detection of each object. A Gaussian filter has been applied to the 2D spectra to visually highlight the detection. The black solid lines are the extracted 1D spectra. The vertical blue line indicated the spectroscopic redshift of the objects. The shaded yellow area is the noise. The grey-shaded regions in all three panels highlight the sky regions, where skylines are prevalent.}
        \label{fig:appendix 3 protocluster members}
\end{figure*}

\section{Table for other objects on slit}\label{sec:other objects}
Table \ref{tab:speczLOW} shows the lines and redshift found for other objects on the {\it DEIMOS} slits.
\begin{table*}
\caption{List of other lines found from other objects present on the slits with their parameters; Columns: (1) COSMOS2020 catalogue ID.; (2) \& (3) R.A. and Dec.; (4) spectroscopic redshift. The lines observed are [OII] at 3726 Å and 3729 Å, H$\beta$ at 4861 Å and [OIII] at 4959 Å and 5007 Å.}
\label{tab:speczLOW}
\centering
\begin{tabular}{l l l l l l l l l}
\hline
\hline
ID & RA & Dec  & $z_{\rm spec}$\\ 
& {\it hh:mm:ss.ss} & {\it dd:mm:ss.ss} & \\ 
(1) & (2) & (3)\\ 
\hline
Other objects on slit\\
\hline
159501-[O\textsc{ii}] & 09:59:16.43 & 02:18:12.28 & 0.8915$^{+0.0001}_{-0.0001}$\\
159501-[O\textsc{ii}]3729 & 09:59:16.43 & 02:18:12.28 & 0.8916$^{+0.0001}_{-0.0001}$\\
159501-H$\beta$ & 09:59:16.43 & 02:18:12.28 & 0.8915$^{+0.0001}_{-0.0001}$\\
159501-[O\textsc{iii}]4959 & 09:59:16.43 & 02:18:12.28 & 0.8910$^{+0.0001}_{-0.0001}$\\
159501-[O\textsc{iii}]5007 & 09:59:16.43 & 02:18:12.28 & 0.8915$^{+0.0001}_{-0.0001}$\\
169710-[O\textsc{ii}]3726 & 09:59:20.37 & 02:13:14.01 & 1.0276$^{+0.0001}_{-0.0001}$\\
169710-[O\textsc{ii}]3729 & 09:59:20.37 & 02:13:14.01 & 1.0276$^{+0.0001}_{-0.0001}$\\
207402-[O\textsc{ii}]3726 & 09:59:15.96 & 02:18:22.39 & 0.8525$^{+0.0001}_{-0.0001}$\\
207402-[O\textsc{ii}]3729 & 09:59:15.96 & 02:18:22.39 & 0.8525$^{+0.0001}_{-0.0001}$\\
207402-[O\textsc{iii}]4959 & 09:59:15.96 & 02:18:22.39 & 0.8527$^{+0.0003}_{-0.0001}$\\
207402-[O\textsc{iii}]5007 & 09:59:15.96 & 02:18:22.39 & 0.8525$^{+0.0001}_{-0.0001}$\\
216791-[O\textsc{iii}]3726 & 09:59:27.42 & 02:08:16.19 & 0.8182$^{+0.0001}_{-0.0001}$\\
216791-[O\textsc{iii}]3729 & 09:59:27.42 & 02:08:16.19 & 0.8181$^{+0.0001}_{-0.0001}$\\
216791-[O\textsc{iii}]4959 & 09:59:27.42 & 02:08:16.19 & 0.8185$^{+0.0001}_{-0.0001}$\\
216791-[O\textsc{iii}]5007 & 09:59:27.42 & 02:08:16.19 & 0.8192$^{+0.0001}_{-0.0001}$\\
356445-[O\textsc{ii}]3726 & 09:59:30.17 & 02:19:03.84 & 0.9250$^{+0.0001}_{-0.0001}$\\
356445-[O\textsc{ii}]3729 & 09:59:30.17 & 02:19:03.84 & 0.9243$^{+0.0001}_{-0.0001}$\\
369301-[O\textsc{ii}]3726 & 09:59:23.41 & 02:14:02.56 & 0.9571$^{+0.0001}_{-0.0001}$\\
369301-[O\textsc{ii}]3729 & 09:59:23.41 & 02:14:02.56 & 0.9572$^{+0.0001}_{-0.0001}$\\
407631-H$\beta$ & 09:59:30.17 & 02:19:19.83 & 0.3732$^{+0.0001}_{-0.0001}$\\
407631-[O\textsc{iii}]4959 & 09:59:30.17 & 02:19:19.83 & 0.3733$^{+0.0001}_{-0.0001}$\\
407631-[O\textsc{iii}]5007 & 09:59:30.17 & 02:19:19.83 & 0.3734$^{+0.0001}_{-0.0001}$\\ 
437549-[O\textsc{ii}]3726 & 09:59:30.73 & 02:09:07.20 & 1.0367$^{+0.0001}_{-0.0000}$\\
437549-[O\textsc{ii}]3729 & 09:59:30.73 & 02:09:07.20 & 1.0363$^{+0.0001}_{-0.0001}$\\
\hline
\end{tabular}
\end{table*} 
\section{Comparison of star formation histories for a constant or non-parametric model when using {\sc Bagpipes}}\label{SFH comparison}
Figure \ref{fig:appendix 2 SFH comparison} shows the SFH for our 10 protocluster galaxies using a constant SFH model and a non-parametric one when fitting with {\sc Bagpipes}.
\begin{figure*}
     \centering
     \begin{subfigure}{\textwidth}
     \centering
         \includegraphics[width=0.49\textwidth]{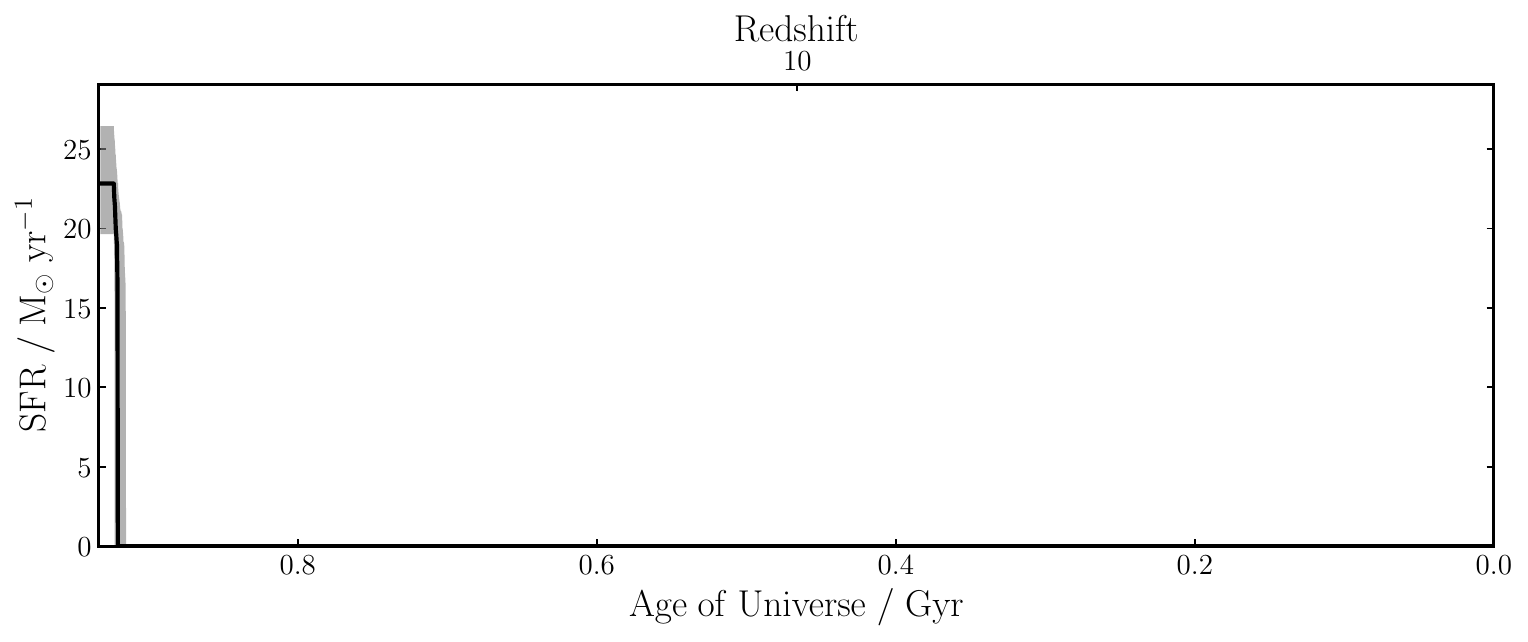}
         \centering
         \includegraphics[width=0.49\textwidth]{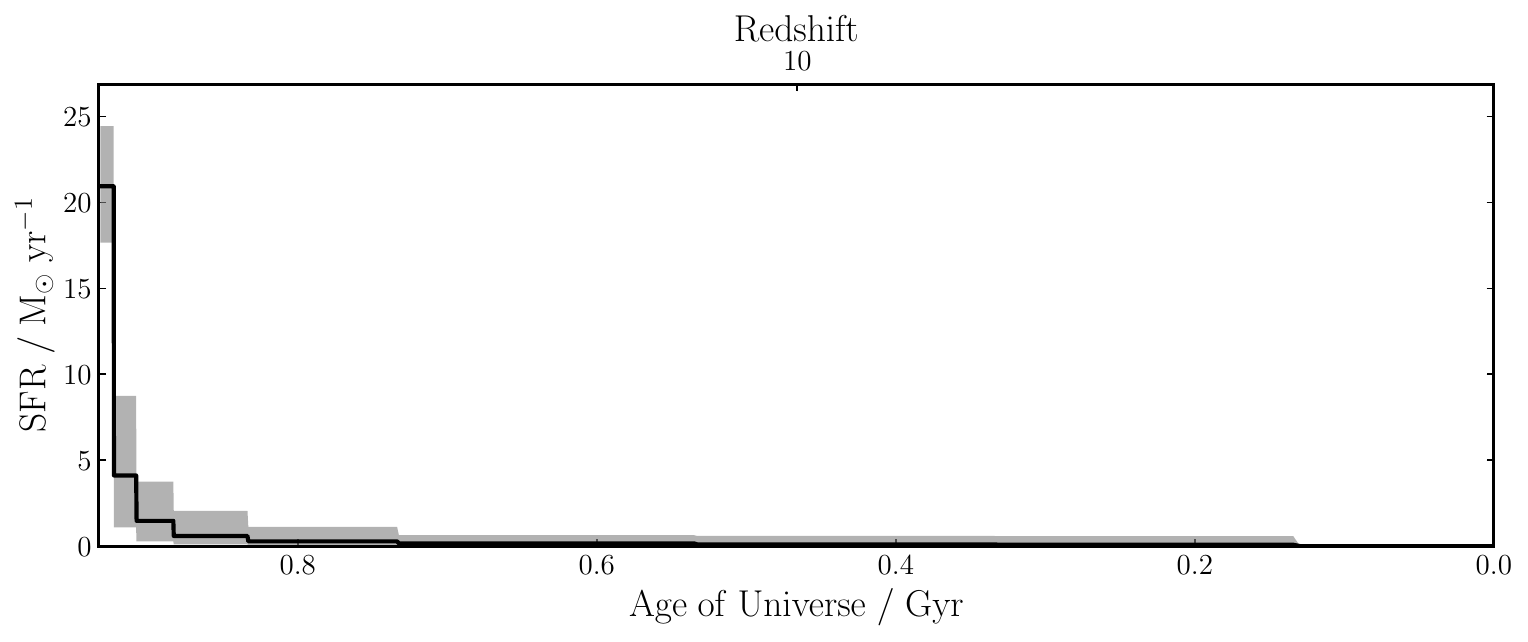}
         \caption{70505}
         \label{fig:70505}
     \end{subfigure}
     \hfill
     \begin{subfigure}{\textwidth}
     \centering
         \includegraphics[width=0.49\textwidth]{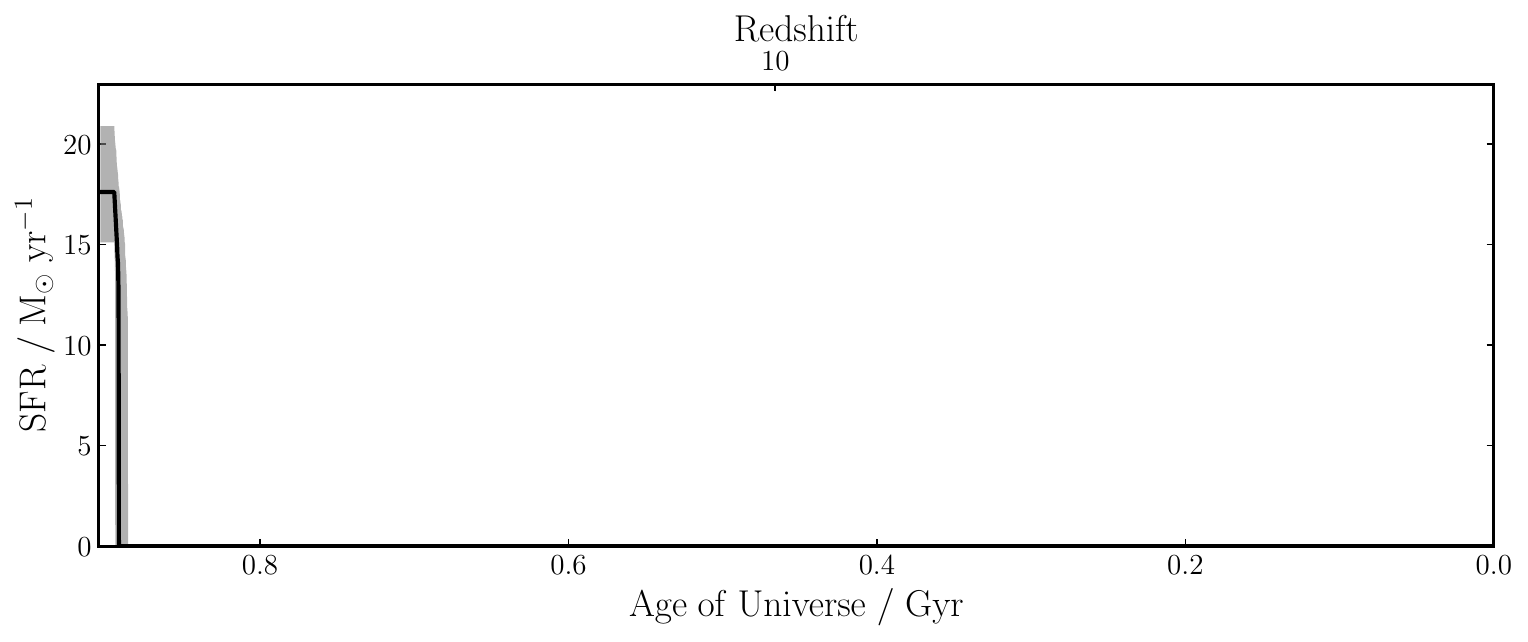}
         \centering
         \includegraphics[width=0.49\textwidth]{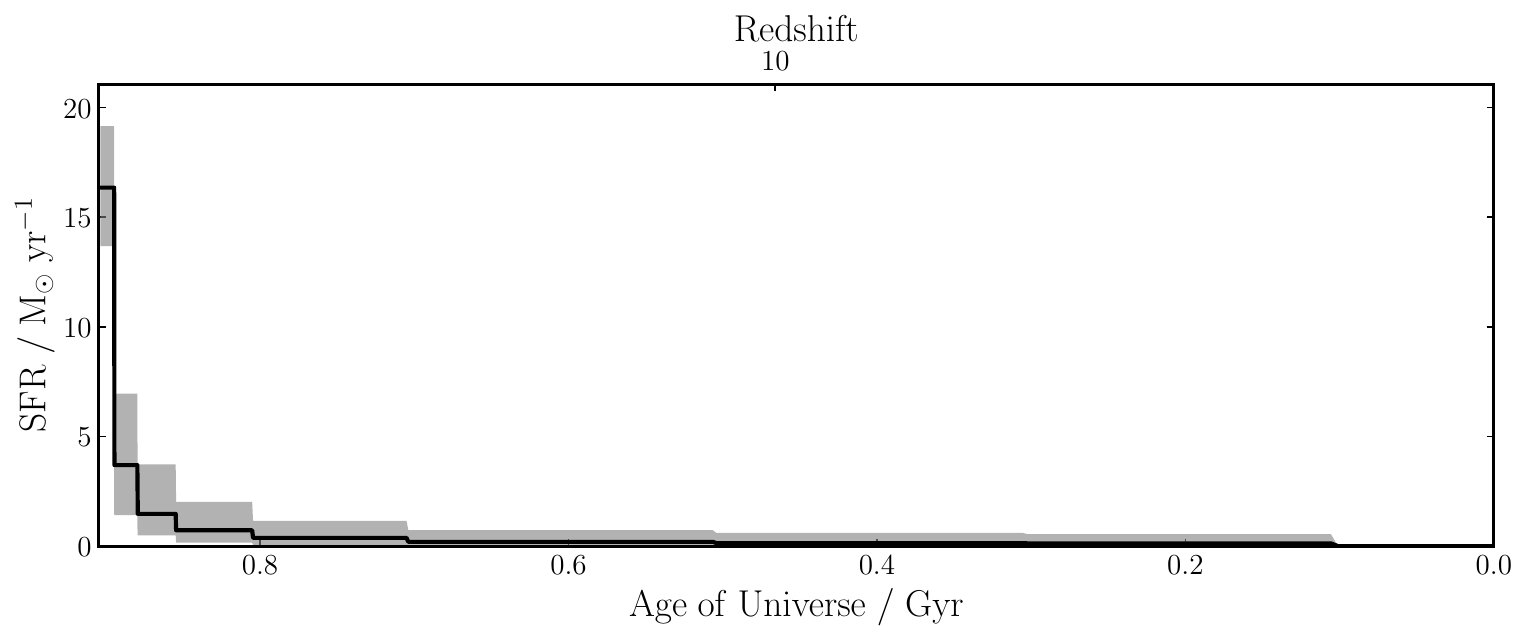}
         \caption{156780}
         \label{fig:156780}
     \end{subfigure}
     \begin{subfigure}{\textwidth}
     \centering
         \includegraphics[width=0.49\textwidth]{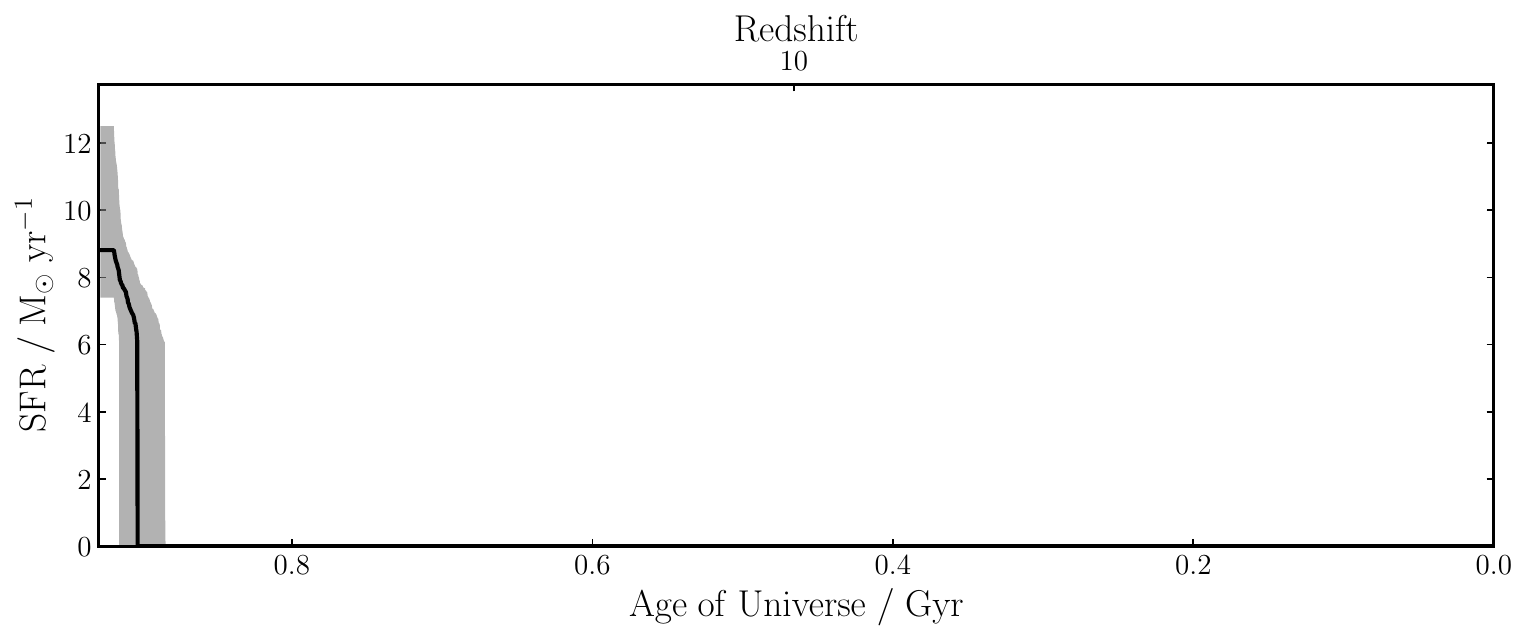}
         \centering
         \includegraphics[width=0.49\textwidth]{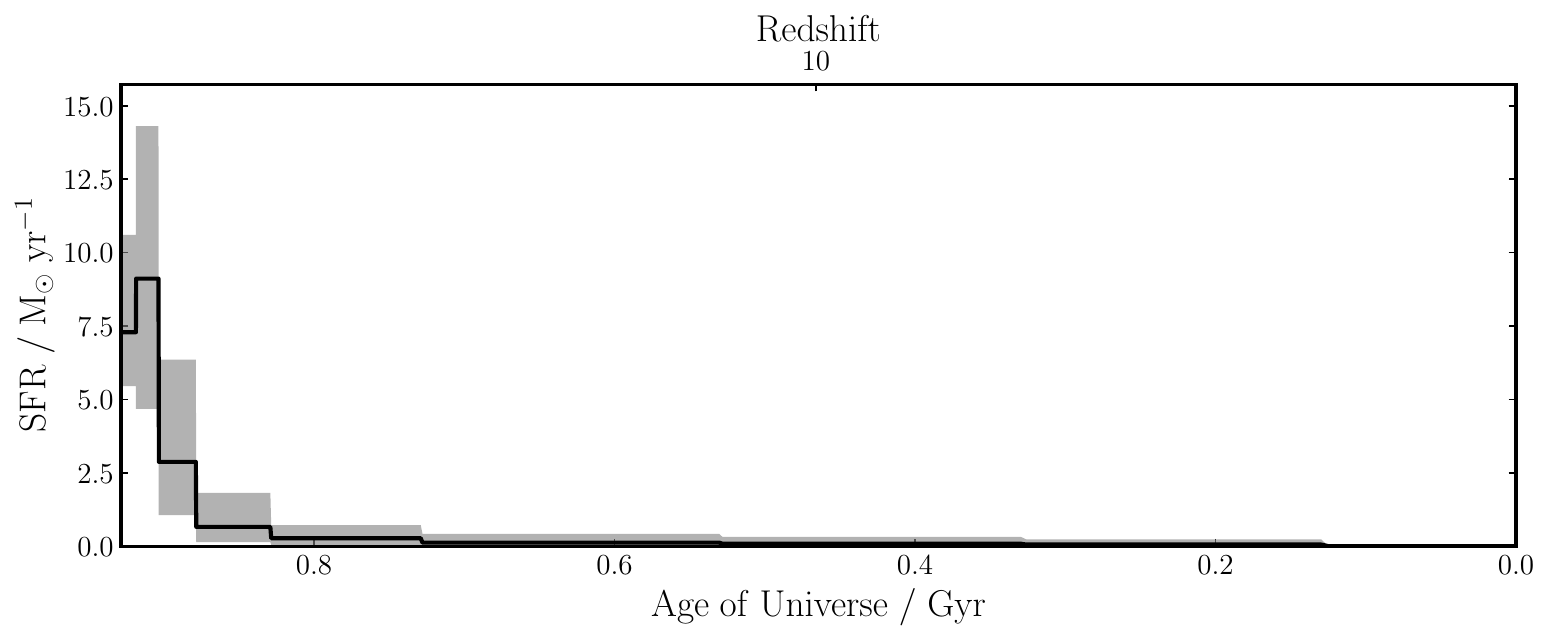}
         \caption{186512}
         \label{fig:186512}
     \end{subfigure}
     \begin{subfigure}{\textwidth}
     \centering
         \includegraphics[width=0.49\textwidth]{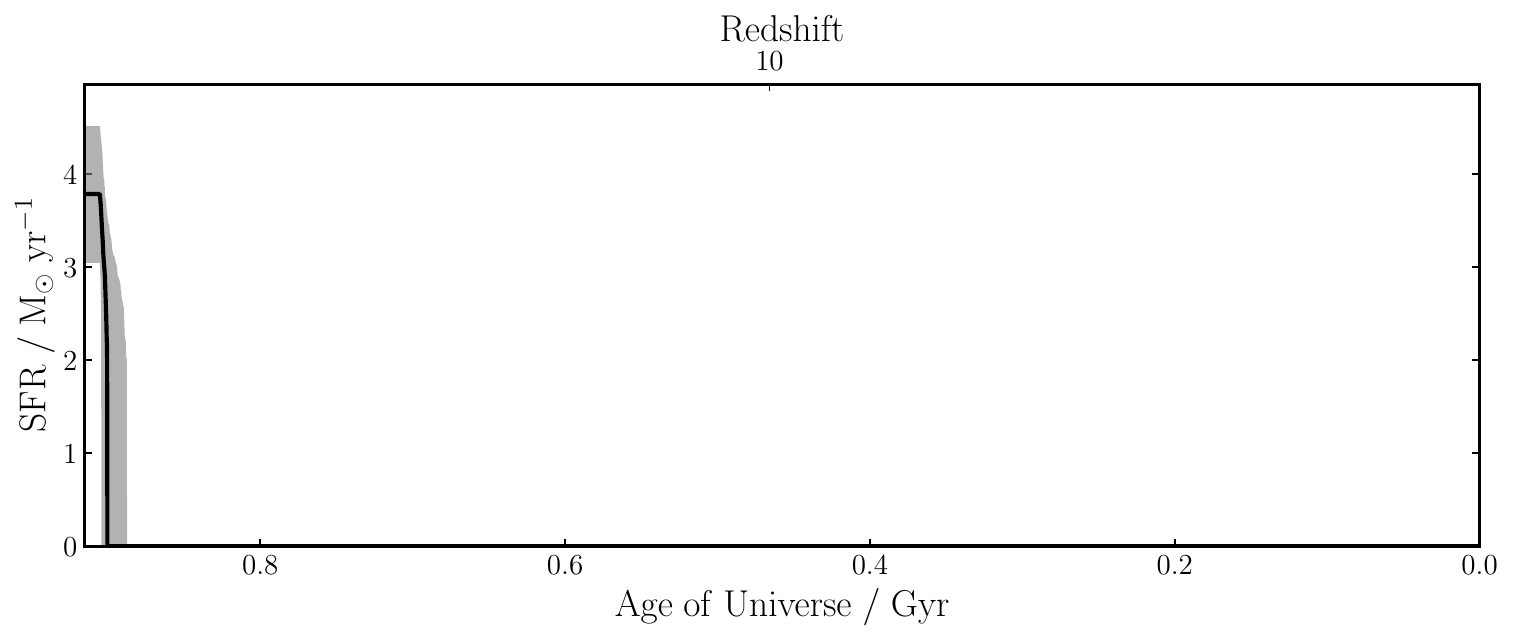}
         \centering
         \includegraphics[width=0.49\textwidth]{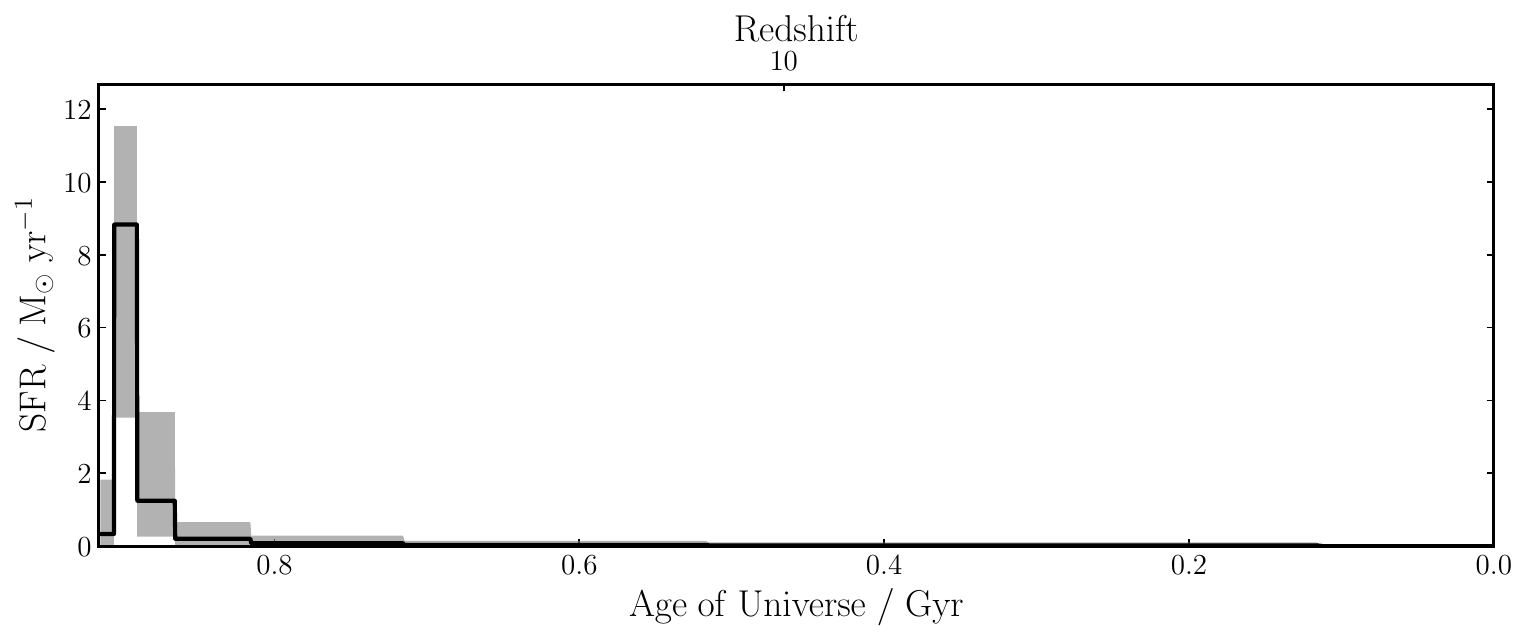}
         \caption{220530}
         \label{fig:220530}
     \end{subfigure}
     \begin{subfigure}{\textwidth}
     \centering
         \includegraphics[width=0.49\textwidth]{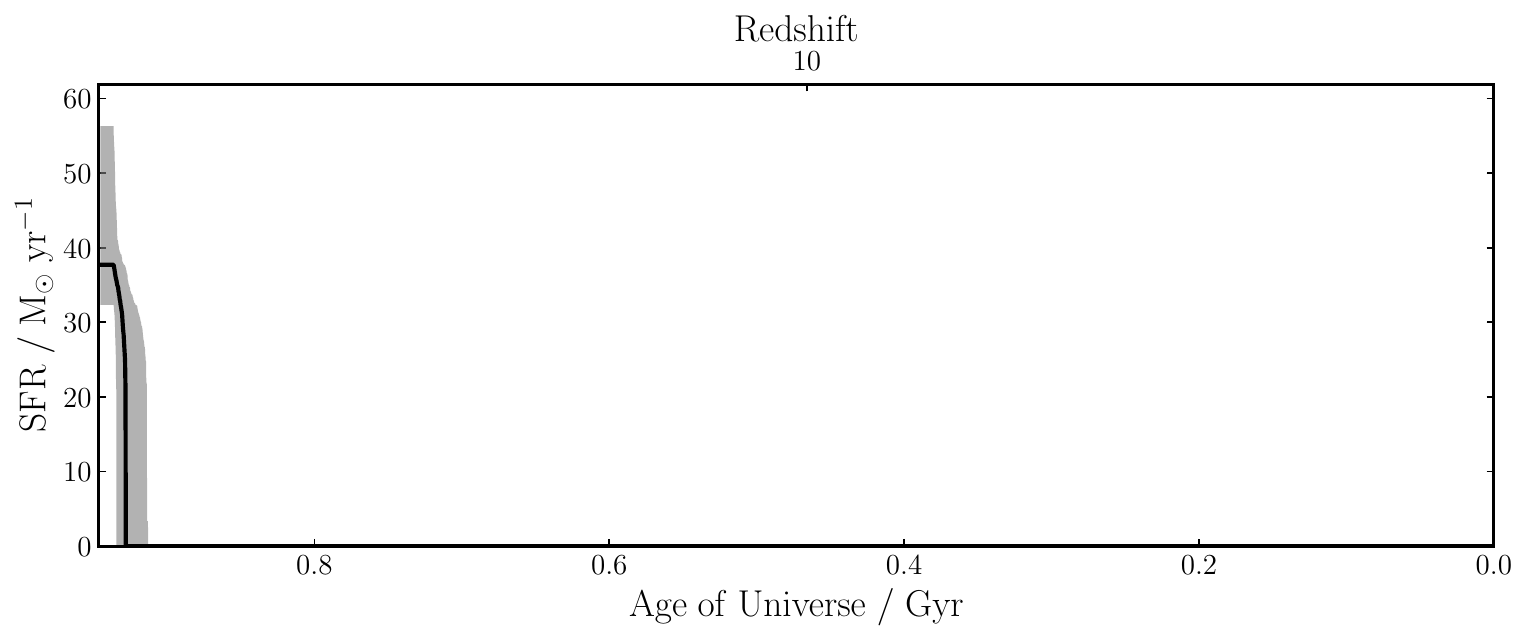}
         \centering
         \includegraphics[width=0.49\textwidth]{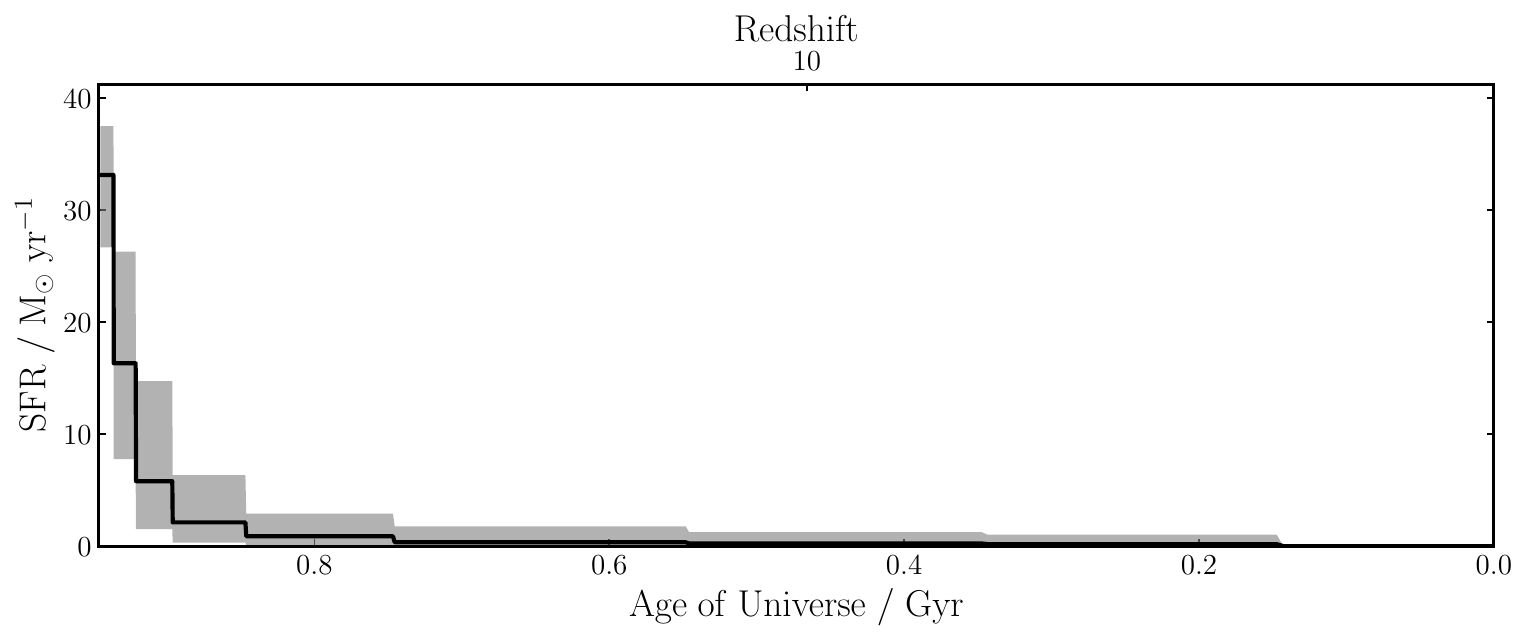}
         \caption{225263}
         \label{fig:225263}
     \end{subfigure}
        \caption{Continued on next page.}
        \label{fig:appendix 1 SFH comparison}
\end{figure*}
\begin{figure*}\ContinuedFloat
     \centering
     \begin{subfigure}{\textwidth}
     \centering
         \includegraphics[width=0.49\textwidth]{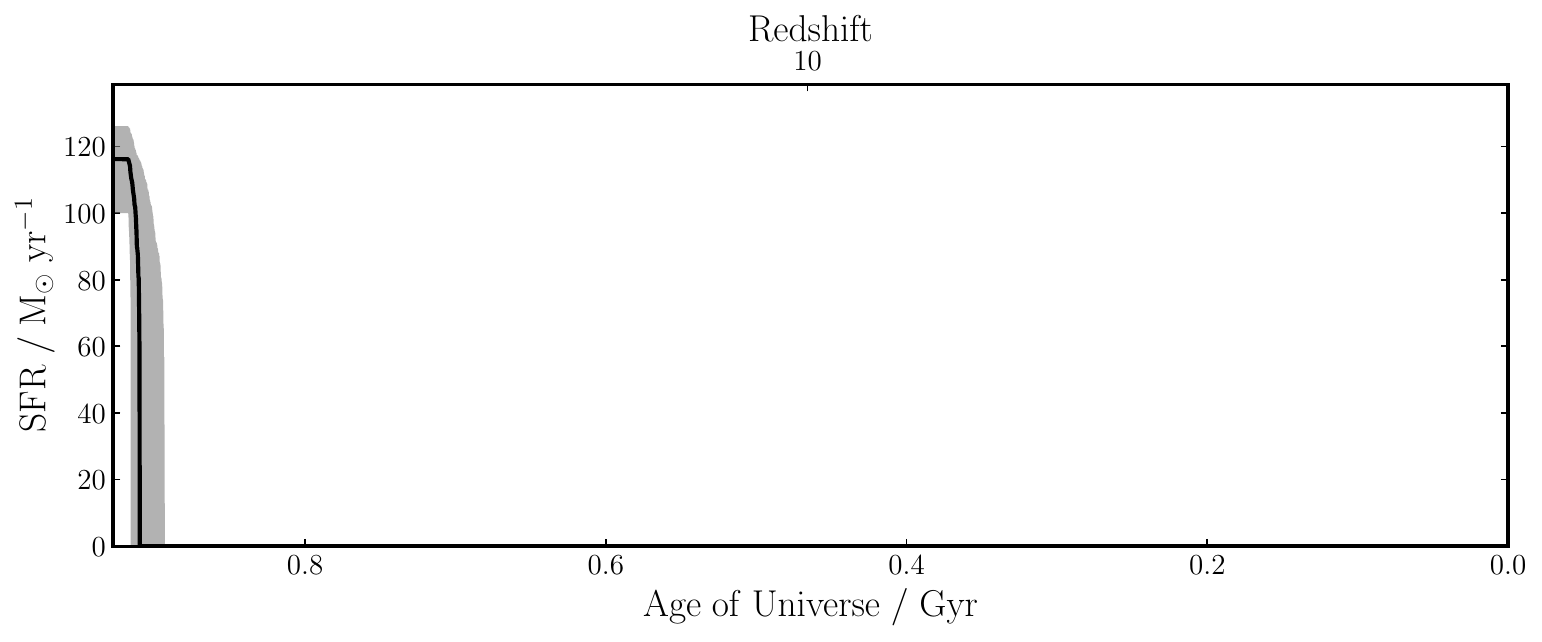}
         \centering
         \includegraphics[width=0.49\textwidth]{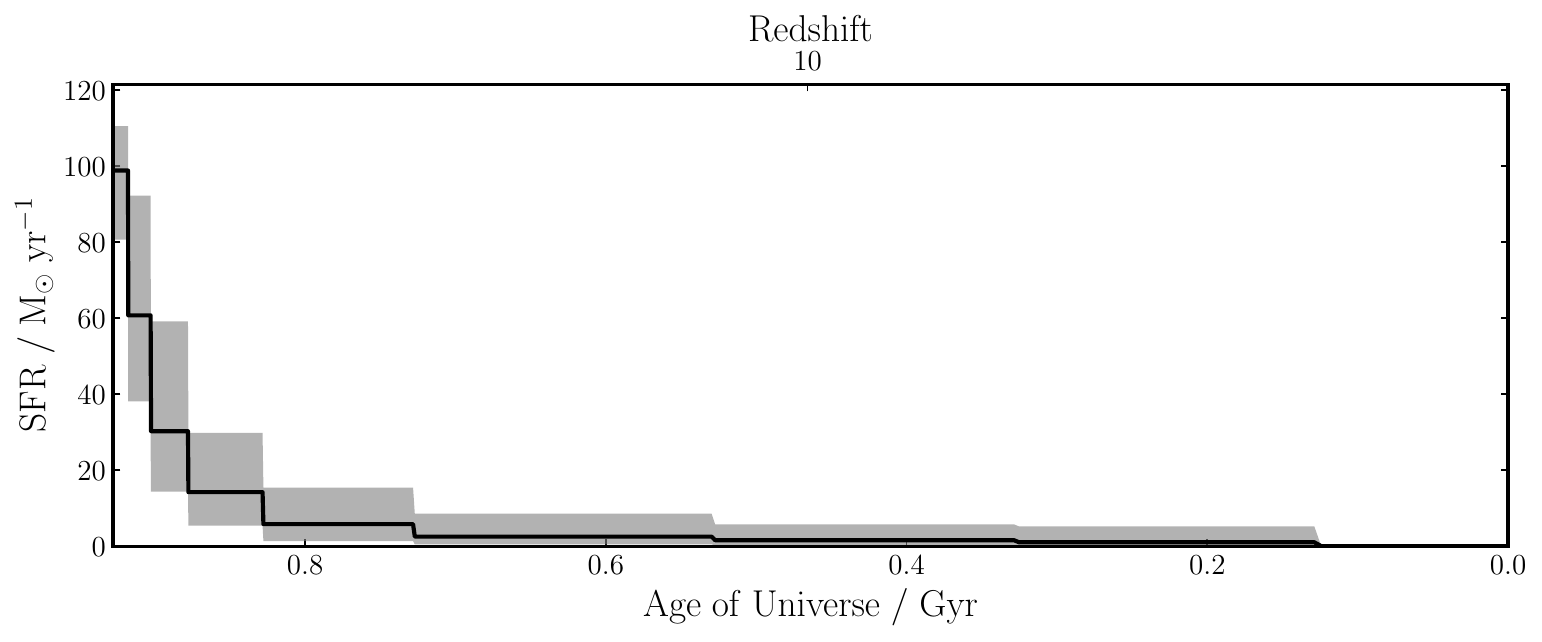}
         \caption{441761}
         \label{fig:441761}
     \end{subfigure}
     \begin{subfigure}{\textwidth}
     \centering
         \includegraphics[width=0.49\textwidth]{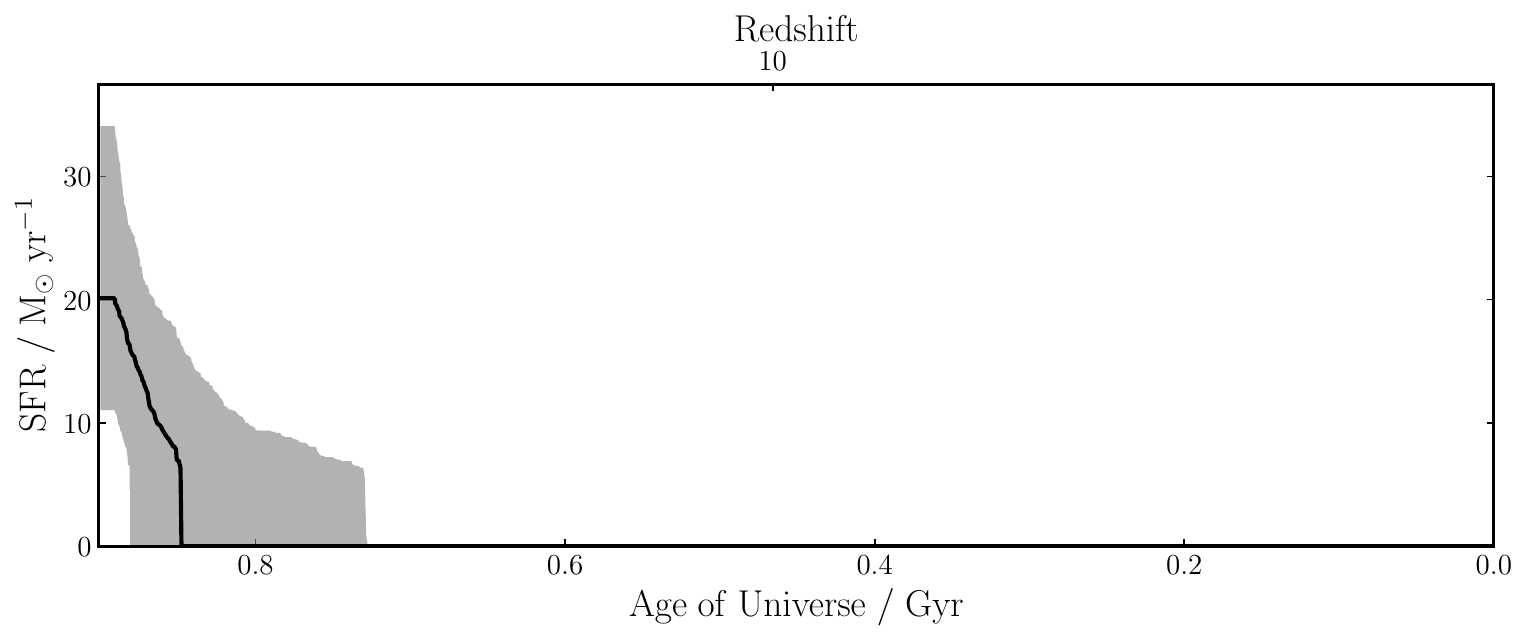}
         \centering
         \includegraphics[width=0.49\textwidth]{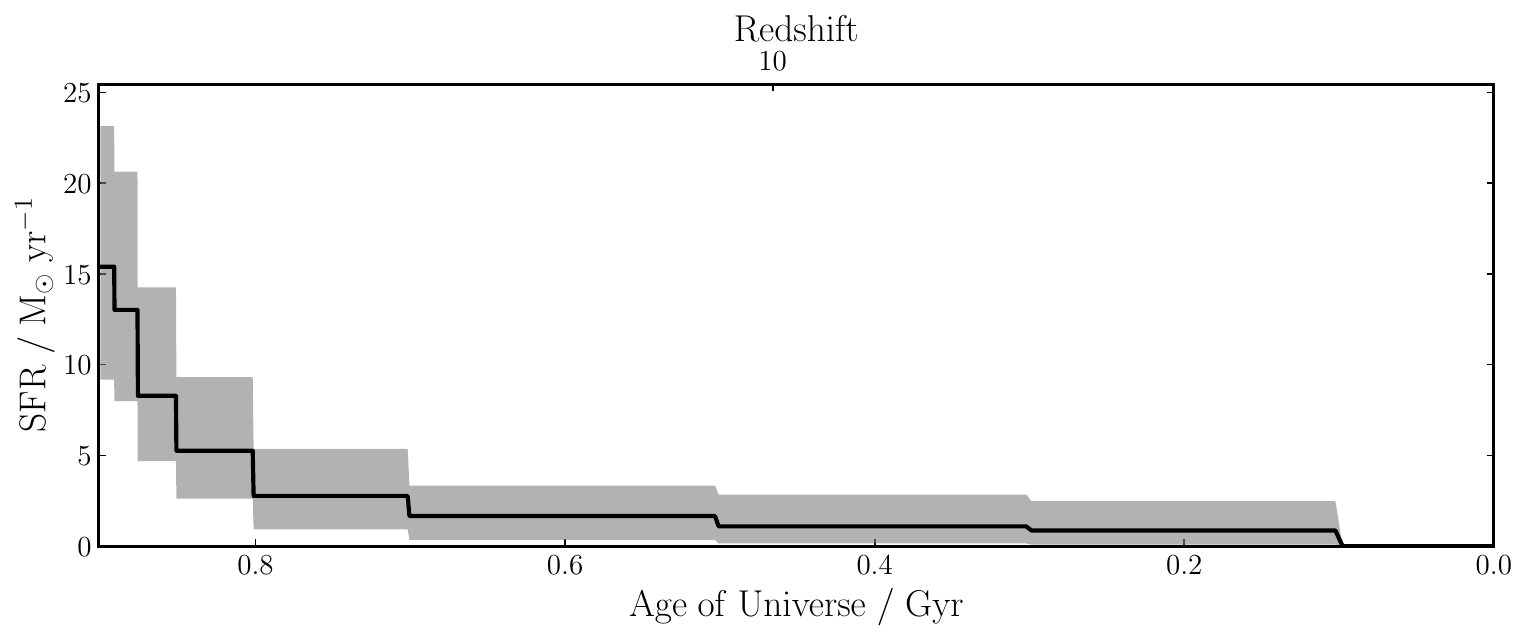}
         \caption{482804}
         \label{fig:482804}
     \end{subfigure}
     \begin{subfigure}{\textwidth}
     \centering
         \includegraphics[width=0.49\textwidth]{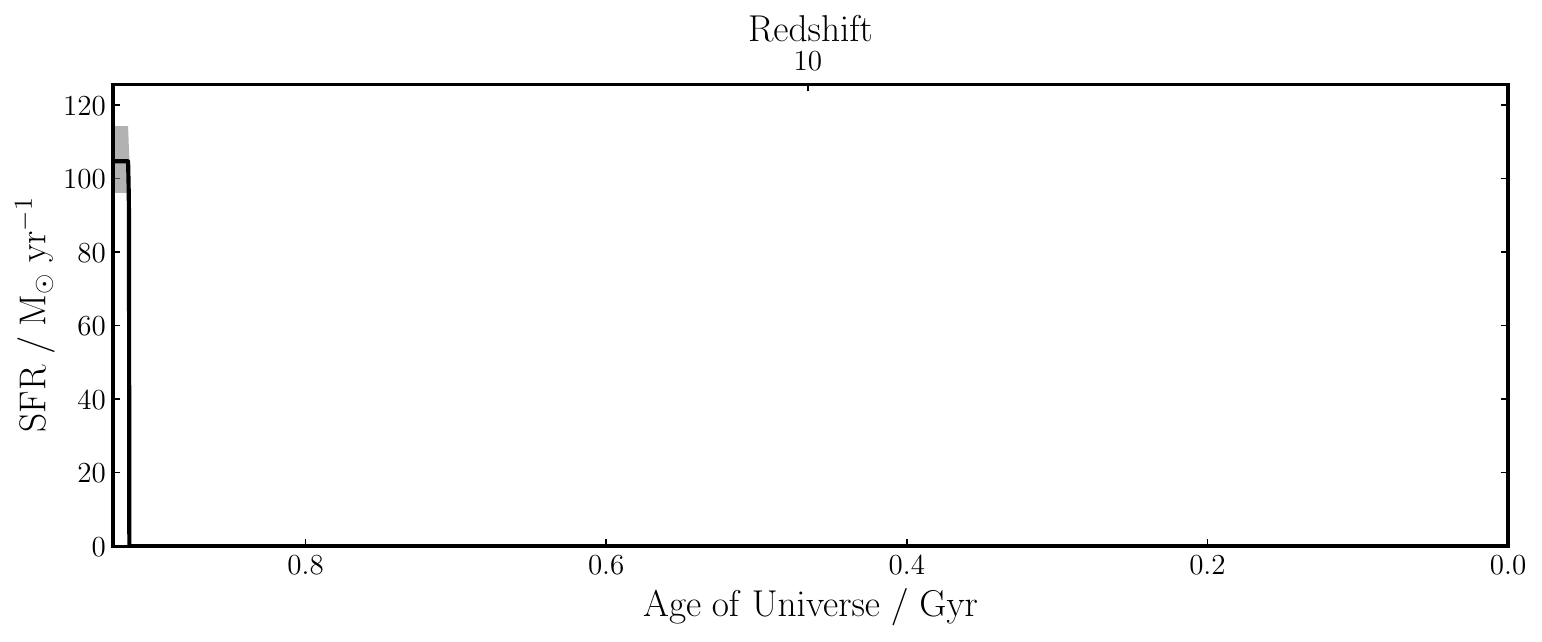}
         \centering
         \includegraphics[width=0.49\textwidth]{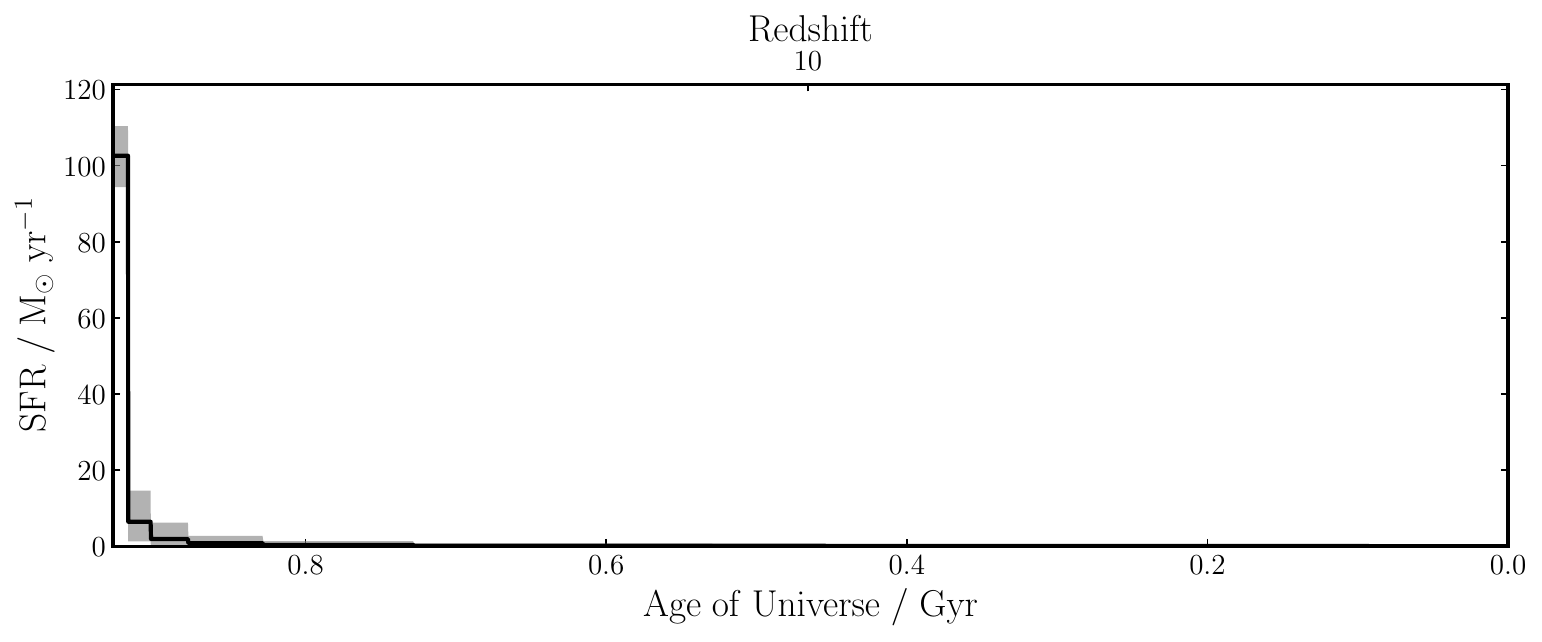}
         \caption{573604}
         \label{fig:573604}
     \end{subfigure}
     \begin{subfigure}{\textwidth}
     \centering
         \includegraphics[width=0.49\textwidth]{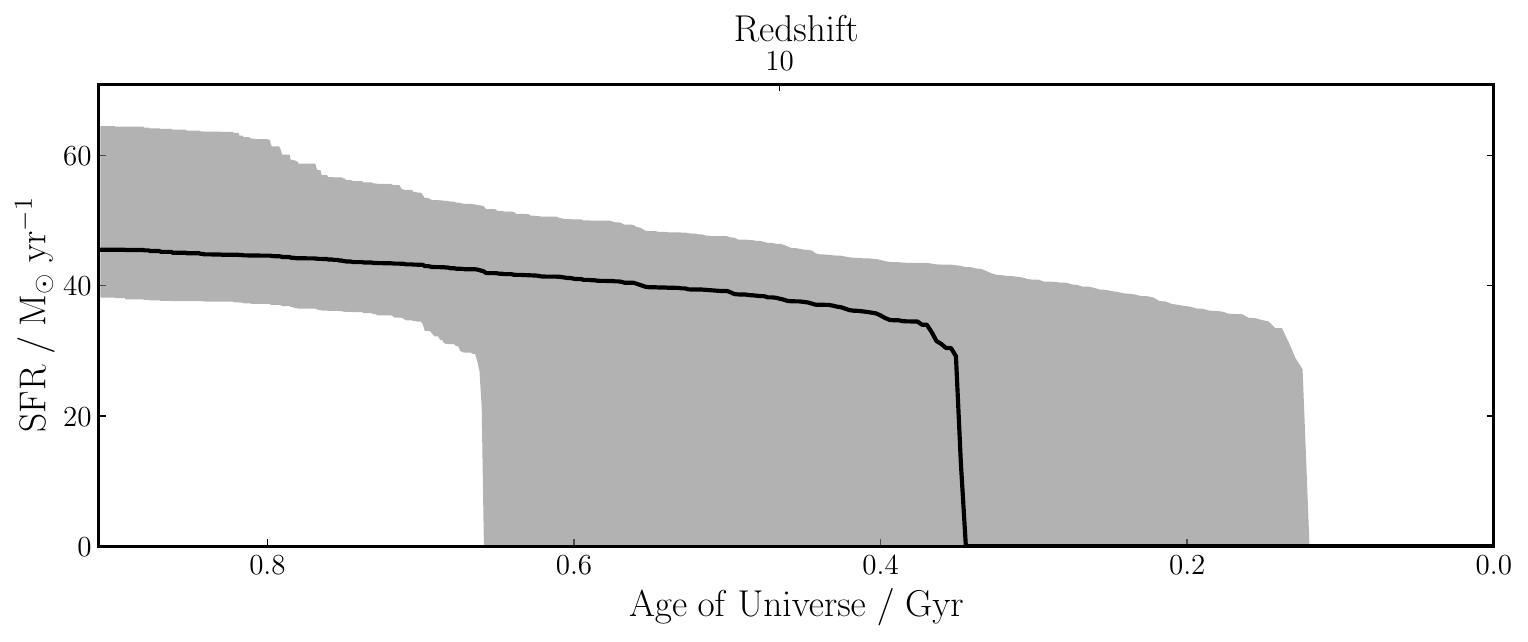}
         \centering
         \includegraphics[width=0.49\textwidth]{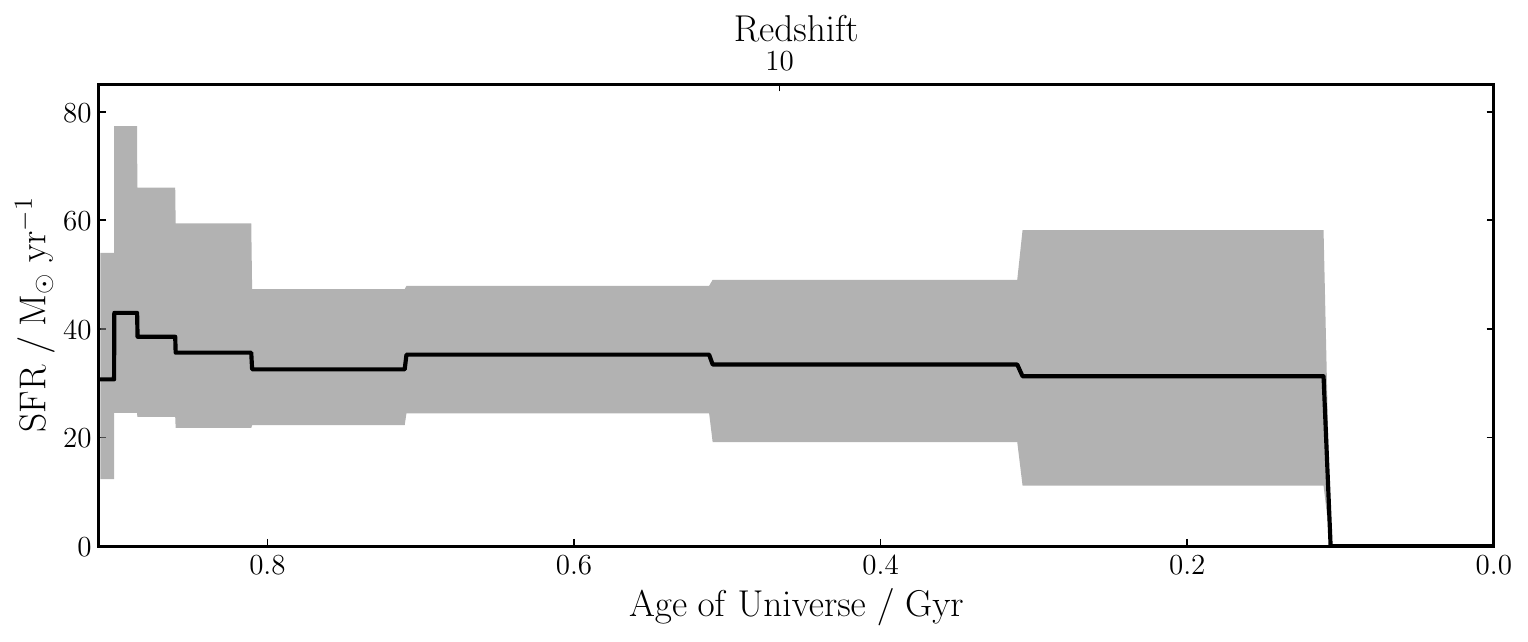}
         \caption{582186}
         \label{fig:582186}
     \end{subfigure}
     \begin{subfigure}{\textwidth}
     \centering
         \includegraphics[width=0.49\textwidth]{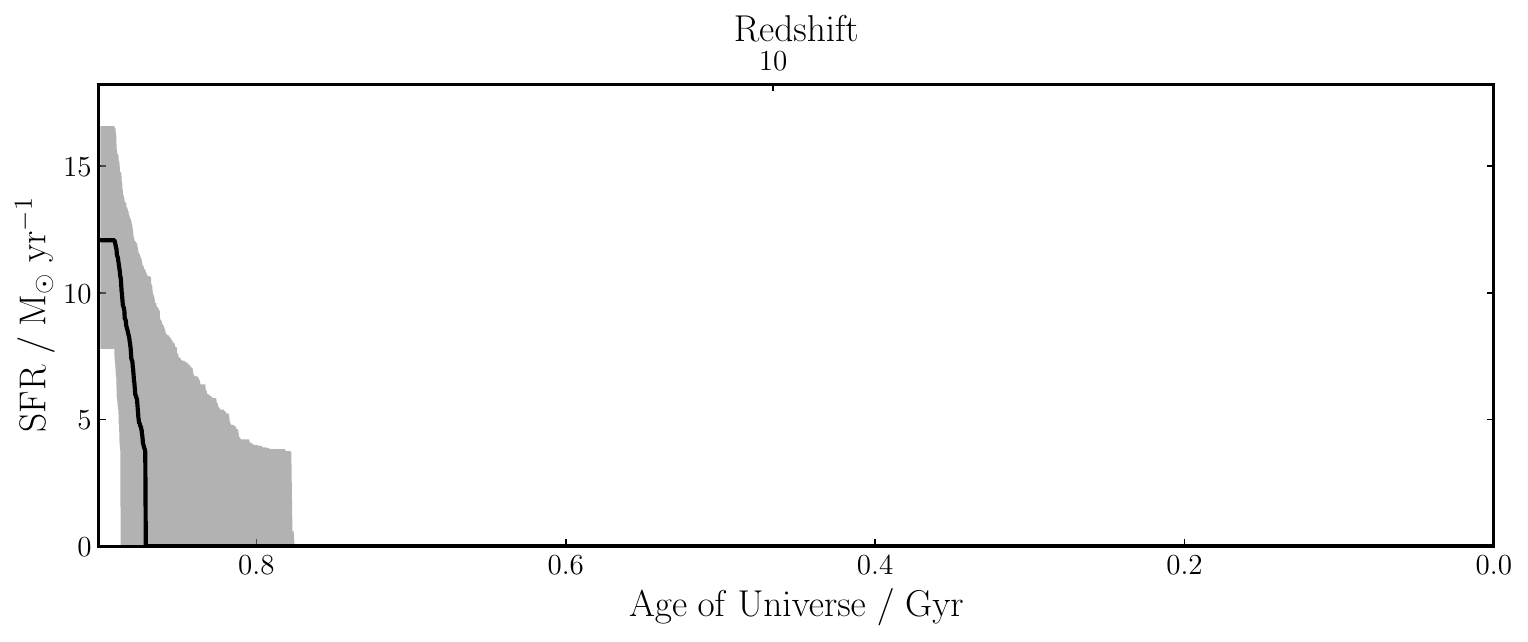}
         \centering
         \includegraphics[width=0.49\textwidth]{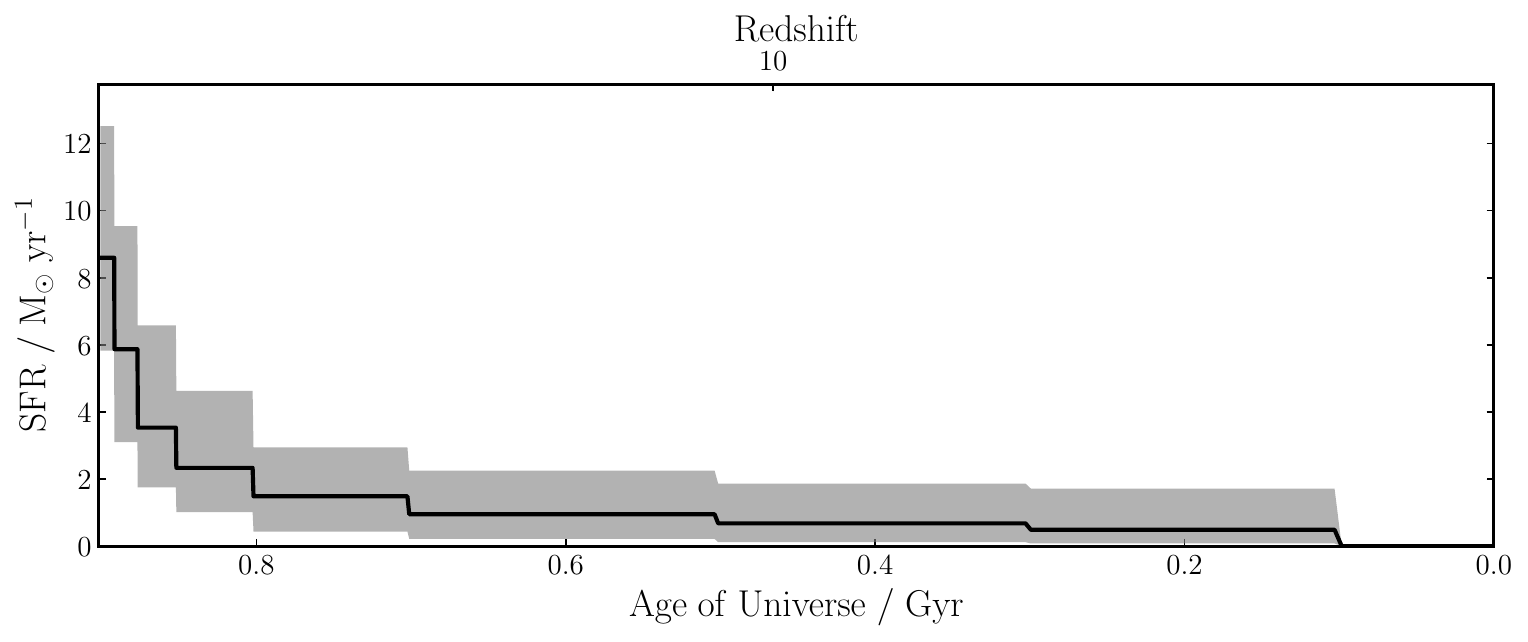}
         \caption{759747}
         \label{fig:759747}
     \end{subfigure}
        \caption{Star formation histories from {\sc Bagpipes} fitting to our protocluster galaxies. (\textbf{left}) a constant SFH model. (\textbf{right}) a non-parametric SFH model.}
        \label{fig:appendix 2 SFH comparison}
\end{figure*}
\section{{\sc Bagpipes} SED fit for protocluster galaxies with spec-z}\label{Bagpipes fits}
Figure \ref{fig:SED fit} shows the {\sc Bagpipes} fit for our 10 protocluster galaxies using a non-parametric SFH.
\begin{figure*}
     \centering
     \begin{subfigure}{\textwidth}
     \centering
         \includegraphics[width=\textwidth]{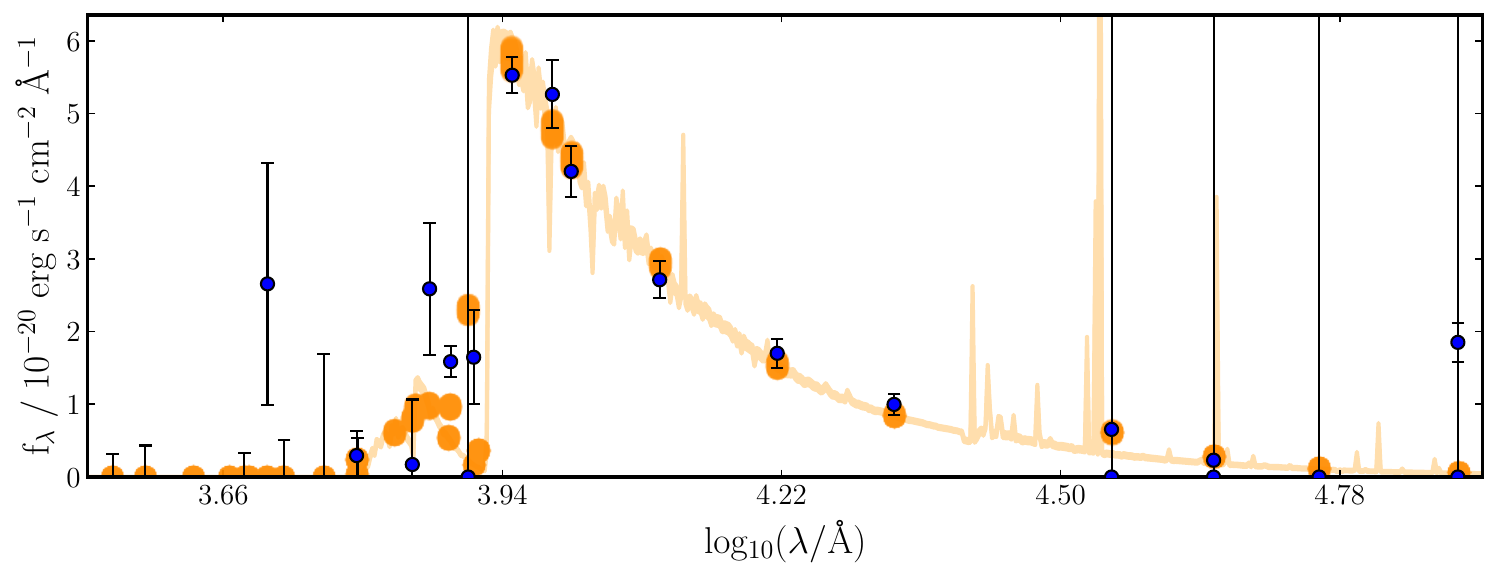}
         \caption{70505}
         \label{fig:70505}
     \end{subfigure}
     \centering
     \hfill
     \begin{subfigure}{\textwidth}
     \centering
         \includegraphics[width=\textwidth]{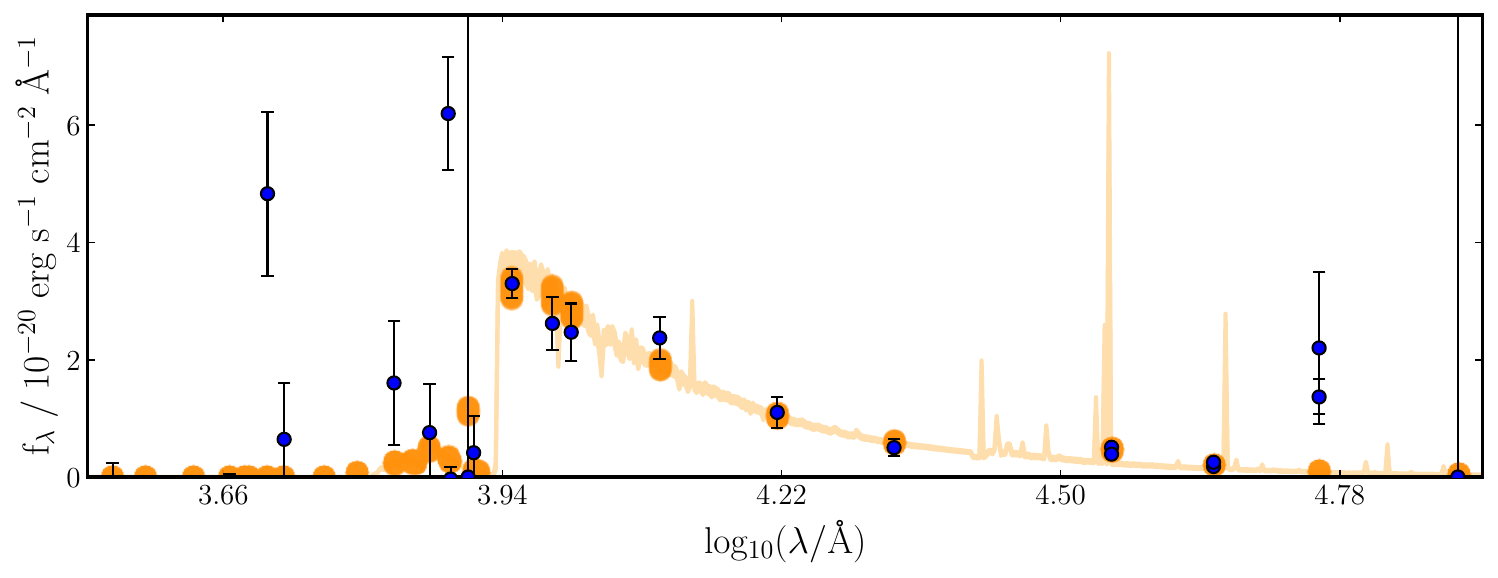}
         \caption{156780}
         \label{fig:156780}
     \end{subfigure}
     \begin{subfigure}{\textwidth}
     \centering
         \includegraphics[width=\textwidth]{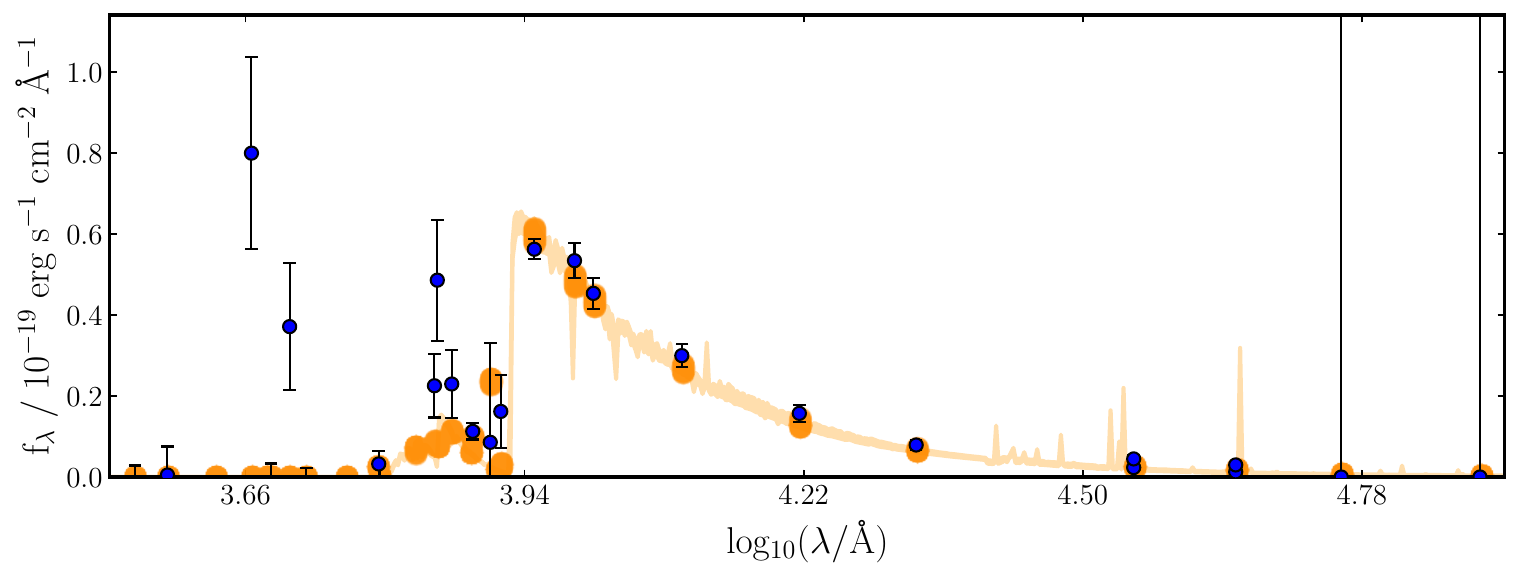}
         \caption{186512}
         \label{fig:186512}
     \end{subfigure}
        \caption{Continued on next page.}
\end{figure*}

\begin{figure*}\ContinuedFloat
     \begin{subfigure}{\textwidth}
     \centering
         \includegraphics[width=\textwidth]{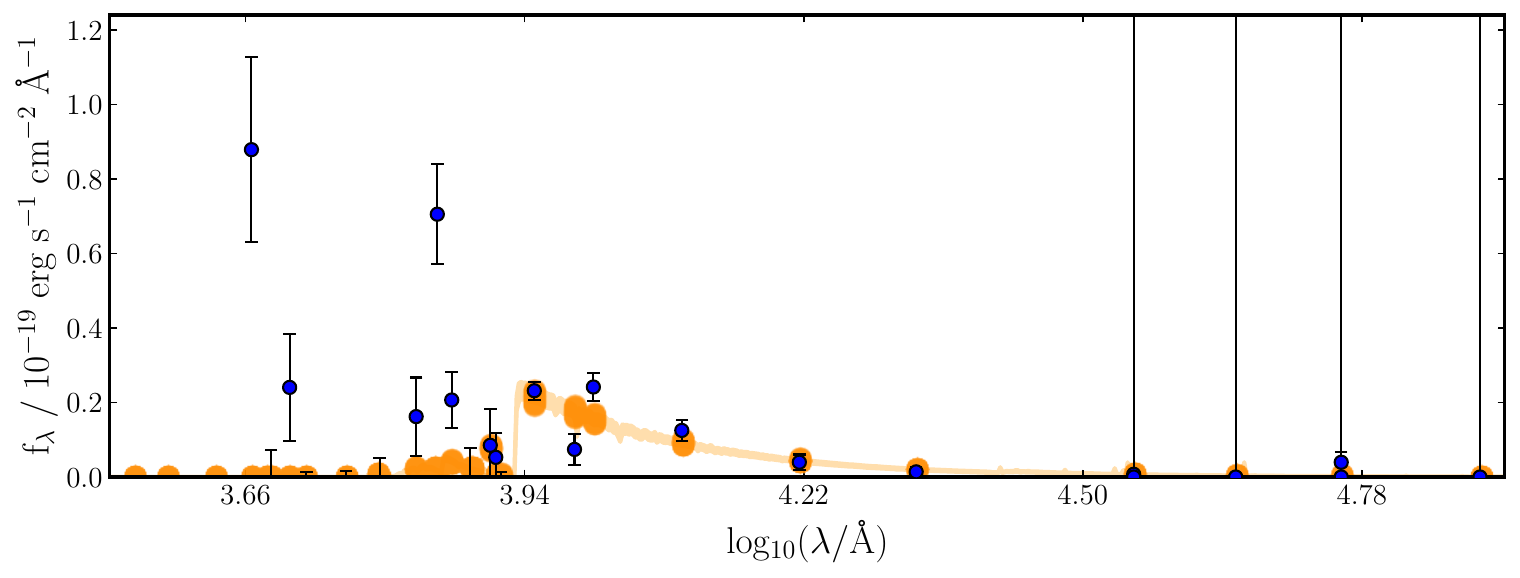}
         \caption{220530}
         \label{fig:220530}
     \end{subfigure}
     \centering
     \begin{subfigure}{\textwidth}
     \centering
         \includegraphics[width=\textwidth]{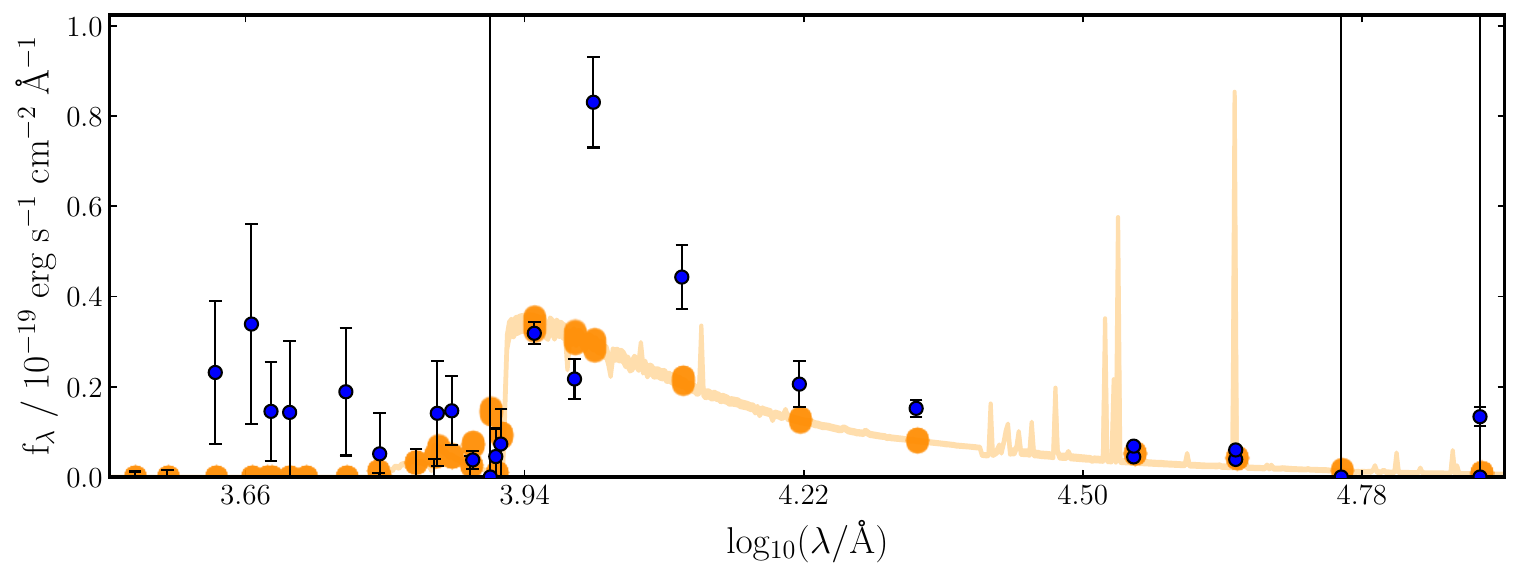}
         \caption{225263}
         \label{fig:225263}
     \end{subfigure}
     \begin{subfigure}{\textwidth}
     \centering
         \includegraphics[width=\textwidth]{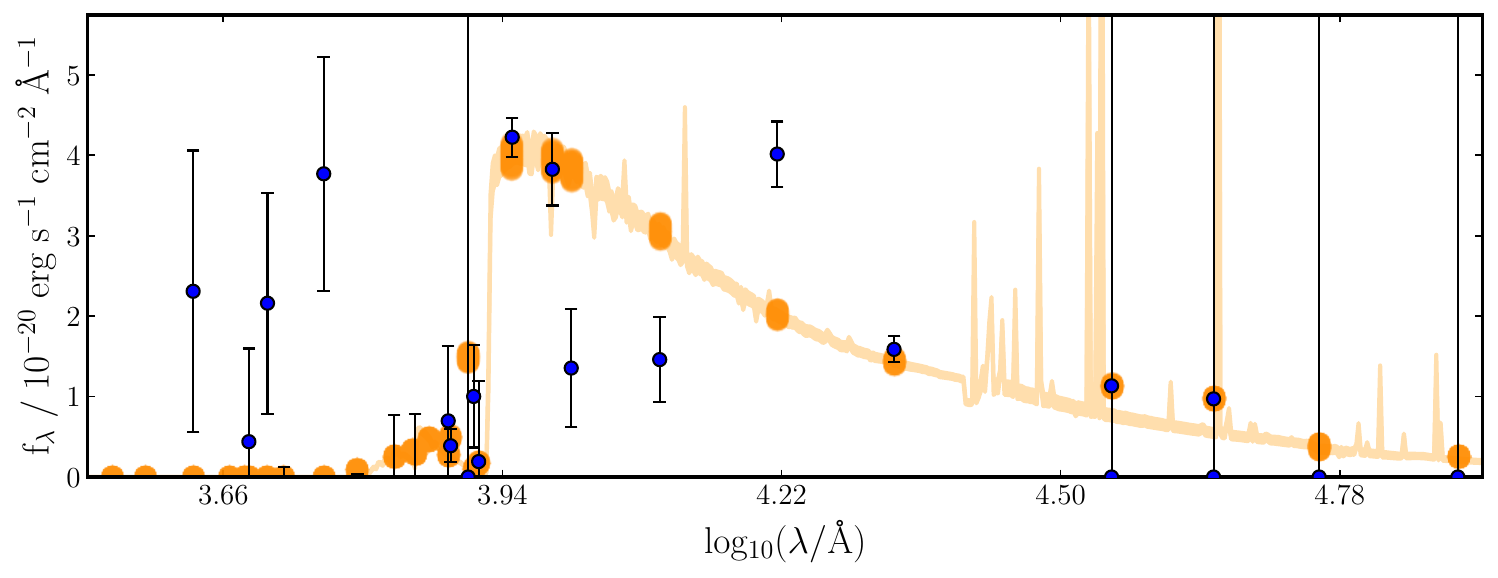}
         \caption{441761}
         \label{fig:441761}
     \end{subfigure}
        \caption{Continued on next page.}
\end{figure*}

\begin{figure*}\ContinuedFloat
     \begin{subfigure}{\textwidth}
     \centering
         \includegraphics[width=\textwidth]{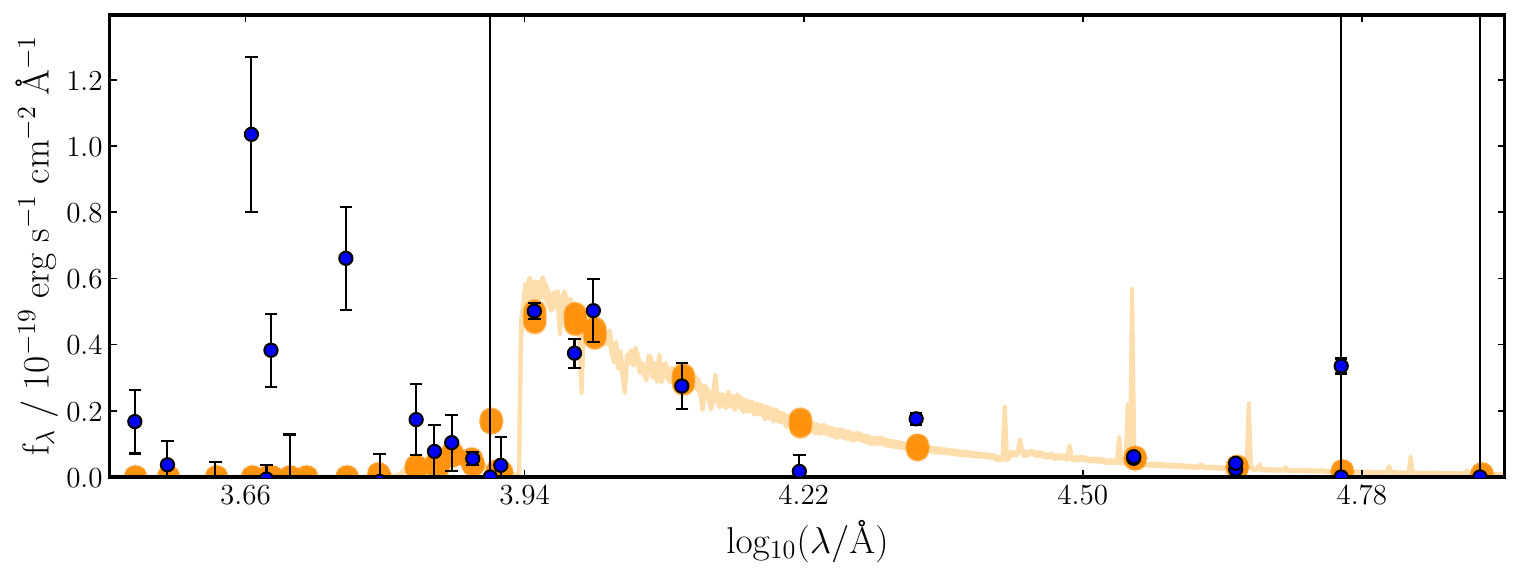}
         \caption{482804}
         \label{fig:482804}
     \end{subfigure}
     \centering
     \begin{subfigure}{\textwidth}
     \centering
         \includegraphics[width=\textwidth]{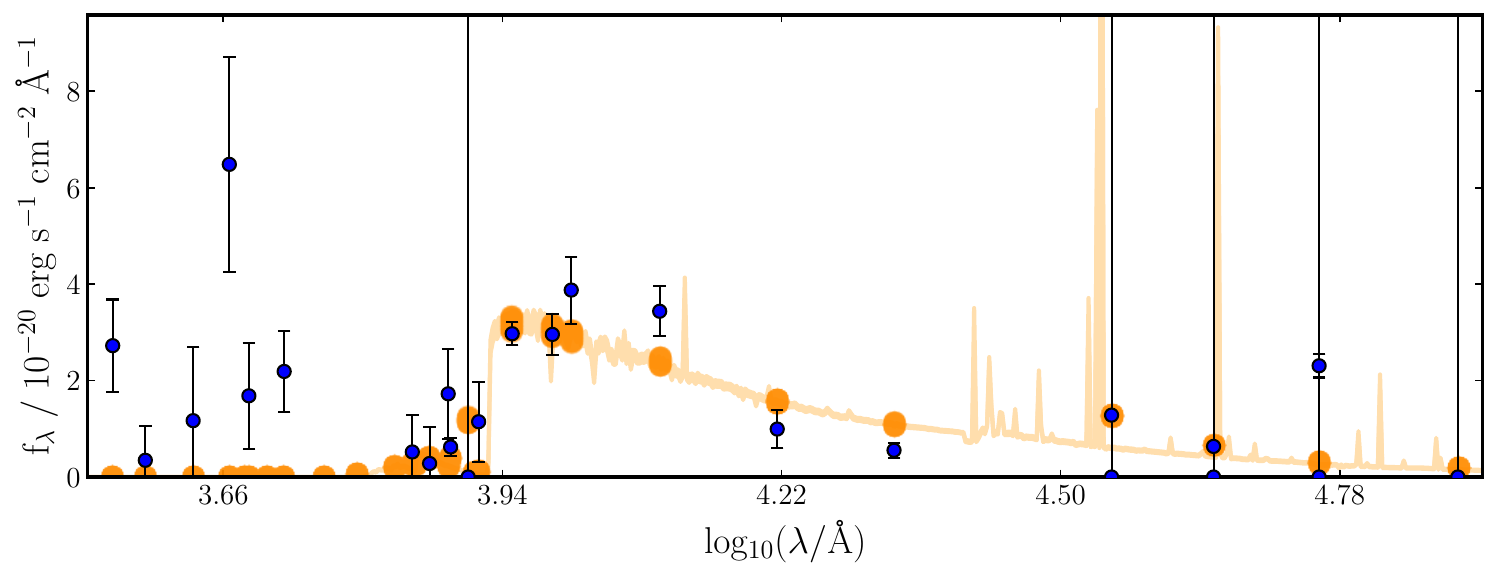}
         \caption{573604}
         \label{fig:573604}
     \end{subfigure}
     \begin{subfigure}{\textwidth}
     \centering
         \includegraphics[width=\textwidth]{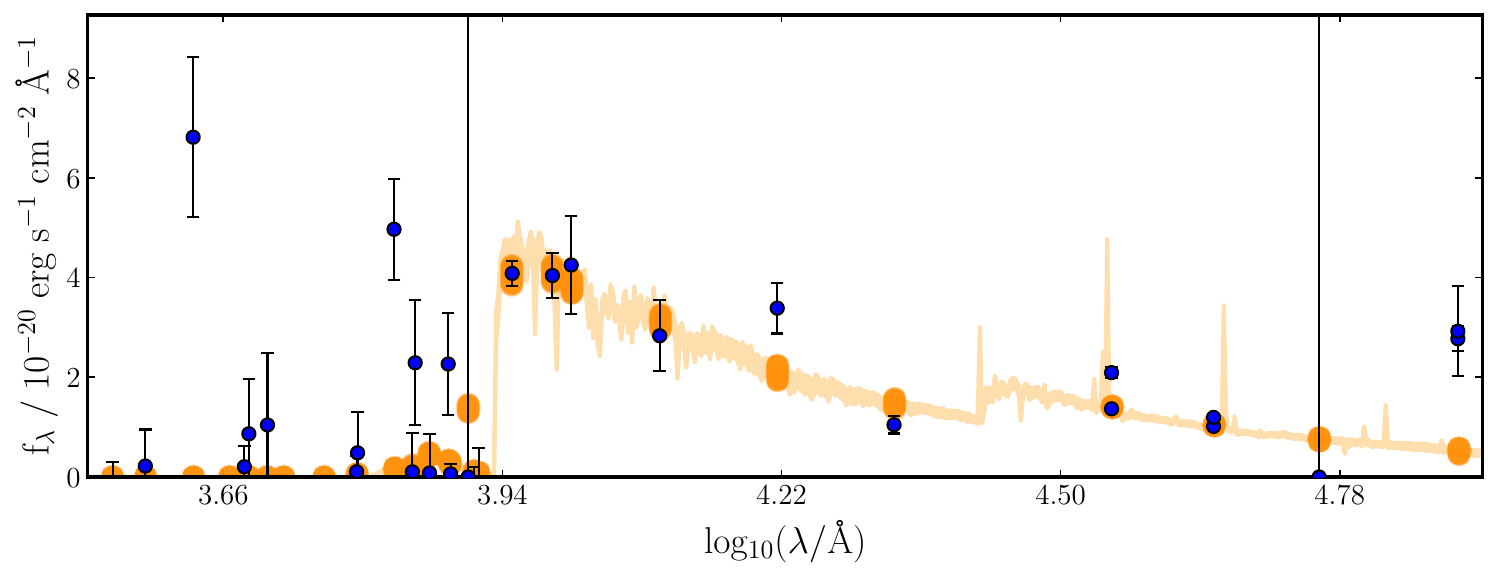}
         \caption{582186}
         \label{fig:582186}
     \end{subfigure}
        \caption{Continued on next page.}
\end{figure*}

\begin{figure*}\ContinuedFloat
     \begin{subfigure}{\textwidth}
     \centering
         \includegraphics[width=\textwidth]{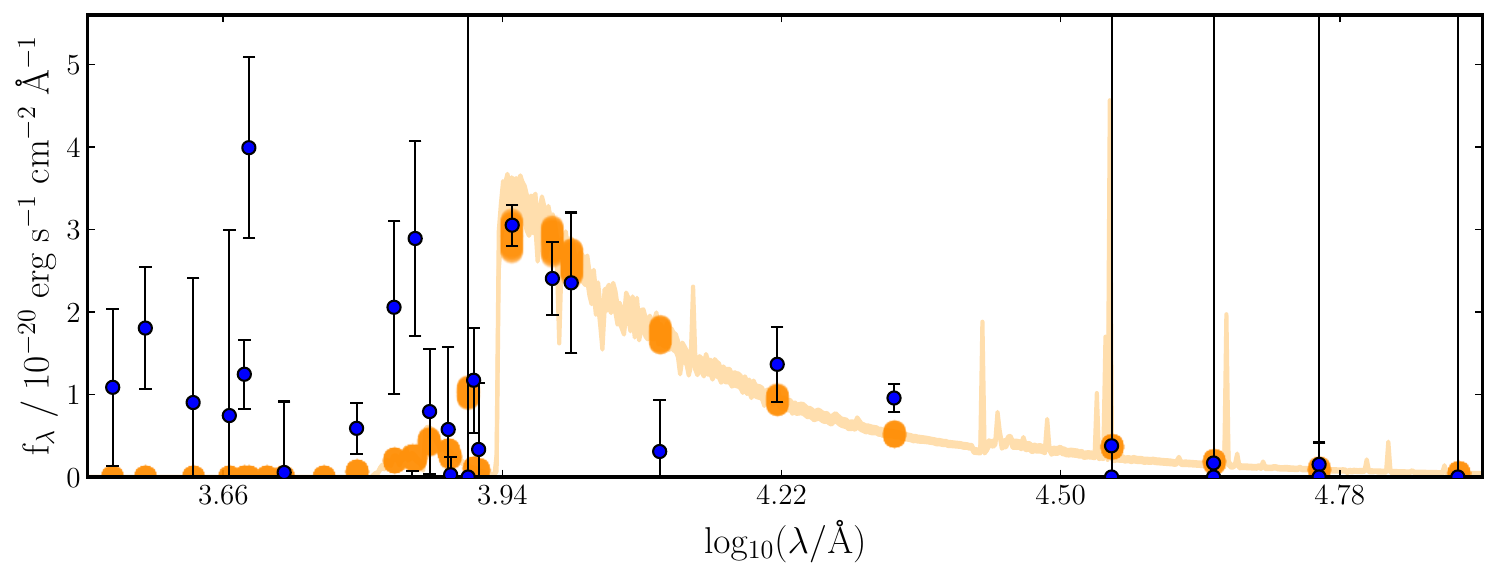}
         \caption{759747}
         \label{fig:759747}
     \end{subfigure}
        \caption{{\sc Bagpipes} SED fit to our protocluster galaxies. The large error bars on some data points are non-detections.}
        \label{fig:SED fit}
\end{figure*}

\bsp	
\label{lastpage}
\end{document}